\documentclass[
aps,showpacs,superscriptaddress,numerical,amsmath,amssymb,floatfix,reprint,longbibliography]{revtex4-2}
\usepackage[T1]{fontenc}
\usepackage[utf8]{inputenc}
\usepackage[
pdffitwindow=true,
colorlinks=true,
frenchlinks=false,
linkcolor=blue,
anchorcolor=blue,
citecolor=blue,
filecolor=blue,
urlcolor=blue,
bookmarks=true,
bookmarksopen=true,
bookmarksnumbered=true,
bookmarksopenlevel=1,
plainpages=false,
pdfpagelayout=TwoPageLeft,
pdfpagelabels=true,
breaklinks]{hyperref}
\usepackage[main=english]{babel}
\usepackage[dvipsnames]{xcolor}
\usepackage[per-mode=symbol,separate-uncertainty]{siunitx}
\usepackage{graphicx}
\usepackage{dcolumn}
\usepackage{color}
\usepackage{cleveref}
\usepackage{graphicx,wrapfig,lipsum}
\usepackage{bm}
\usepackage{physics}
\usepackage{color}
\usepackage[normalem]{ulem}
\usepackage{float}
\usepackage{chemformula}
\usepackage{todonotes}
\usepackage{mathtools}
\usepackage{bbold}
\usepackage{todonotes}
\usepackage{booktabs}  % For high-quality table rules
\usepackage{microtype} % Improves text justification and layout density
\usepackage{booktabs,tabularx,array,makecell}
\usepackage{tikz}
\usetikzlibrary{calc}
\usepackage{xspace}

\makeatletter
\def\switch@array{}%
\let\rt@orig@tabular\tabular
\let\rt@orig@endtabular\endtabular
\renewenvironment{ruledtabular}{%
  \def\tabular{\@ifnextchar[{\rt@tabular}{\rt@tabular[c]}}%
  \def\rt@tabular[##1]##2{%
    \def\tabular{\let\endtabular\rt@orig@endtabular\rt@orig@tabular}%
    \let\endtabular\rt@endtabular
    \csname tabular*\endcsname{\hsize}[##1]%
      {@{\hspace{\tabcolsep}\extracolsep{\fill}}##2}%
    \hline\hline\noalign{\vskip\doublerulesep}}%
  \def\rt@endtabular{\noalign{\vskip\doublerulesep}\hline\hline
    \rt@orig@endtabular}%
}{}
\makeatother
\usepackage{verbatim}
\newcommand{\TW}[1]{\textcolor{blue}{#1}}
\newcommand{\cut}[1]{\textcolor{red}{#1}}
\renewcommand{\arraystretch}{1.2}
\newcommand{\Ns}{N_\mathrm{S}}
\newcommand{\Nb}{N_\mathrm{B}}
\newcommand{\Cq}{C_\mathrm{q}}
\newcommand{\ar}{\hat{a}_\mathrm{R}}
\newcommand{\ai}{\hat{a}_\mathrm{I}}
\newcommand{\as}{\hat{a}_\mathrm{S}}
\newcommand{\Perr}{P_\mathrm{e}}
\newcommand{\Rq}{R_\mathrm{Q}}
\newcommand{\Rc}{R_\mathrm{C}}

\newcommand*{\cwo}{CaWO\textsubscript{4}\xspace}
\newcommand*{\hatH}{\hat{\mathcal{H}}}
\newcommand{\npk}[1]{[\textbf{\textcolor{blue}{NK: #1}}]}
\newcommand{\gwm}[1]{\textbf{\textcolor{purple}{GM: #1}}}
\DeclareMathOperator{\sgn}{sgn}

\newcommand*{\erion}{Er\textsuperscript{3+}\xspace} % Er3+ ion
\newcommand*{\eriononesixseven}{\textsuperscript{167}Er\textsuperscript{3+}\xspace} % 167Er3+ ion
\newcommand*{\ybion}{Yb\textsuperscript{3+}\xspace} % Yb3+ ion
\newcommand*{\ybiononesevenone}{\textsuperscript{171}Yb\textsuperscript{3+}\xspace} % 171Yb3+ ion
\newcommand*{\ybiononeseventhree}{\textsuperscript{173}Yb\textsuperscript{3+}\xspace} % 173Yb3+ ion

\Crefname{figure}{Fig.}{Figs.} % {<type>}{<singular>}{<plural>}
\Crefname{equation}{Eq.}{Eqs.} % {<type>}{<singular>}{<plural>}

\begin{document}

%\title{Direct observation of high-rank hyperfine interactions of erbium and ytterbium in \cwo and their implications on spin coherence
\title{Observation of high-rank hyperfine interactions of rare earths in the solid state
%Direct observation of high-rank hyperfine interactions of erbium and ytterbium in \cwo
%Old: Hyperfine and quadrupole interactions of \erion and \ybion in \cwo
}
\author{Georg~W.~Mair}
%\email[]{Georg.Mair@wmi.badw.de}
\affiliation{Walther-Mei{\ss}ner-Institut, Bayerische Akademie der Wissenschaften, 85748 Garching, Germany} 
\affiliation{School of Natural Sciences, Technical University of Munich, 85748 Garching, Germany}

\author{Naoya~Iwahara}
\affiliation{Graduate School of Engineering, Chiba University, 1-33 Yayoi-cho, Inage-ku, Chiba-shi, Chiba 263-8522, Japan}

\author{Thierry~Chaneli\`{e}re}
\affiliation{Univ. Grenoble Alpes, CNRS, Grenoble INP, Institut N\'eel, 38000 Grenoble, France}

\author{Ana~Strini\'{c}}
\affiliation{Walther-Mei{\ss}ner-Institut, Bayerische Akademie der Wissenschaften, 85748 Garching, Germany}

\author{Kirill~G.~Fedorov}
\affiliation{Walther-Mei{\ss}ner-Institut, Bayerische Akademie der Wissenschaften, 85748 Garching, Germany}
\affiliation{School of Natural Sciences, Technical University of Munich, 85748 Garching, Germany}
\affiliation{Munich Center for Quantum Science and Technology (MCQST), 80799 Munich, Germany}

\author{Hans~Huebl}
\affiliation{Walther-Mei{\ss}ner-Institut, Bayerische Akademie der Wissenschaften, 85748 Garching, Germany}
\affiliation{School of Natural Sciences, Technical University of Munich, 85748 Garching, Germany}
\affiliation{Munich Center for Quantum Science and Technology (MCQST), 80799 Munich, Germany}

\author{Rudolf~Gross}
\affiliation{Walther-Mei{\ss}ner-Institut, Bayerische Akademie der Wissenschaften, 85748 Garching, Germany}
\affiliation{School of Natural Sciences, Technical University of Munich, 85748 Garching, Germany}
\affiliation{Munich Center for Quantum Science and Technology (MCQST), 80799 Munich, Germany}
\affiliation{Senior Excellence Faculty, Technical University of Munich, 85748 Garching, Germany} 

\author{Nadezhda~P.~Kukharchyk}
\email[]{Nadezhda.Kukharchyk@wmi.badw.de}
\affiliation{Walther-Mei{\ss}ner-Institut, Bayerische Akademie der Wissenschaften, 85748 Garching, Germany}
\affiliation{School of Natural Sciences, Technical University of Munich, 85748 Garching, Germany}
\affiliation{Munich Center for Quantum Science and Technology (MCQST), 80799 Munich, Germany}

\date{\today}
\pacs{}
\keywords{} 

\begin{abstract}

%Rare-earth ions in doped solids are promising platforms for hybrid quantum technologies in the microwave regime, where performance is strongly interlinked with the properties of spin transitions governed by the microscopic Hamiltonian. We used broadband electron paramagnetic resonance spectroscopy to probe ground-state hyperfine transitions of \textsuperscript{167}Er\textsuperscript{3+}, \textsuperscript{171}Yb\textsuperscript{3+}, and \textsuperscript{173}Yb\textsuperscript{3+} dopants in \cwo. We determined the ground-state Hamiltonian parameters, including the first direct measurement of the \textsuperscript{173}Yb\textsuperscript{3+} quadrupole interaction in \cwo. At zero magnetic field, we uncovered avoided crossings that reveal interactions beyond the conventional spin Hamiltonian, arising from a pseudospin-nuclear octupole interaction and local symmetry-breaking strain. To our knowledge, this constitutes the first direct observation of a rare-earth octupole interaction in a solid-state system. Remarkably, these interactions also generate new zero-field ZEFOZ transitions with substantially reduced magnetic-field sensitivity compared with those predicted by the conventional Hamiltonian used in literature. Our results demonstrate that low-field broadband EPR can reveal otherwise hidden higher-rank interactions while identifying new operating regimes for rare-earth-based quantum technologies.

Rare-earth ions in solids are widely employed as quantum memories and microwave-to-optical transducers. Their performance depends on the precise knowledge of the microscopic spin Hamiltonian and the existence of transitions with reduced magnetic field sensitivity. While conventional effective-spin descriptions are typically limited to Zeeman, hyperfine, and quadrupole interactions, weak higher-rank terms remain largely unexplored. Here we perform broadband electron paramagnetic resonance spectroscopy (EPR) of \eriononesixseven, \ybiononesevenone, and \ybiononeseventhree ions in \cwo at millikelvin temperatures and magnetic fields below \SI{50}{\milli\tesla}. We determine the ground-state hyperfine Hamiltonian parameters, including the first direct measurement of the quadrupole interaction of \ybiononeseventhree in \cwo. High-resolution spectra provide the first direct observation of higher-rank octupole and hexadecapole hyperfine interactions of a rare-earth ion in a solid-state environment. Along with local lattice deformations, these interactions generate previously unknown zero first-order Zeeman (ZEFOZ) transitions in zero magnetic field. We observe a more than 10-fold longer coherence time than that of the previously predicted ZEFOZ transition. Our results establish higher-rank interactions as a relevant resource for engineering protected spin transitions and demonstrate low-field broadband EPR as a powerful tool for revealing weak interactions hidden in conventional spectroscopy.

%More importantly, high-resolution spectra reveal reproducible zero-field avoided crossings that cannot be explained by the conventional spin Hamiltonian. By combining symmetry analysis, isotope dependence, and quantitative modeling, we show that the splittings of the $m_F=\pm2$ states originate from a pseudospin-nuclear octupole interaction, whereas avoided crossings involving $m_F=\pm1$ states are governed by strain-induced symmetry lowering. The higher-rank interaction produces zero-field splittings on the order of tens of kilohertz and, to our knowledge, constitutes the first direct observation of an octupole interaction of a rare-earth ion in a solid-state environment. Remarkably, both octupole and strain-induced interactions generate previously unknown zero-field ZEFOZ transitions. Coherence measurements confirm substantially enhanced coherence times at these operating points compared with the ZEFOZ transition predicted by the conventional Hamiltonian. Our results establish higher-rank interactions as a relevant resource for engineering protected spin transitions and demonstrate low-field broadband EPR as a powerful tool for revealing weak interactions hidden in conventional spectroscopy.

\end{abstract}

\maketitle

%__________________________________________________________________

\section{Introduction}

\begin{figure*}[th!]
% \\badwwmi\user\groups\Quantum-Speicher\Manuscripts\GMair_CWO_ham\Figures\Multipoles_Fig1_v2\Multipoles.png
    \includegraphics[width=1\textwidth]{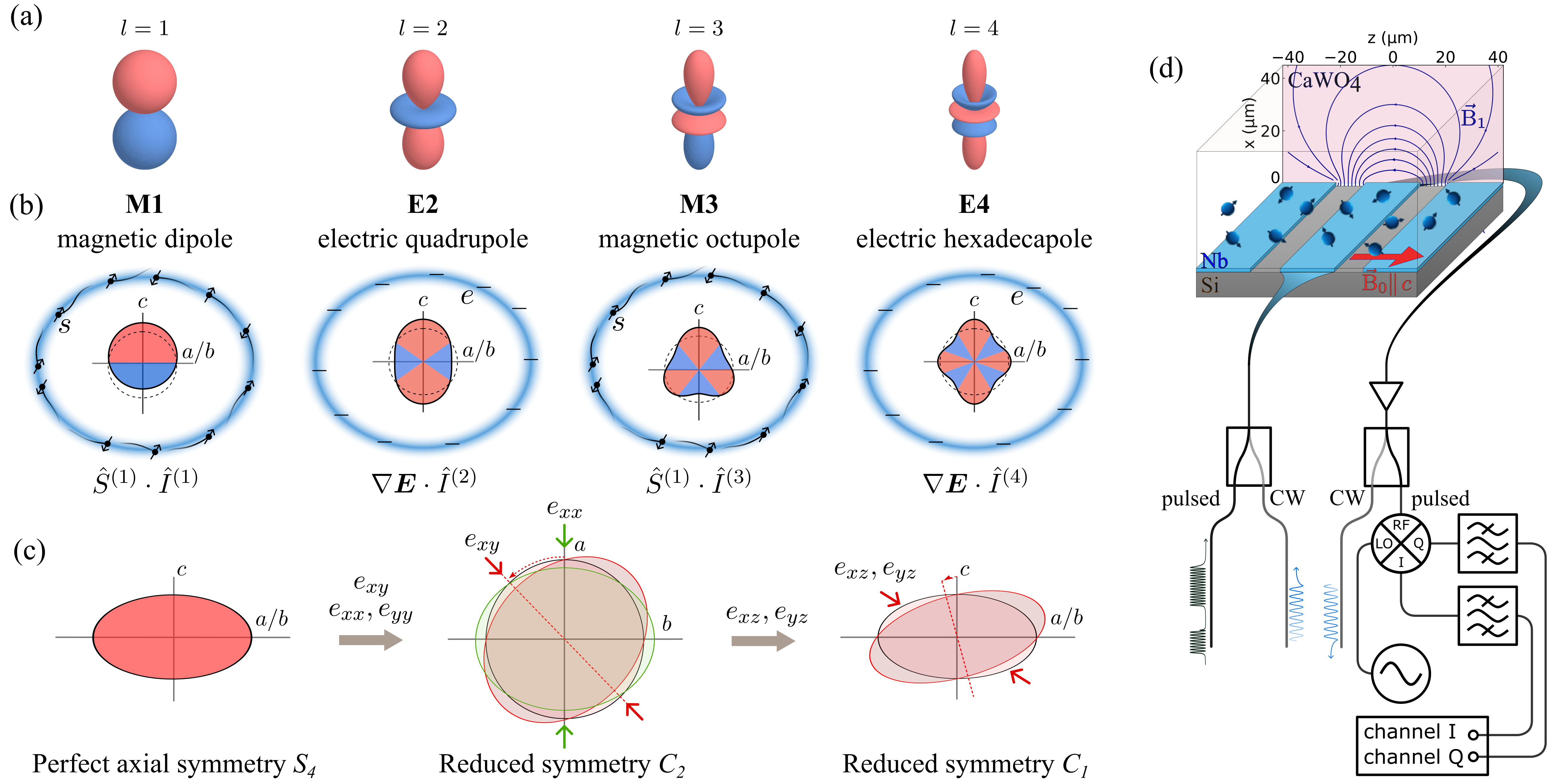}
    \caption{
    (a)~Visual representations of real-valued spherical harmonic functions $Y_{l m}(\theta,\phi)$ for $l=1,2,3$ and $4$ and $m = 0$. Red and blue lobes correspond to the regions with positive and negative function values, respectively.
    (b)~Schematic picture of higher-rank moments of a nucleus within a paramagnetic ion. Dashed circles depict a non-deformed spherically symmetric nucleus with radius $R_0$. Colored shapes correspond to the deformation of the nucleus in the respective rank, $R(\theta,\phi)=R_0(1+\sum_{l m} \alpha_{l m} Y_{l m}(\theta,\phi))$~\cite{Bohr.1953,Verney.2025}.
    (c)~Axially symmetric dipole magnetic moment undergoes deformations by random strain of the host crystal, which leads to a microscopic reduction of symmetry. While normal strain within the symmetry plane $xy$ preserves the symmetry axis $z$, shear strains in planes $xz$ and $yz$ lead to the complete removal of symmetry, creating the lowest symmetry point group $C_1$. Tetragonal axial strain along the $z$-axis is not shown, as it does not change the symmetry group.  
    (d)~Schematics of the experimental setup. The \cwo crystal is placed on top of a superconducting coplanar waveguide with an external static magnetic field applied in-plane to the superconducting thin film. For spectroscopy, the absorption of a continuous wave (CW) broadband signal through the transmission line is detected with a vector network analyzer. For time-domain measurements, a shaped pulse sequence is sent to the sample. The output signal undergoes frequency down-conversion and is captured by $I$ and $Q$ quadrature components using an analog-to-digital converter. 
    }
    \label{fig:intro}
\end{figure*}

Electron paramagnetic resonance (EPR) provides a sensitive probe of magnetic properties and local environments of unpaired electrons. In a paramagnetic solid, the electronic degrees of freedom are coupled to nuclei, neighboring spins, and the crystalline environment. Consequently, EPR can reveal not only the electronic Zeeman response, but also nuclear structure, electron-nuclear interactions, local symmetry, and lattice-induced perturbations~\cite{AbragamBleaney.1970,Schweiger.2001}.
%Among these contributions, the interaction between electronic and nuclear magnetic moments within the same ion is typically among the most important. 
Among these contributions, the interaction between electronic and nuclear magnetic moments, superimposed by crystal-field symmetry, is the primary EPR probe. 
To first order, the presence of a nuclear magnetic moment gives rise to a conventional hyperfine structure~\cite{AbragamBleaney.1970,Bleaney.1963}. For nuclei with sufficiently large spin $I$, additional higher-rank electric and magnetic multipole corrections are allowed, such as the electric quadrupole for $I\geq 1$, the magnetic octupole for $I\geq \sfrac{3}{2}$, and the electric hexadecapole for $I\geq 2$~\cite{Bohr.1953,Garrett.2022,Baker.1969,Baker.1970}. These higher ranks are associated with non-spherical charge distribution within the nucleus, which is commonly expressed through spherical harmonics and denoted as nuclear shapes in the generalized model of the nucleus~\cite{Bohr.1953,Garrett.2022,Verney.2025}, as illustrated schematically in \Cref{fig:intro}(a,b). 

To date, considerable theoretical and experimental effort has been devoted to determining higher-rank nuclear moments, particularly in atomic and trapped-ion systems~\cite{Lewty.2012,Beloy.2008a,Beloy.2008b,Gerginov.2003,Jaccarino.1954}. In such systems, precision spectroscopy can provide access to nuclear multipole moments through their coupling to well-characterized electronic states. For ions placed in a crystalline environment, the situation is more complex. Measured coefficients of various interaction terms depend not only on intrinsic nuclear properties and electronic wave functions, but also on the strength and symmetry of lattice-induced electric-field gradients. 
Moreover, assisted by the crystal-field interaction, second-order combinations of lower-rank interactions produce pseudo-multipolar interaction terms, which are experimentally indistinguishable from genuine higher-rank terms~\cite{Gupta.1973,Erickson.1986,Sternheimer.1986,Schwartz.1955}. Therefore, the spectroscopy of ions in a solid-state environment only allows for determining effective multipole couplings, whose microscopic origin must be assessed using symmetry, perturbative scaling, isotope dependence, and comparison with complementary measurements.
They manifest in the form of modified energy-level splittings and the lifting of degeneracies, which become significant in the low magnetic field range~\cite{Baker.1969,Baker.1970,Schwartz.1955}.

%Until today, considerable theoretical and experimental effort has been devoted to determining higher-order nuclear moments, particularly in atomic and trapped-ion systems~\cite{Lewty.2012,Beloy.2008a,Beloy.2008b,Gerginov.2003,Jaccarino.1954}. In such systems, the precision spectroscopy can provide access to nuclear multipole moments through their coupling to well-characterized electronic states. For ions placed in a crystalline environment, the situation is more complex. Measured coefficients of various interaction terms depend not only on intrinsic nuclear properties, but also on the electronic wave functions, crystal-field admixture, local symmetry, and lattice-induced electric-field gradients. Moreover, second-order combinations of lower-rank interactions may transform in the same way as genuine higher-rank terms. Therefore, the spectroscopy of ions in a solid state environment only allows for determining effective multipole couplings, whose microscopic origin must be assessed using symmetry, perturbative scaling, isotope dependence, and comparison with complementary measurements.

Crystals doped with rare-earth ions (REIs) are particularly suitable for investigation of these interactions. The partially filled $4f$ shell provides large total angular momentum and strong electron-nuclear interactions, while its shielding by the filled $5s$ and $5p$ shells mitigates, but does not eliminate, coupling to the crystalline environment. REIs can consequently exhibit long-lived electronic and hyperfine states together with narrow optical transitions~\cite{Kunkel.2018}. These properties have motivated their use in microwave and optical quantum memories~\cite{Afzelius.2013,Probst.2015} and in microwave-to-optical transduction schemes~\cite{Zhou.2023,Rochman.2023}. For these applications, useful operating points are frequently found at low magnetic fields~\cite{Rakonjac.2020,Strinic.2025}, where electron and nuclear spin states are strongly mixed and selected hyperfine transitions can exhibit vanishing first-order sensitivity to magnetic-field fluctuations~\cite{Longdell.2006,McAuslan.2012}. An accurate modeling of this regime requires consideration of weak high-rank interactions.

\cwo is an appealing host for REIs due to the relatively low abundance of magnetic nuclei present in the host crystal, suppressing an important source of magnetic decoherence~\cite{LeDantec.2021,An.2025}, while the substitutional incorporation of trivalent REIs provides electron-nuclear hyperfine manifolds with transition frequencies in the microwave-frequency range. Erbium ions are especially attractive dopants because of their optical transition in the telecommunications wavelength band~\cite{Saglamyurek.2015,Ourari.2023}. Ytterbium has likewise received substantial attention, with \ybiononesevenone exhibiting exceptionally long electron-nuclear spin coherence times in suitable host crystals~\cite{Ortu.2018,Tiranov.2026}. \erion and \ybion in \cwo are promising systems for quantum memory and hybrid quantum devices including applications in quantum microwave communication~\cite{Fesquet.2024, Yam.2026}.

Previous optical and high-field EPR studies established the dominant electronic Zeeman and magnetic dipole hyperfine interactions of REIs in \cwo~\cite{Antipin.1968,Sattler.1970a}. Interactions beyond these leading terms, however, have hardly been studied. The quadrupole tensor of \eriononesixseven:\cwo was determined only recently~\cite{Marsh.2026}, whereas previous investigations of \ybiononeseventhree:\cwo provided only an upper bound on the quadrupole coupling coefficient~\cite{Ranon.1964,Sattler.1970a}. However, systematic investigation of higher-rank multipolar interactions and weak symmetry-breaking terms in these systems has been lacking, while it remains an essential ingredient for identification of optimal working points with long coherence times, so-called zero first-order Zeeman (ZEFOZ) or clock transitions.

This limited knowledge is partly a consequence of the spectroscopic regimes used in earlier measurements. In optical spectroscopy, inhomogeneous broadening and excited-state linewidths can prevent a complete resolution of the hyperfine structure. Optical transitions also connect hyperfine manifolds belonging to both the electronic ground and excited states, complicating the unambiguous extraction of ground-state Hamiltonian parameters. 

Conventional EPR is commonly performed at comparatively high frequencies and magnetic fields, where the electronic Zeeman interaction dominates. Higher-rank interactions are generally much weaker and are particularly difficult to identify and quantify unambiguously using conventional single-frequency EPR spectroscopy in high fields. In contrast, broadband EPR techniques provide a fine-grained understanding of all magnetic resonance transition frequencies at any selected magnetic field~\cite{Strinic.2025, Strinic.2026, Clauss.2013}. Such flexibility grants access to the unique fingerprints of weak interactions~\cite{Travesedo.2025}, which are especially pronounced in near-zero magnetic fields due to strong electron-nuclear state mixing. %Near zero magnetic field, strong electron-nuclear state mixing produces transitions with substantial changes in their electronic and nuclear character due to domination of field-independent higher-rank interactions and allow to take advantage of their presence\gwm{Reformulate this sentence}.

%Broadband low-field EPR spectroscopy offers a complementary approach~\cite{Clauss.2013,Strinic.2025}. 
%Near zero magnetic field, strong electron-nuclear state mixing produces transitions with substantial changes in their electronic and nuclear character. %Weak interactions can then determine splittings of otherwise degenerate or nearly degenerate states, open avoided crossings, and modify transition frequencies and intensities~\cite{Strinic.2025,Marsh.2026}. 
%High-resolution measurements in this regime are particularly sensitive to quadrupolar and higher-order terms. %At the same time, because different microscopic interactions can generate second-order effective operators of the same tensor rank, assigning a spectral feature to an intrinsic nuclear multipole requires caution. Symmetry analysis, isotope dependence, perturbative estimates, and the reproducibility of the spectral features must all be considered.

In this work, we deploy broadband EPR spectroscopy to investigate the low-field hyperfine structure within the lowest Stark doublet of the electronic ground state for the three nuclear-spin-carrying isotopes \eriononesixseven, \ybiononesevenone, and \ybiononeseventhree in \cwo at millikelvin temperatures using superconducting coplanar waveguides to probe hyperfine transitions directly.
%By comparing experimentally measured transition maps with spin-Hamiltonian calculations, we determine the magnetic-dipole hyperfine and nuclear quadrupole parameters and assess the contributions of additional weak interactions.
We show that the hyperfine quadrupole, octupole and hexadecapole interactions are spectroscopically significant for \eriononesixseven:\cwo and \ybiononeseventhree:\cwo in the low-field regime. In addition, we analyze strain-induced symmetry lowering, as evidenced by line broadening and avoided crossings in the vicinity of zero magnetic field.
%We further observe reproducible avoided crossings near zero magnetic field that are not captured by the conventional effective-spin Hamiltonian containing only electronic Zeeman, magnetic-dipole hyperfine, and nuclear quadrupole terms. We analyze these features using symmetry-allowed higher-rank effective operators and strain-induced symmetry lowering, while distinguishing phenomenological spectral assignments from claims about intrinsic nuclear moments. \gwm{Unclear what you mean.} 
Thus, we formulate the extended Hamiltonian, which provides an improved description of the measured spectra and reveals previously unexplored ZEFOZ transitions with enhanced coherence properties. %These results establish low-field broadband EPR as a sensitive method for resolving weak interactions in rare-earth electron-nuclear spin systems and provide parameters relevant to the design of rare-earth-based quantum memories and hybrid quantum devices.
% \gwm{Can we join the last two paragraphs? I would also like to directly mention octupole and hexadecapole interactions here.}

\section{Spin system} \label{sec:theory}

% \subsection{Rare-earth ions in \CWO}

The investigated spin systems are diluted ensembles ($\qtyrange{34}{50}{ppm}$) of \erion ([Xe]\,4$f^{11}$) and \ybion ([Xe]\,4$f^{13}$) ions embedded in a tetragonal \cwo crystalline matrix, where REIs are reported to occupy Ca\textsuperscript{2+} sites of local $S_4$ symmetry~\cite{Mims.1965, Enrique.1971}. The substitution of Ca\textsuperscript{2+} by a trivalent ion requires charge compensation~\cite{Nassau.1963} and can result in the formation of additional low-symmetry REI sites and inhomogeneous broadening~\cite{Ranon.1964,Tiranov.2026,Billaud.2025}. Owing to the odd number of $4f$ electrons, the spin-orbit coupled ground states of \erion (\textsuperscript{4}I\textsubscript{15/2}) and \ybion (\textsuperscript{2}F\textsubscript{7/2}) are split by the crystal field into 8 and 4 Kramers doublets, respectively, which are degenerate in the absence of external magnetic fields and separated by $\sim\SI{600}{\giga\hertz}$~\cite{Becker.2025} and $\sim\SI{6.6}{\tera\hertz}$~\cite{Zheng.2004}, respectively. At sub-Kelvin temperatures, only the lowest Kramers doublet is populated, forming an effective spin-$\sfrac{1}{2}$ system. For the isotopes with intrinsic nuclear spin, $I=\sfrac{7}{2}$ for the \eriononesixseven isotope, and $I=\sfrac{1}{2}$ and $I=\sfrac{5}{2}$ for \ybiononesevenone and \ybiononeseventhree isotopes, respectively, an additional hyperfine structure appears.

\subsection{Conventional spin Hamiltonian in axial symmetry}

% Effective spin Hamiltonian
Spin systems at millikelvin temperatures can be well described by an effective spin $S=\sfrac{1}{2}$ Hamiltonian~\cite{AbragamBleaney.1970,Rudowicz.2015,Rudowicz.2001,Mostafanejad.2014}. For diluted spin ensembles, it commonly contains contributions from the electronic Zeeman interaction $\hatH_\mathrm{eZ}$, the nuclear Zeeman term $\hatH_\mathrm{nZ}$, the hyperfine interaction $\hatH_\mathrm{hf}$ for nuclear spin $I \ne 0$, and the electric quadrupole interaction $\hatH_\mathrm{q}$ for nuclei with spin $I > \sfrac{1}{2}$ arising from the coupling between the nuclear quadrupole moment and the local electric field gradient:
\begin{equation}
\begin{split}
    \hatH_\mathrm{eff,0} = & \hatH_\mathrm{eZ} + \hatH_\mathrm{hf} + \hatH_\mathrm{q} + \hatH_\mathrm{nZ} \\
    = & \mu_\mathrm{B}\bm{B}_0 \cdot \boldsymbol{g} \cdot \hat{\bm{S}}
    + \hat{\bm{S}} \cdot \boldsymbol{A} \cdot \hat{\bm{I}} \\
    &+ P_\parallel\left(\hat{I}_z^2 - \tfrac{1}{3}\hat{\bm{I}}^2\right)
    - \mu_\mathrm{N}g_\mathrm{N}\bm{B}_0 \cdot \hat{\bm{I}}.
\end{split}
\label{eq:conventional_ham}
\end{equation}
Here, $\mu_\mathrm{B}$ and $\mu_\mathrm{N}$ denote the Bohr and nuclear magnetons, respectively, $\hat{\bm{S}}=(\hat{S}_x, \hat{S}_y, \hat{S_z})$ and $\hat{\bm{I}}=(\hat{I}_x, \hat{I}_y, \hat{I_z})$ represent the electronic and nuclear spin operators, $g_\mathrm{N}$ is the nuclear $g$-factor, and $\hat{\bm{I}}^2=I(I+1)$. Furthermore, $\bm{B}_0$ is the vector of the external static magnetic field. This Hamiltonian is truncated at nuclear tensor rank $k=2$, although higher ranks are generally permitted up to $k=7$ for \eriononesixseven and $k=5$ for \ybiononeseventhree.

In the tetragonal crystal lattice of \cwo with $S_4$ site symmetry, the electronic $g$-tensor $\boldsymbol{g}$ and the hyperfine tensor $\boldsymbol{A}$ take a diagonal form with principal components parallel and perpendicular to the crystallographic $c$-axis, namely,  $g_{xx}=g_{yy}=g_\perp$, $g_{zz}=g_\parallel$, $A_{xx}=A_{yy}=A_\perp$, and $A_{zz}=A_\parallel$. The quadrupole term is sometimes equivalently written as $\hatH_\mathrm{q}=\hat{\bm{I}}\cdot\boldsymbol{P}\cdot\hat{\bm{I}}$. The traceless tensor $\boldsymbol{P}$ has diagonal elements $P_{xx}=P_{yy}=-P_{zz}/2$ and $P_{zz} = \frac{2}{3}P_{_\parallel}$~\cite{AbragamBleaney.1970}. At zero magnetic field, the system is fully described by three independent parameters: $A_\perp$, $A_\parallel$, and $P_{\parallel}$. In earlier works, $\hatH_\mathrm{q}$ has often been neglected, since, at typical experimental conditions, it was not possible to reliably separate it from the hyperfine term~\cite{Sattler.1970a,Antipin.1968,Bertaina.2009,Ranon.1964}. This issue has been discussed in Ref.~\cite{Strinic.2025} in detail, where the value of the quadrupole coefficient has been determined for a \eriononesixseven:LiYF\textsubscript{4} spin system.

\begin{figure}[t]
% latest version: \\badwwmi\user\groups\Quantum-Speicher\Austausch\Georg Mair\Figures\level_plot\level_plot_167Er_v5_NK
    \includegraphics[width=0.95\linewidth]{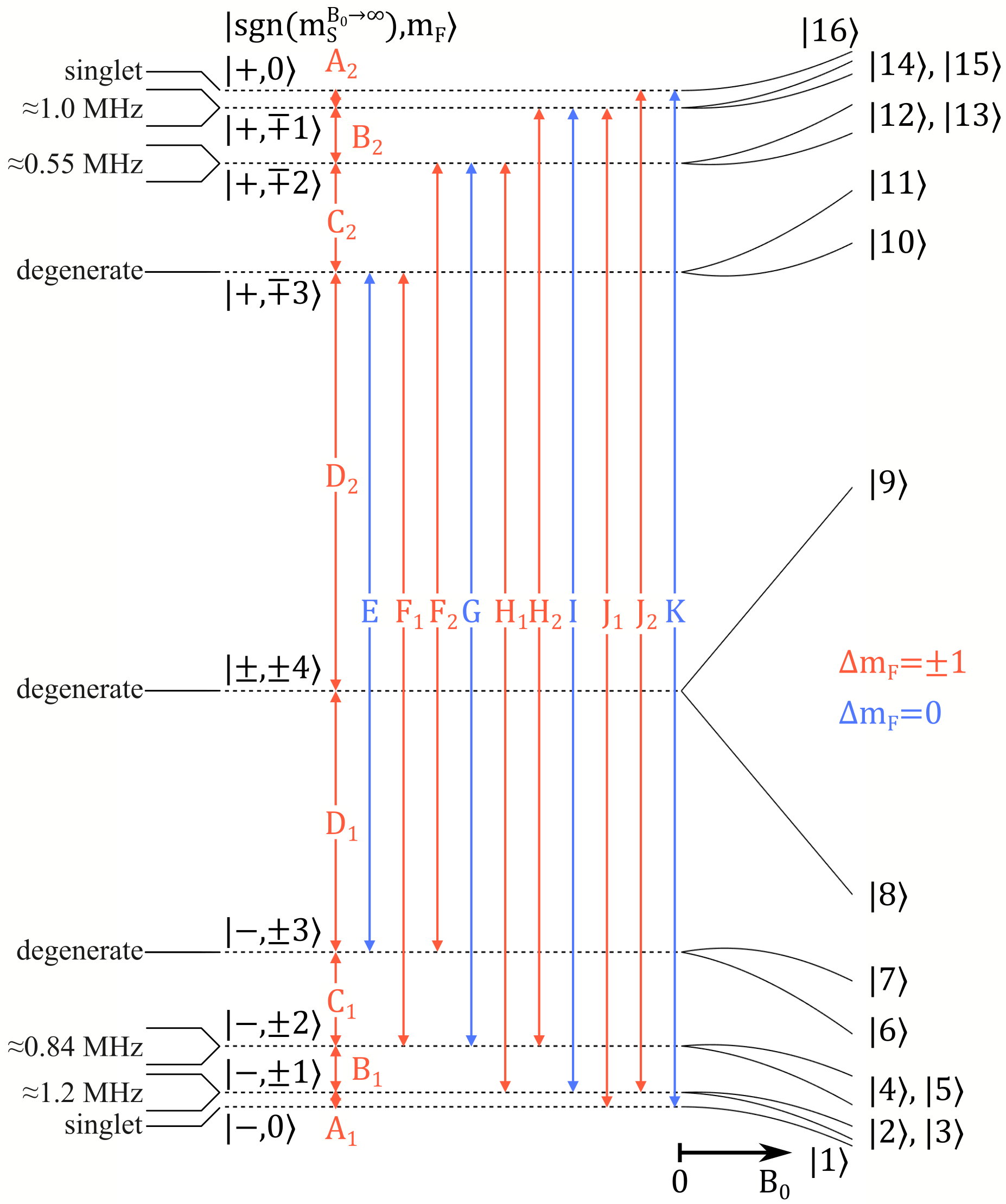}
    \caption{\textbf{Low-field hyperfine level structure of \eriononesixseven in \cwo:} According to the effective spin Hamiltonian given in~\Cref{eq:conventional_ham}, 16 energy levels form seven doublets ($m_F\neq0$) and two singlets ($m_F=0$) in the absence of magnetic field. The indicated transition groups follow the selection rules for mixed states. Transitions with $\Delta m_F=\pm1$ (orange arrows) are driven by oscillatory fields $B_1$ orthogonal to the anisotropy axis of the crystal ($\bm{B}_1\perp c$), while transitions with $\Delta m_F=0$ (blue arrows) require $\bm{B}_1\parallel c$. The degeneracy of transition frequencies within groups labeled with the same letter is lifted by the quadrupole term, except for D\textsubscript{1} and D\textsubscript{2}, where the anisotropic (diagonal) hyperfine components generate the main asymmetry. Lifting of the degeneracy of the $m_F=\pm1$ and $m_F=\pm2$ doublets results from local lattice strain and higher-rank interactions, respectively, which are considered in the extended Hamiltonian (\Cref{eq:full_ham}).
    }
    \label{fig:levelplot}
\end{figure}

% Quantum numbers and exact diagonalization
In the high-field limit, the eigenstates of the Hamiltonian can be described by the uncoupled Zeeman basis $\ket{m_S, m_I}$ using the electronic and nuclear spin quantum numbers $m_S$ and $m_I$, which strongly mix in the field regime $\lesssim\SI{100}{\milli\tesla}$, as depicted in \Cref{fig:levelplot}. In zero magnetic field, we can represent the eigenstates in the coupled basis $\ket{\pm,m_F}$, where corresponding energy levels are labeled by the sign of $m_S$ in the high-field limit and the total angular momentum $m_F=m_S+m_I$, which also holds for magnetic fields applied along the $c$-axis. Spin transitions then obey the selection rules $\Delta m_F=\pm1$ for magnetic driving fields $B_1$ oriented perpendicular to the $c$-axis ($\bm{B}_1\perp c$) and $\Delta m_F=0$ for components along $c$ ($\bm{B}_1\parallel c$). Within the coupled basis, the Hamiltonian can be diagonalized analytically, following the approach in Ref.~\cite{Mohammady.2010}, such that knowledge of the transition frequencies in zero-field uniquely determines the magnitude of the free parameters entering $\boldsymbol{A}$ and $\boldsymbol{P}$. The sign of the parameters can be inferred from the measurement of relative intensities of the transitions governed by the thermal population difference and transition probability between each pair of energy levels~\cite{Supplement}.

% Naming convention for zero-field transition groups
%\subsection{Transition naming convention}
In this work, we introduce a nomenclature convention for zero-field spin transition groups of \eriononesixseven:\cwo, as depicted in \Cref{fig:levelplot}, where the frequencies of transition groups sharing the same letter are expected to be identical in the absence of the quadrupole interaction (except for D\textsubscript{1} and D\textsubscript{2}). In addition, we specify sequential level numbering from $\ket{1}$ to $\ket{2I+1}$ based on the order of levels from lowest to highest energy state at any finite magnetic field applied in this work. Designed for the highest nuclear spin $I=\sfrac{7}{2}$ among the stable isotopes of REIs, this labeling scheme can be conveniently adapted for the other REIs, maintaining the labeling in connection to the $\ket{\mathrm{sgn}(m_S),\pm m_F}$ pairs. Its adaptation to \ybiononesevenone and \ybiononeseventhree is provided in the Supplementary Information~\cite{Supplement}.

% multipoles_table_prx_logbars_coefficients.tex
% Include in a REVTeX/PRX manuscript with:
%   \input{multipoles_table_prx_logbars_coefficients}
%
% Required in the main preamble:
%   \usepackage{array}
%   \usepackage{makecell}
%   \usepackage{tikz}
%   \usetikzlibrary{calc}
%
% The common logarithmic scale spans 10^{-1} to 10^{9} Hz.
% Each filled bar indicates the range stated in the original draft.

% ---------- Common logarithmic strength scale ----------
% Coordinate map: x = 0.30*(log10(f/Hz)+1), so 10^{-1} Hz -> 0 cm
% and 10^{9} Hz -> 3.0 cm.
\newcommand{\StrengthAxis}{%
\begin{tikzpicture}[x=1cm,y=1cm,baseline=-0.55ex]
  \path[use as bounding box] (0,-0.34) rectangle (3.05,0.12);
  \draw[black!55,line width=0.35pt] (0,0)--(3.0,0);
  \foreach \x in {0,0.3,...,3.0}{\draw[black!45,line width=0.3pt] (\x,-0.075)--(\x,0.075);}
  \node[anchor=north,font=\tiny] at (0,-0.09) {$10^{-1}$};
  \node[anchor=north,font=\tiny] at (0.9,-0.09) {$10^{2}$};
  \node[anchor=north,font=\tiny] at (1.8,-0.09) {$10^{5}$};
  \node[anchor=north,font=\tiny] at (2.7,-0.09) {$10^{8}$};
\end{tikzpicture}}

% #1=start x, #2=end x, #3=fill color, #4=range label,
% #5=outline color, #6=outline width
\begin{comment}
    
\newcommand{\StrengthBar}[6]{%
\begin{tikzpicture}[x=1cm,y=1cm,baseline=-0.55ex]
  \path[use as bounding box] (0,-0.18) rectangle (3.05,0.37);
  \draw[black!40,line width=0.35pt] (0,0)--(3.0,0);
  \foreach \x in {0,0.3,...,3.0}{\draw[black!35,line width=0.25pt] (\x,-0.055)--(\x,0.055);}
  \fill[#3,rounded corners=0.45pt] (#1,-0.105) rne terms, parity of purely nuclear terms is odectangle (#2,0.105);
  \draw[#5,line width=#6,rounded corners=0.45pt] (#1,-0.105) rectangle (#2,0.105);
  \node[anchor=south,font=\tiny,inner sep=0.6pt] at ({(#1+#2)/2},0.13) {#4};
\end{tikzpicture}}

\end{comment}

\newcommand{\StrengthBar}[6]{% 
\begin{tikzpicture}[x=1cm,y=1cm,baseline=-0.55ex] \path[use as bounding box] (0,-0.18) rectangle (3.05,0.37); % 
axis \draw[black!40,line width=0.35pt] (0,0)--(3.0,0); \foreach \x in {0,0.3,...,3.0}{ \draw[black!35,line width=0.25pt] (\x,-0.055)--(\x,0.055); } 
% actual interaction bar 
\fill[#3,opacity=0.55,rounded corners=0.45pt] (#1,-0.105) rectangle (#2,0.105); \draw[#5,line width=#6,rounded corners=0.45pt] (#1,-0.105) rectangle (#2,0.105); \node[anchor=south,font=\tiny,inner sep=0.6pt] at ({(#1+#2)/2},0.13) {#4};
% linewidth window (10^4--10^5 Hz) 
%\fill[blue!60,opacity=0.20] (1.50,-0.105) rectangle (1.80,0.105); 
\fill[blue!60,opacity=0.15] (1.50,-0.105) rectangle (1.80,0.105); 
\draw[blue!80,dotted,line width=0.4pt] (1.50,-0.105) rectangle (1.80,0.105);
\end{tikzpicture}}

% Original ranges mapped onto the shared logarithmic axis.
% First-order terms: red fill. Second-order terms: orange fill.
% Rank k >= 3: thick teal outer line.
\newcommand{\BarMone}{\StrengthBar{2.70}{3.00}{black!65}{$10^{8}$--$10^{9}$}{black}{0.35pt}}
\newcommand{\BarEtwo}{\StrengthBar{2.10}{2.70}{cyan!50!black}{$10^{6}$--$10^{8}$}{black}{0.35pt}}
\newcommand{\BarMoneMone}{\StrengthBar{1.80}{2.40}{orange!70}{$10^{5}$--$10^{7}$}{orange}{0.35pt}}
\newcommand{\BarMthree}{\StrengthBar{1.20}{1.80}{purple!70!black}{$10^{3}$--$10^{5}$}{black}{.35pt}}
\newcommand{\BarMoneEtwo}{\StrengthBar{0.90}{1.50}{purple!70!black}{$10^{2}$--$10^{4}$}{black}{.35pt}}
\newcommand{\BarEtwoEtwo}{\StrengthBar{0.00}{0.60}{purple!70!black}{$10^{-1}$--$10^{1}$}{black}{.35pt}}
\newcommand{\BarEfour}{\StrengthBar{0.30}{0.90}{purple!70!black}{$10^{0}$--$10^{2}$}{black}{.35pt}}

\begin{table*}[t]
\caption{\label{tab:multipoles}%
Summary of the nuclear multipole interaction hierarchy, second-order corrections, and pseudo-moments for an effective electronic doublet ($S=\sfrac{1}{2}$) coupled to the nuclei of \eriononesixseven ($I=\sfrac{7}{2}$), \ybiononesevenone ($I=\sfrac{1}{2}$) and \ybiononeseventhree ($I=\sfrac{5}{2}$) under ideal tetragonal $S_4$ symmetry at $\abs{\bm{B}_0}=\SI{0}{\tesla}$. The ``Min. $I$ (Term)'' column specifies the minimal nuclear spin number required to add the respective term in the Hamiltonian. However, for pseudo-terms to form level shifts in spectra similar to the true terms, a higher minimal nuclear spin number is required, as listed in the ``Min. $I$ (Split.)'' column.
Bar widths of the interaction scale represent the stated order-of-magnitude range, not an uncertainty distribution.
%Parity $\boldsymbol{\mathcal T}$ is given for the total hyperfine terms, parity of purely nuclear terms is odd. 
}
\centering
\scriptsize
\setlength{\tabcolsep}{1.8pt}
\renewcommand{\arraystretch}{1.22}
\begin{ruledtabular}
\begin{tabular}{
  l c c c c
  >{\centering\arraybackslash}p{0.174\textwidth}
  >{\centering\arraybackslash}p{0.070\textwidth}
  >{\centering\arraybackslash}p{0.092\textwidth}
  >{\centering\arraybackslash}p{0.092\textwidth}
  >{\centering\arraybackslash}p{0.092\textwidth}}
\textbf{Interaction term} &
\shortstack{\textbf{Rank}\\\textbf{($k$)}} &
\shortstack{\textbf{Min. $I$}\\\textbf{(Term)}} &
\shortstack{\textbf{Min. $I$}\\\textbf{(Split.)}} &
\shortstack{\textbf{Time-reversal} \\ \textbf{parity}} &
\makecell{\textbf{Interaction scale, Hz}\\[-1pt]\StrengthAxis} &
\textbf{Coeff.} &
\multicolumn{3}{c}{\textbf{Values, MHz}} \\
%\cline{8-10}
& & & & & & &
\makecell{\textbf{\eriononesixseven} \\$I=\sfrac{7}{2}$} &
\makecell{\textbf{\ybiononesevenone} \\$I=\sfrac{1}{2}$} &
\makecell{\textbf{\ybiononeseventhree} \\$I=\sfrac{5}{2}$} \\
\hline
\makecell[l]{Magnetic dipole\\$(M1)$}
& 1 & $1/2$ & $1/2$ & Odd & \BarMone
& \makecell{$A_{\perp}$${}^{(1)}$\\$A_{\parallel} $${}^{(1)}$}
& \makecell{$-871.275$\\$-127.013$}
& \makecell{$3083.844$\\$786.85$}
& \makecell{$-849.65$\\$-217.03$} \\

\makecell[l]{Electric quadrupole\\$(E2)$}
& 2 & $1$ & $1$ & Even & \BarEtwo
& \makecell{ $P_{\parallel} $${}^{(1)}$  \\ $P_{\mathrm{true}}$ ${}^{(2)}$ }
& \makecell{ $-5.104% P_zz = -3.404
$ \\ $\sim-2.21$ }
& \makecell{\textemdash${}^{(3)}$} 
& \makecell{ $13.99 % P_zz = 9.330
$ \\ $\sim-8.68$ } \\

\makecell[l]{Pseudo-quadrupole\\$(M1\times M1)$\\}
& 2 & $1/2$ & $1$ & Even & \BarMoneMone
& $P_{\mathrm{pseudo}}$ ${}^{(2)}$
& $\mathit{\sim-2.71}$
& \textemdash
& $\mathit{\sim-0.53}$ \\

\makecell[l]{Magnetic octupole\\$(M3)$ \\ }
& 3 & $3/2$ & $3/2$ & Odd & \BarMthree
& \makecell{$\alpha_{11} $${}^{(1)}$\\$\alpha_{31} $${}^{(1)}$}
& \makecell{$6.3\cdot10^{-3}$\\$29.7\cdot10^{-3}$}
& \makecell{\textemdash${}^{(3)}$}
& \makecell{\textemdash\\$16\cdot10^{-3}$} \\

\makecell[l]{Pseudo-octupole\\$(M1\times E2)$ \\ }
& 3 & $1$ & $3/2$ & Odd & \BarMoneEtwo
& $\alpha_{\mathrm{pseudo}}$${}^{(2)}$
& $\mathit{\sim10^{-2}}$  
& \textemdash${}^{(3)}$
& $\mathit{\sim10^{-2}}$   \\

\makecell[l]{Electric hexadecapole\\$(E4)$}
& 4 & $2$ & $2$ & Even & \BarEfour
& \makecell{$d_\perp$${}^{(1)}$ \\ $d_\parallel$${}^{(1)}$}
& \makecell{ $4.6\cdot 10^{-4}$ \\ $4.9\cdot 10^{-4}$ }
& \textemdash${}^{(3)}$
& \textemdash \\

\makecell[l]{Pseudo-hexadecapole\\$(E2\times E2)$}
& 4 & $1$ & $2$ & Even & \BarEtwoEtwo
& $d_{\mathrm{pseudo}}$${}^{(2)}$
& $\mathit{\sim10^{-5}}$ 
& \textemdash${}^{(3)}$
& $\mathit{\sim10^{-5}}$  \\

\end{tabular}
\end{ruledtabular}

\vspace{3pt}
\parbox{0.97\textwidth}{\scriptsize
\textit{Reading the scale:} %All TikZ pictures use the same fixed bounding box and the same $\log_{10}(f/\mathrm{Hz})$ axis from $10^{-1}$ to $10^{9}$ Hz. 
%Bar widths represent the stated order-of-magnitude range, not an uncertainty distribution. 
\begin{tikzpicture}[x=1cm,y=1cm,baseline=-0.55ex] 
\fill[blue!60,opacity=0.15] (1.50,-0.105) rectangle (1.80,0.105); 
\draw[blue!80,dotted,line width=0.4pt] (1.50,-0.105) rectangle (1.80,0.105);
\end{tikzpicture}
marks the experimentally identified linewidth. \\
${}^{(1)}$ Values identified in this work, see \Cref{table:HamParametersErYb}. \\
${}^{(2)}$ Estimated upper bound for the value. \\
${}^{(3)}$ The corresponding irreducible nuclear tensor is inactive for $I=1/2$ at this rank.}
\end{table*}

\subsection{Higher-rank and higher-order terms}
\label{ssec:higher_rank_terms}

In the effective spin $S=\sfrac{1}{2}$ model, the spin operator $\hat S$ appears only in the first rank. Therefore, all higher ranks are produced by nuclear spin operators $\hat I^{k}$, so that the general form of the Hamiltonian can be written in terms of irreducible tensor operators $\hat{T}_{kq}$
\begin{equation} 
\hatH_\mathrm{eff} = \sum_{k=1,3,...} c_{kqq^\prime} \, \hat{T}^{S}_{1q^\prime} \, \hat{T}^{I}_{kq} 
+ 
\sum_{k=2,4,...} c_{kq} \, \hat{T}^{I}_{kq},
\label{eq:higher-order_general}
\end{equation}
where index $q$ implies the sum from $-k$ to $k$. For all ranks with $k>1$, higher-rank terms arise either due to pure higher-rank hyperfine interactions of the nucleus or due to higher-order contributions arising from coupling to higher-lying crystal field states within the ground state manifold. The higher-rank terms originating from second-order interactions are commonly denoted as ``pseudo''. 
%With the true higher-rank interaction comprising of higher-rank nuclear magnetic moment interaction with $4f$ electrons, e.g. $P_{4f}$ for the quadrupole interaction, and crystalline fields, $P_\mathrm{CF}$, and pseudo higher-order interaction, $P_\mathrm{pseudo}$, the final terms is experimentally described by an effective coefficient, s.a. $P_\mathrm{eff}=P_\mathrm{true}+P_\mathrm{pseudo}$, with $P_\mathrm{true}=P_{4f}+P_\mathrm{CF}$. 
The complete higher-rank hyperfine terms comprise true first-order components arising from interactions with fields of $4f$ electrons and shielded crystal fields, i.e., $P_\mathrm{true}=P_{4f}+P_\mathrm{CF}$, which are directly proportional to nuclear higher-rank moments, such as $Q$ for quadrupole and $\Omega$ for octupole, and a higher-order (pseudo) component, which is inversely proportional to the energy splitting of the Stark levels ($\propto E_{CF}$). Empirical equations developed for the quadrupole term in earlier works~\cite{Ram.1985} allow us to estimate these coefficients separately, which are summarized in \Cref{tab:multipoles} with details on calculation provided in the Supplementary Information~\cite{Supplement}. The obtained numbers suggest that the intrinsic contribution to the second-rank quadrupole interaction is expected to be of a similar magnitude as the pseudo-quadrupole interaction $\propto  \sfrac{A^2_{hf}}{\Delta E_\mathrm{CF}}$ for \eriononesixseven, while for \ybiononeseventhree the intrinsic quadrupole interaction is expected to dominate. This is a direct consequence of the crystal field splitting $\Delta E_\mathrm{CF}$ of the ground multiplet, which is by one order of magnitude larger for ytterbium.

The form of the rank $k=3$ pseudo-octupole and rank $k=4$ pseudo-hexadecapole terms for tetragonal symmetry has been proposed by Baker and Blake~\cite{Baker.1970}. We have revisited the problem of the exact term for the case of the true first-order interaction~\cite{Supplement}, which is equivalent to the field-independent part of the pseudo-term:
%
%\begin{widetext}
\begin{equation}
\begin{split}
    \hatH_{\mathrm{oct}}
    &=
    \frac{\alpha_{00}}{2}\hat{S}_z
    \left(
        5\hat{I}_z^3 - 
        \left(
            3\hat{\bm{I}}^2 - 1
        \right)
        \hat{I}_z
    \right)
    \\
    &\quad
    -
    \alpha_{11}
    \sqrt{\frac{3}{32}}
    \left(
        \hat{S}_x
        \left\{
            \hat{I}_x, \Xi
        \right\}
        +
        \hat{S}_y
        \left\{
            \hat{I}_y, \Xi
        \right\}
    \right)
    \\
    &\quad
    +
    \alpha_{31}
    \sqrt{\frac{5}{8}}
    \bigg[
        \hat{S}_x
        \left(
            \hat{I}_x^3
            -
            3\overline{\hat{I}}_{xyy}
        \right)
        +
        \hat{S}_y
        \left(
            \hat{I}_y^3
            -
            3\overline{\hat{I}}_{yxx}
        \right)
    \bigg]
    \label{eq:HAMoctupole}
\end{split}
\end{equation}
%\end{widetext}
%
Here, we use the anticommutator $\{A,B\}=AB+BA$, the symmetrized cubic product $
    \overline{\hat{I}}_{ijj}
    =
    \tfrac{1}{3}
    \left(
        \hat{I}_i\hat{I}_j\hat{I}_j
        +
        \hat{I}_j\hat{I}_i\hat{I}_j
        +
        \hat{I}_j\hat{I}_j\hat{I}_i
    \right)
$, and $\Xi=5\hat{I}_z^2-\hat{\bm{I}}^2-\tfrac{1}{2}$.

Despite minor differences in the form, both our expression in \Cref{eq:HAMoctupole} and the term by Baker et al. generate similar changes in the energy levels at zero magnetic field.
When expressed in terms of the ladder operators
\(\hat{S}_{\pm}\) and \(\hat{I}_{\pm}\), the part proportional to
\(\alpha_{31}\) contains terms of the form
\(\hat{S}_{\pm}\hat{I}_{\pm}^{3}\). These terms obey
$\Delta m_S=\pm1$, $\Delta m_I=\pm3$ and create mixing between the states with $\Delta m_F =\pm4$, thereby removing the degeneracy in the $\ket{-,\pm2}$ and $\ket{+,\pm2}$ doublets. The magnitude of the splitting is linear in \(\alpha_{31}\). 
For the pseudo-octupole term, additional field-dependent terms are possible, but can be neglected under low-magnetic-field experimental conditions.

The next rank $k=4$ hexadecapole interaction can be written as~\cite{Supplement,Baker.1970}
\begin{equation}
\begin{split}
    \hatH_{\mathrm{hdp}}
    &=
    \frac{d_{\parallel}}{8}
    \left[
        35\hat{I}_z^4
        -
        30\hat{\bm{I}}^2\hat{I}_z^2
        +
        25\hat{I}_z^2
        +
        3\bigl(\hat{\bm{I}}^2\bigr)^2
        -
        6\hat{\bm{I}}^2
    \right]
    \\
    &\quad
    +
    d_{\perp}
    \sqrt{\frac{35}{128}}
    \left(
        \hat{I}_{+}^{4}
        +
        \hat{I}_{-}^{4}
    \right).
    \label{eq:HAMhexadecapole}
\end{split}
\end{equation}
Here, the first term, being proportional to $\hat{I}_z^4$ and $\hat{I}_z^2$, adds a correction to the zero-field energy-level structure without mixing any states, as the quadrupole interaction does. The second term in the equation is the Hermitian combination of the \(q=\pm4\) components. It can mix nuclear-spin states satisfying \(\Delta m_I=\pm4\), thus allowing for further correction of the splitting in the $m_F=\pm2$ doublets.

%Lower-energy contributions expected from terms with ranks $k\ge 5$ are at least one order below the experimental linewidth of the spin transitions and thus cannot be resolved in the current experimental data. However, obeying point symmetry $S_4$, they would similarly couple states with $\Delta m_F=\pm4$. Only in presence of reduced symmetry, e.g. $C_2$, rank-5 and rank-6 can couple states with $\Delta m_F=\pm6$ and thus lift degeneracy within $\ket{+,\pm3}$ and $\ket{-,\pm3}$ doublets. \gwm{How did you check?}

\subsection{Local strain}
\label{ssec:local_strain}

We consider symmetry-lowering (\Cref{fig:intro}(c)) by elastic lattice deformations arising from point defects as a possible source of the observed level splittings and spectral broadening. In particular, we focus on in-plane symmetry-breaking strain components for the REI site. Their influence on the level structure and spectral line shapes has been investigated for non-Kramers doublets within the hyperfine manifolds of Tb\textsuperscript{3+} and Ho\textsuperscript{3+} dopants in scheelite hosts~\cite{Kirton.1965, Shakurov.2014, Boldyrev.2022}. Although the pure electronic doublets of the even \erion isotopes are strictly degenerate in the absence of time-reversal symmetry-breaking interactions, the doublets formed by the hyperfine states of \eriononesixseven are not protected by Kramers' theorem. Consequently, their degeneracy can be lifted, allowing the application of the theory developed for random strain in an elastically anisotropic continuum~\cite{Malkin.2017}.

%For the scheelite structure with point group $C_{4h}$, we consider strain leading to an orthorhombic distortion $e_1=(e_{xx}-e_{yy})/2$ and shear strains $e_2=e_{xy}$ transforming according to the irreducible representation $B_g$. 
For the $S_4$ site symmetry of the REI, the two in-plane symmetry-breaking strain components $e_1=(e_{xx}-e_{yy})/2$ and shear strain $e_2=e_{xy}$ both transform according to the irreducible representation $B_g$.
The component $e_1$ describes an orthorhombic distortion along the crystallographic $a$ and $b$ axes, whereas $e_2$ describes an in-plane shear distortion.
Rather than explicitly incorporating the strain tensor elements $e_i$ into the crystal-field Hamiltonian, we model their effect by a distortion of the principal axes of the hyperfine interaction tensor. The resulting corrections $\delta A_i=A_\perp e_i$ ($i=1,2$) therefore describe, respectively, an asymmetry between $A_{xx}$ and $A_{yy}$ and off-diagonal $A_{xy}$ components:
\begin{equation}
    \hatH_\mathrm{strain} = \frac{\delta A_1}{2}\left(\hat{S}_x\hat{I}_x - \hat{S}_y\hat{I}_y\right) + \delta A_2\left(\hat{S}_x\hat{I}_y + \hat{S}_y\hat{I}_x\right) .
    \label{eq:HAMstrain}
\end{equation}
Expressed in terms of the raising and lowering operators $\hat{S}_\pm$ and $\hat{I}_\pm$ of the electronic and nuclear spins, this interaction contains terms proportional to $\hat{S}_\pm\hat{I}_\pm$, and couples states with $\Delta m_F=\pm2$. Perturbation theory shows that the energy shifts of the $m_F=\pm1$ states scale linearly in the strain magnitude, whereas shifts of the remaining levels arise only in second order. Consequently, the $m_F=\pm1$ states are particularly sensitive to lattice strain.

At zero magnetic field, the strain interaction lifts the degeneracy of the $\ket{-,\pm1}$ and $\ket{+,\pm1}$ doublets in \eriononesixseven and \ybiononeseventhree, as well as the $\ket{\pm,\pm1}$ levels in \ybiononesevenone. In an ensemble containing a distribution of inhomogeneous local strain, these level splittings give rise to avoided crossings and pronounced asymmetric line shapes for transitions involving the affected states. The resulting spectral response consists of a broadened envelope with a characteristic reduction in intensity at its center. Similar line shapes have been reported for several REI–host systems~\cite{Klimin.2010, Malkin.2012, Shakurov.2014, Malkin.2017, Boldyrev.2022}.

The strain components follow a generalized two-dimensional Lorentzian distribution,
\begin{equation}
    g\left(e_1, e_2\right) = \frac{\nu_B\xi}{2\pi}\left(\nu_B^2 {e_1^\prime}^2 + {e_2^\prime}^2 + \xi^2\right)^{-3/2}
    \label{eq:dist_function}
\end{equation}
where $e_1^\prime=e_1\cos{\theta_B}+e_2\sin{\theta_B}$ and $e_2^\prime=-e_1\sin{\theta_B}+e_2\cos{\theta_B}$. The distribution parameters $\nu_B=2.44$ and $\theta_B=\SI{-48.7}{\degree}$ are given for \cwo in Ref.~\cite{Malkin.2017}. The parameter $\xi$ determines the width of the distribution and is proportional to the concentration of defects. The strain-induced spectral response of the ensemble is then obtained by integrating the spectra $I_{e_1,e_2}$ calculated for individual strain configurations over the distribution.
\begin{equation}
    I_\mathrm{strain} = \iint I_{e_1,e_2}g(e_1, e_2)\,\mathrm{d}e_1\mathrm{d}e_2
\end{equation}
Each spectrum is obtained by convolving the corresponding transition frequencies with a Lorentzian of narrow homogeneous linewidth and weighting by their calculated transition intensities.

For a realistic crystal, the inhomogeneous local strain distribution may deviate from the idealized Lorentzian form used here. Strain components in the $z$ direction can also shift the energy levels, but do not generate new avoided crossings. Although such strains may therefore influence the line shapes of other transitions through these frequency shifts, they are not required to reproduce the avoided crossings considered here. We consequently restrict the analysis to the $x$- and $y$-components, which constitute the minimal model required to reproduce the observed spectral features.

\section{High-resolution microwave spectroscopy}

\begin{figure*}[tbh!]
% \\badwwmi\user\groups\Quantum-Speicher\Austausch\Georg Mair\Figures\broadband_ESR_sim_overlay\broadband_ESR_sim_overlay_v5\broadband_ESR_XLD1_r46_167Er-corr_171Yb-corr_173Yb-corr_labeled.png
    \includegraphics[trim={{.02\textwidth} 0 {.11\textwidth} 0},clip,width=1\textwidth]{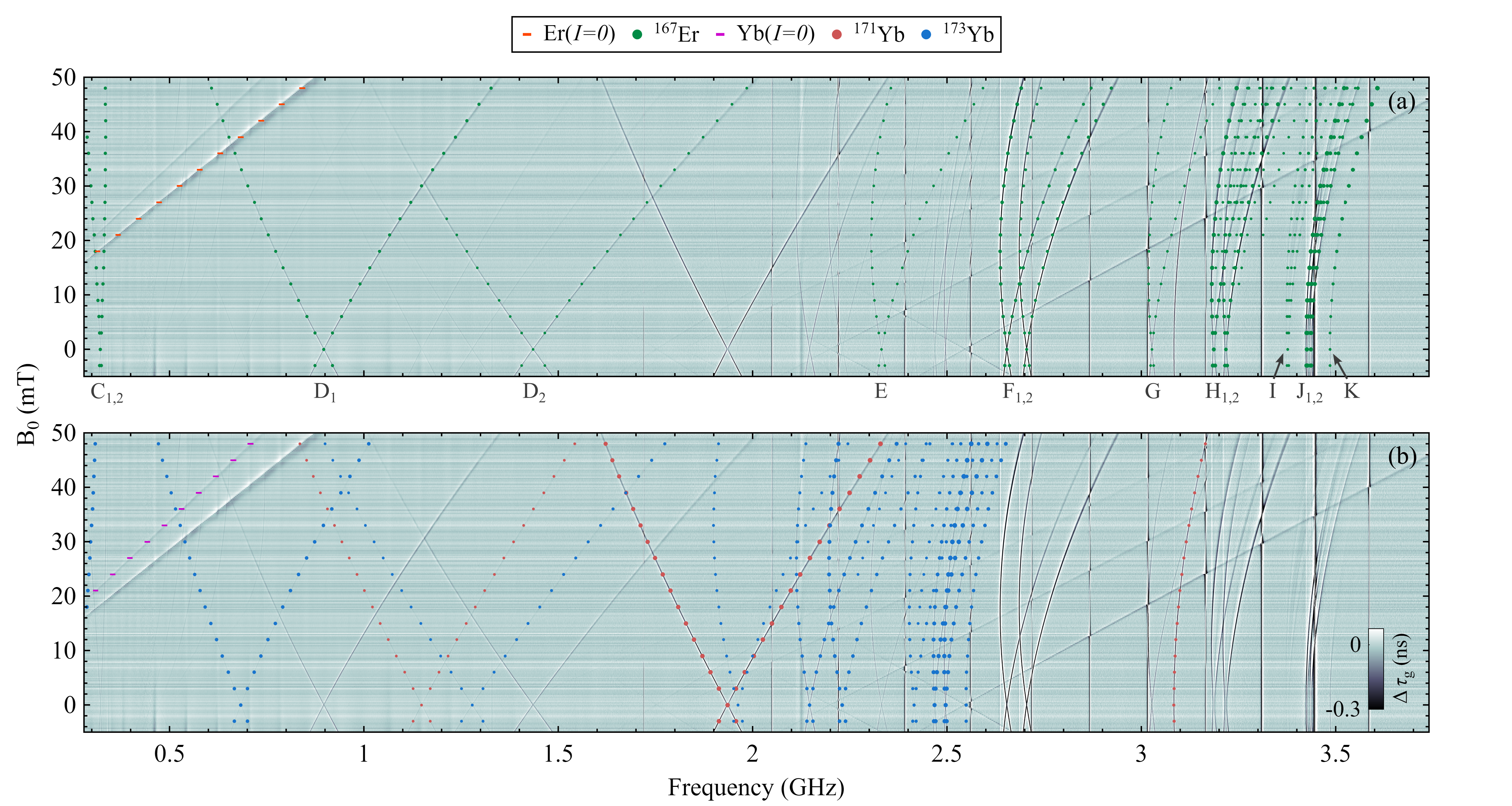}
    \caption{\textbf{Broadband microwave spectroscopy:}
    The normalized group delay $\Delta\tau_g$ reveals hyperfine transitions of paramagnetic spin species in Sam3 as a function of frequency and magnetic field. The external magnetic field $B_0$ is oriented along the crystalline $c$-axis, with a misalignment angle of $\SI{0.76}{\degree}$. In addition to erbium dopants, trace impurities of ytterbium and manganese (see~\cite{Supplement}) are identified. The measured spectrum is overlaid with the simulation data for \erion (a) and \ybion (b) ions. The hyperfine parameters for \eriononesixseven and \ybiononeseventhree are obtained using least-squares fitting, while those for \ybiononesevenone are calculated by the exact diagonalization of the Hamiltonian given in \Cref{eq:conventional_ham}. The electronic $g$-factors are taken from Refs.~\cite{Antipin.1968,Sattler.1970a}. Parasitic background resonances originating from the experimental setup appear as vertical lines in the spectrum. The symbol size represents the relative transition intensity at a spin temperature of \SI{13}{mK}. An intensity threshold is applied in the simulation to exclude transitions below the experimental detection limit.}
    \label{fig:broadbandErYb}
\end{figure*}

% General remarks on broadband spectroscopy
Broadband EPR spectroscopy allows a comprehensive characterization of the spin systems. Mechanically attached to a superconducting coplanar waveguide transmission line, the REI-doped crystal is probed by absorption of a propagating microwave signal, which generates a magnetic field $B_1$ driving spin transitions in resonance. The setup is schematically shown in \Cref{fig:intro}(d). Absorption is detected as a drop in the magnitude of the scattering parameter $S_{21}$. Alternatively, the shift of the phase $\phi$ of $S_{21}$ is converted to a group delay $\tau_g=-\tfrac{1}{2\pi}\sfrac{\mathrm{d}\phi}{\mathrm{d}f}$. This allows a convenient identification of absorption features without further normalization, provided the background phase response is sufficiently linear. Background-subtracted data are denoted by $\Delta\tau_g$. The samples used in this study comprise nominal doping of \SI{50}{ppm} of \textsuperscript{n.a.}\erion (Sam1), \SI{34}{ppm} of \textsuperscript{n.a.}\erion (Sam2), \SI{50}{ppm} of \textsuperscript{n.a.}\erion with trace amounts of \textsuperscript{n.a.}\ybion (Sam3), and \SI{34}{ppm} of isotopically enriched \ybiononesevenone (Sam4)~\footnote{{n.a.}: Natural isotopic abundance.}. Further details on the experimental setup and samples are described in the Supplementary Information~\cite{Supplement}.

% Broadband spectrum of Sam3
The broadband spectrum of Sam3 in the $\bm{B}_0\parallel c$ configuration is shown in \Cref{fig:broadbandErYb}. Additional spectra for all samples are provided in the Supplementary Information~\cite{Supplement}. The orientation of the propagating microwave mode relative to the crystalline $c$-axis ($k\perp c$) was chosen such that the corresponding selection rules permit the observation of all allowed magnetic dipole transitions. For erbium, this configuration allows detection of all transitions from group C\textsubscript{1} to K (ordered by increasing transition frequency), as depicted by the overlaid simulation in \Cref{fig:broadbandErYb}(a). The effective spin temperature is determined to be $T_\mathrm{eff}=\SI{13}{\milli\kelvin}$~\cite{Supplement}, which is in agreement with earlier experiments under similar experimental conditions~\cite{Strinic.2025}. Transitions C\textsubscript{2} and D\textsubscript{2} remain undetected at weak probing powers due to small thermal population differences between the involved energy levels. This can be mitigated by applying a strong saturating pump tone to transition group F\textsubscript{1} (D\textsubscript{1}) prior to each frequency sweep on transition group C\textsubscript{2} (D\textsubscript{2}) to populate levels $\ket{+,\mp3}$ ($\ket{\pm,\pm4}$). In contrast, transitions from A\textsubscript{1} to B\textsubscript{2} remain unobserved due to a limited transmission outside the operational bandwidth of the employed microwave components. In addition to the intentional doping of sample Sam3 with erbium, weak absorptions associated with ytterbium and manganese ions are observed (see overlaid simulation in \Cref{fig:broadbandErYb}(b)). The broadband spectrum also contains field-independent background resonances arising from standing waves caused by impedance mismatches at interfaces in the microwave circuitry.

% High-resolution zero-field spectroscopy
While broadband field-dependent spectra allow us to gain information on the complete Hamiltonian, including field-dependent terms, we have additionally taken high-resolution spectra of hyperfine transitions in the vicinity of zero magnetic field, up to \SI{2.5}{\milli\tesla}, which allow us to extract field-independent spin Hamiltonian parameters with high precision. %The spectra of all detected hyperfine transition groups of \eriononesixseven and \ybiononesevenone are provided in the Supplementary Information~\cite{Supplement}. 
Characterizing samples with different doping concentrations, we observed a narrower linewidth at lower dopant concentrations. For the $\ket{-,+2}\rightarrow\ket{+,+2}$ transition (group G), the full width at half maximum (FWHM) decreases from $\Gamma_\mathrm{FWHM}\simeq\SI{50}{\kilo\hertz}$ at \SI{50}{ppm} to $\Gamma_\mathrm{FWHM}\simeq\SI{20}{\kilo\hertz}$ at \SI{34}{ppm}. We associate this with the reduction of magnetic noise amplitude and spectral diffusion at lower concentrations. Therefore, we take data from lower-concentration samples Sam2 and Sam4 to precisely determine transition frequencies of \eriononesixseven and \ybiononesevenone, respectively. The zero-field frequencies of \ybiononeseventhree are obtained from the residual doping in Sam3.

% Fitting hyperfine parameters of conventional Hamiltonian
Considering the conventional Hamiltonian (\Cref{eq:conventional_ham}), the nuclear spin $I=\sfrac{1}{2}$ of \ybiononesevenone results in four hyperfine levels, and the two independent Hamiltonian parameters ($A_\perp$ and $A_\parallel$) can be determined directly from the two observed transition frequencies~\cite{Sattler.1970b,Supplement}. The associated uncertainties are primarily limited by the accuracy with which the center frequency of each spectral line is determined. In contrast, for \eriononesixseven and \ybiononeseventhree, the system is overdetermined due to the larger number of observable transitions. In these cases, the optimal parameter set is obtained by a global weighted least-squares fit of \Cref{eq:conventional_ham} to the energy levels, which are, in turn, calculated from the detectable transition frequencies. The uncertainties are calculated from the covariance matrix obtained from the Jacobian of the weighted residuals of the fit. The resulting hyperfine and quadrupole parameters for \eriononesixseven, \ybiononesevenone, and \ybiononeseventhree are summarized in \Cref{table:HamParametersErYb}.

% Hyperfine parameter sign
%Low-field spectroscopy at ultra-low temperatures allows unambiguous determination of the signs of the hyperfine coefficients. Owing to the strong thermal population gradient between hyperfine levels, the measured transition intensities provide direct information on the relative ordering of the energy levels~\cite{Supplement}.

Owing to the strong thermal population gradient between hyperfine levels at ultra-low temperatures, the measured transition intensities provide direct information on the relative ordering of the energy levels. We identify the sign of $A_\parallel$ to be negative for \eriononesixseven and \ybiononeseventhree and positive for \ybiononesevenone. Applying the symmetry relation between the principal values of $\bm{A}$ and $\bm{g}$~\cite{AbragamBleaney.1970}, we further infer the signs of $A_\perp$.

% Misalignment angle
The electronic and nuclear $g$-factors are taken from established and confirmed literature values~\cite{Antipin.1968,Sattler.1970a,Stone.2014}, allowing us to derive a misalignment angle of $\SI{0.76}{\degree}$ between the applied magnetic field $B_0$ and the crystallographic $c$-axis based on the slope of the transitions associated with the even isotopes.

% Discussion of fitting results
The extracted quadrupole tensor for \eriononesixseven:\cwo is in agreement with recently reported values~\cite{Marsh.2026}. For \eriononesixseven, the root-mean-square (RMS) deviation of the fit is \SI{0.47}{\mega\hertz}, which exceeds the narrowest observed linewidths and uncertainties of the transition frequencies. We conclude that the conventional Hamiltonian requires additional interaction terms to explain the observed spectral features.

\section{Zero-field avoided crossings}

\begin{figure*}[tbh!]
% \\badwwmi\user\groups\Quantum-Speicher\Austausch\Georg Mair\Figures\zero_field_features_167Er_171Yb_173Yb\zero_field_features_167Er_171Yb_173Yb_v5\zero_field_features_167Er_171Yb_173Yb_labeled.png
    \centering
    \includegraphics[width=\linewidth]{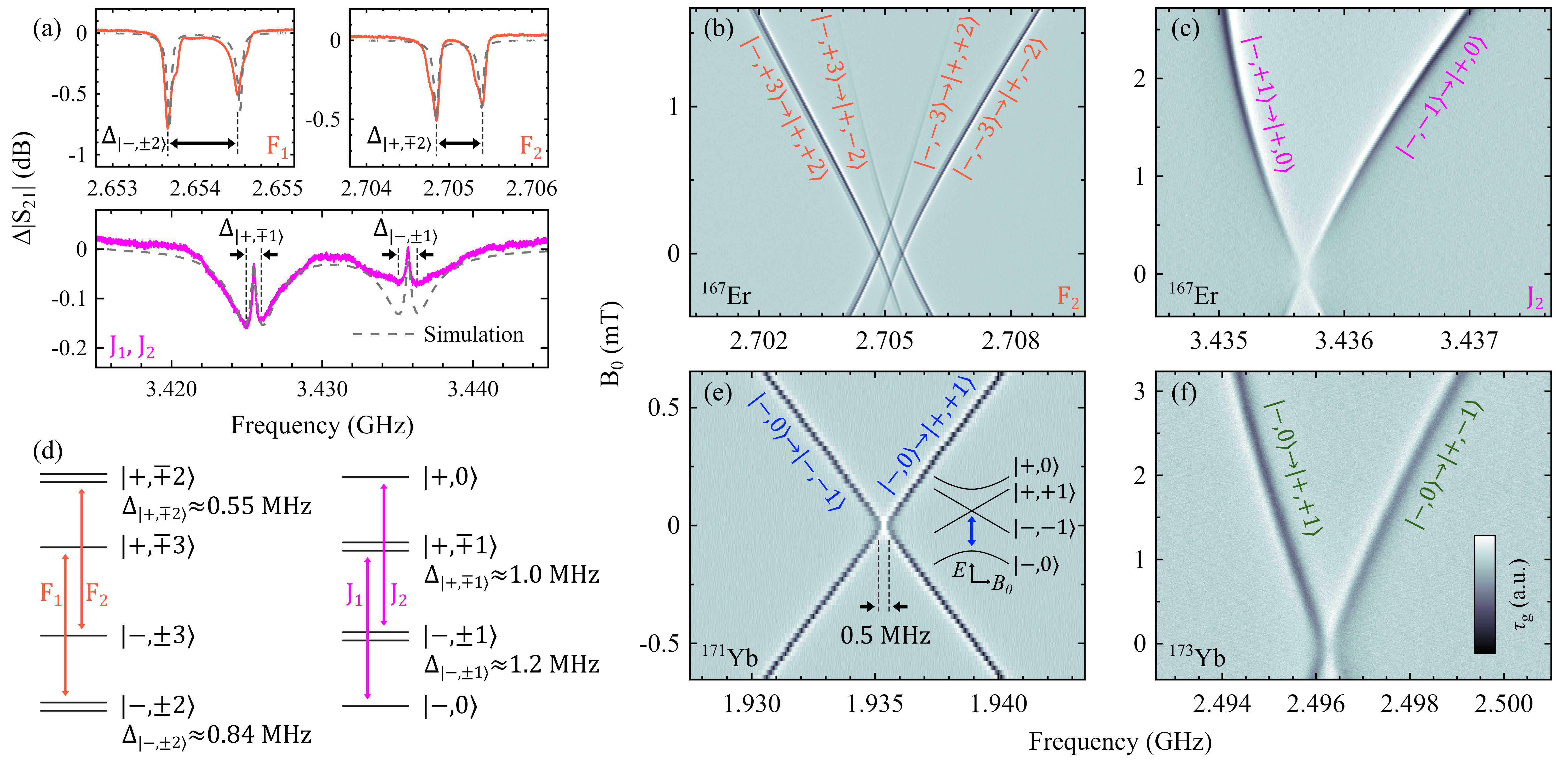}
    \caption{\textbf{Zero-field avoided crossings:}
    (a)~Observed zero-field splittings of selected \eriononesixseven hyperfine transitions, arising from lifted degeneracies of the doublets $\ket{-,\pm2}$, $\ket{+,\mp2}$, $\ket{-,\pm1}$, and $\ket{+,\mp1}$. These splittings are not accounted for by the conventional model Hamiltonian (\Cref{eq:conventional_ham}). Including the octupole and hexadecapole interactions, as well as random strain, into the model Hamiltonian allows us to reproduce the observed splitting of the $m_F=\pm2$ doublets and the distribution (and resulting line shapes) of the $m_F=\pm1$ states, respectively. Data are normalized by subtraction of a spectrum with detuned spin transitions. The simulation is scaled to match the maximum intensity of each transition group.
    (b), (c)~High-resolution magnetic field sweeps (Sam2) around \SI{0}{\milli\tesla} reveal pronounced avoided crossings in the C\textsubscript{1,2} and F\textsubscript{1,2} transition groups (only F\textsubscript{2} shown), indicating coupling to transitions that are expected to be forbidden by selection rules. The doublet-to-singlet transition group J\textsubscript{2} exhibits an avoided crossing.
    (d)~Schematic energy level diagrams corresponding to the transitions shown in panels (a)–(c).
    (e)~Previously unobserved avoided crossing between the hyperfine levels $\ket{\pm,\pm1}$ in \ybiononesevenone (Sam4), which are expected to be degenerate within the conventional Hamiltonian. The corresponding level structure is indicated.
    (f)~Unpredicted avoided crossing between $m_F=\pm1$ hyperfine levels in \ybiononeseventhree (Sam3), which are likewise expected to be degenerate according to the conventional Hamiltonian.}
    \label{fig:splitting_Er_Yb}
\end{figure*}

% Observation of additional avoided crossings
For all isotopes under investigation, at least one transition group exhibits avoided crossings in the vicinity of zero magnetic field, as illustrated in Fig.~\ref{fig:splitting_Er_Yb}, which result from additional interactions not accounted for in the spin Hamiltonian specified in \Cref{eq:conventional_ham}. At zero magnetic field, the eigenstates with $m_F\neq0$ of erbium and ytterbium ions are expected to form degenerate pairs. For the \eriononesixseven spin system, transition groups D\textsubscript{1}, D\textsubscript{2} and E do not exhibit any avoided crossing, implying that the relevant level pairs $\ket{-,\pm3}, \ket{\pm,\pm4}$ and $\ket{+,\mp3}$ remain degenerate and consistent with the conventional Hamiltonian. In contrast, transition groups F\textsubscript{1} and F\textsubscript{2} (analogously C\textsubscript{1} and C\textsubscript{2}) exhibit clear zero-field avoided crossings (ZFACs), which can be associated with lifting of the degeneracy within levels $\ket{-,\pm2}$ and $\ket{+,\mp2}$, respectively. Furthermore, the transition groups J\textsubscript{1} and J\textsubscript{2} reveal ZFACs arising from coupling between the states $\ket{+,\mp1}$ and $\ket{-,\pm1}$, respectively. Similarly, both ytterbium isotopes exhibit ZFACs, as shown in Figs.~\ref{fig:splitting_Er_Yb}(e) and \ref{fig:splitting_Er_Yb}(f). These observations are likewise not accounted for by the Hamiltonian in \Cref{eq:conventional_ham}, indicating the presence of additional perturbative interactions.

% Reproducibility aspects
Potential influences of the experimental setup on the creation of ZFACs can be ruled out by the reproducibility of the results across independent cryogenic systems and measurement configurations. This includes any contributions from uncompensated bias fields. A possible effect from the superhyperfine interaction to nearby \textsuperscript{183}W nuclei can be ruled out on the basis of qualitative and quantitative arguments. The observed splittings indicate uniform interactions that affect all spins in the ensemble, as evidenced by the narrow linewidths of the transitions in the high-resolution spectra. The reproducibility across all investigated samples, including crystals from different growth processes and suppliers, excludes crystal quality as the underlying origin. Furthermore, the absence of detectable dependence on dopant concentration suggests that interactions between dopant ions do not dominate. Detailed considerations of the aforementioned experimental and sample aspects are provided in the Supplementary Information~\cite{Supplement}. These observations lead to the conclusion that the observed splittings originate from an intrinsic property of the individual ions.

% Short summary and segue to following sections
Since the additional ZFACs are confined to a narrow magnetic field range of $|\bm{B}_0|\lesssim\SI{2}{\milli\tesla}$, the conventional spin Hamiltonian in \Cref{eq:conventional_ham} remains a simple yet sufficiently accurate approximation for many practical applications. However, our results demonstrate the significance of interactions beyond those conventionally included in the spin Hamiltonian. In particular, the non-negligible contributions to the transition frequencies and line shapes in zero magnetic field provide strong evidence for the presence of higher-rank interactions and rhombic strain as discussed in the following sections. Hence, analyzing the experimental data requires a more elaborate model Hamiltonian including the octupole interaction (\Cref{eq:HAMoctupole}), the hexadecapole interaction (\Cref{eq:HAMhexadecapole}), and random lattice deformations (\Cref{eq:HAMstrain})
\begin{equation}
    \hatH_\mathrm{full} = \hatH_\mathrm{eff,0} + \hatH_\mathrm{oct} + \hatH_\mathrm{hdp} + \hatH_\mathrm{strain}.
    \label{eq:full_ham}
\end{equation}

\begin{table*}[tbh]
    \centering
    \caption{Published experimental hyperfine parameters for \eriononesixseven~(a), \ybiononesevenone and \ybiononeseventhree~(b) in \cwo in frequency units (MHz). Despite being distributed, $\delta A_2$ here corresponds to the maximum of the strain-broadened line shape and depends on the defect concentration of the crystal. Inclusion of $\alpha_{00}$ and $\alpha_{11}$ does not improve the fit for \ybiononeseventhree. The signs of $A_\perp$, $\delta A_2$, and $\alpha_{31}$ are indeterminate.}
    
    \smallskip
    
    (a) \\
    \begin{tabular}{ llcccccccc } 
    \hline \hline
    Source & Method* & $A_\perp$  & $A_\parallel$ & $P_\parallel$ & $\delta A_2$ & $\alpha_{31}$ & $\alpha_{11}$ & $d_\perp$ & $d_\parallel$ \\
    \hline
    
    Antipin et al.~\cite{Antipin.1968} & XEPR & 878(9) & 127(2) & - & - & - & - & - & -\\
    
    Marsh et al.~\cite{Marsh.2026} & bbEPR & -871.09(4) & -128.3(2) & -5.04(6) & - & - & - & - & - \\
    
    This work, \Cref{eq:conventional_ham} & bbEPR & -871.303(6) & -127.043(28) & -5.103(3) & - & - & - & - & - \\
    
    This work, \Cref{eq:full_ham} & bbEPR & -871.275(7) & -127.013(36) & -5.104(3) & 0.46(4) & 29.7(11)$\cdot10^{-3}$ & 6.3(8)$\cdot10^{-3}$  & 4.6(13)$\cdot10^{-4}$ & 4.9(26)$\cdot10^{-4}$ \\
    
    \hline \hline
    \end{tabular}

    \medskip

    (b) \\
    \begin{tabular}{ llcccccccc } 
    \hline \hline
    Source & Method* & $^{171}A_\perp$ & $^{171}A_\parallel$ & $^{171}\delta A_2$ & $^{173}A_\perp$& $^{173}A_\parallel$ & $^{173}P_\parallel$ & $^{173}\delta A_2$ & $^{173}\alpha_{31}$ \\
    \hline
    
    Ranon \& Volterra$^\dagger$~\cite{Ranon.1964} & XEPR & 3100(9) & 836(9) & - & 854(3) & 231(3) & <60 & - & - \\
    
    Kirton \& Newman~\cite{Kirton.1964} & XEPR & - & 809(3) & - & 854(9) & 216(3) & - & - & - \\
    
    Sattler \& Nemarich$^\dagger$~\cite{Sattler.1970a} & XEPR & 3082(3) & 789.1(15) & - & 850.8(25) & 217(1) & <21 & - & - \\
    
    Tiranov et al.~\cite{Tiranov.2026} & Opt. & 3083.84(1) & 787(1) & -  & & & & & \\
    
    This work, \Cref{eq:conventional_ham} & bbEPR & 3083.844(3) & 786.85(40) & - & -849.65(6) & -217.03(23) & 13.99(5) & - & - \\
    
    This work, \Cref{eq:full_ham} & bbEPR & 3083.844(3) & 786.85(40) & 0.7(2) & -849.65(6) & -217.03(23) & 13.99(5) & 0.2(1) & 16(31)$\cdot10^{-3}$ \\
    
    \hline \hline
    \end{tabular}

    \medskip
    *XEPR: X-band EPR, Opt.: Optical spectroscopy, bbEPR: Broadband EPR.
    \\ $^\dagger$Authors specify the magnitude of the parameters.
    
    \label{table:HamParametersErYb}
\end{table*}

\section{Avoided crossings caused by high-rank hyperfine interactions}

% Spectral observation
The pseudospin-nuclear octupole and hexadecapole hyperfine interactions, introduced in~\Cref{eq:HAMoctupole} and~\Cref{eq:HAMhexadecapole}, lift the degeneracy between the pairs of states with $m_F= \pm2$ of \eriononesixseven (likewise states with $m_F= \pm2$ in \ybiononeseventhree) by creating avoided crossings between these energy levels. In \eriononesixseven, the produced ZFACs are directly accessible via the F\textsubscript{1} and F\textsubscript{2} (or C\textsubscript{1} and C\textsubscript{2}) transition groups. These doublet-to-doublet transition groups involve one degenerate pair of $m_F=\pm3$ states. The two superimposed avoided crossings create a characteristic diamond-shaped spectral line pattern in an external magnetic field parallel to the $c$-axis (see \Cref{fig:splitting_Er_Yb}(b)). Transition group G links both octupole-split doublets ($\ket{-,\pm2} \leftrightarrow \ket{+,\pm2}$) and forms a field-symmetric avoided crossing, see \Cref{fig:new_zefoz_G}, which is discussed in more detail in \Cref{sec:new_zefoz_points}.

% Extended Hamiltonian: Octupole
We form an extended Hamiltonian as $ \hatH_\mathrm{eff,0} +  \hatH_\mathrm{oct}$, to derive the $\alpha_{00}$, $\alpha_{11}$ and $\alpha_{31}$ parameters and refine hyperfine and quadrupole parameters of \eriononesixseven. As described in \Cref{ssec:higher_rank_terms}, the zero-field splitting of the $m_F=\pm2$ states directly relates to the magnitude of $\alpha_{31}$. The sign of $\alpha_{31}$ does not influence transition frequencies or intensities. Both $\alpha_{00}$ and $\alpha_{11}$ independently yield a satisfactory match to the observed transition frequencies. 
However, when varied simultaneously, their fitted values exhibit a strong correlation. Although the corresponding $q=0$ and $q=\pm1$ rank-3 operators are linearly independent over the complete spin Hilbert space, their projections onto the set of experimentally probed transitions produce nearly proportional frequency shifts. The present dataset therefore constrains primarily one linear combination of $\alpha_{00}$ and $\alpha_{11}$, rather than both coefficients independently. To avoid reporting unstable covariance-derived uncertainties, we retain only $\alpha_{11}$, to which the measured transitions are more sensitive, and fix $\alpha_{00}$ to zero. This choice should be regarded as a parametrization of the identifiable interaction, not as evidence that the omitted tensor component is physically absent.
%\npk{ Remove: \sout{However, $\alpha_{11}$ allows for matching the hyperfine and quadrupole parameters from the fit to \Cref{eq:conventional_ham} more accurately. Thus, we omit $\alpha_{00}$ and determine the $\alpha_{31}$ and $\alpha_{11}$ parameters to most precisely describe the experimental spectra in terms of the octupole interaction.} }
The RMS error of the fit thereby improves to \SI{0.08}{\mega\hertz}.

% Extended Hamiltonian: Hexadecapole
The inclusion of the octupole interaction generates a fixed ratio of the splittings $\Delta_{\ket{-,\pm2}}$ and $\Delta_{\ket{+,\mp2}}$, which, however, does not precisely match the observation. This fine mismatch (along with other small deviations) can be further reduced by the addition of the hexadecapole term and fitting the zero-field spectrum to the extended Hamiltonian $ \hatH_\mathrm{eff,0} + \hatH_\mathrm{oct} + \hatH_\mathrm{hdp}$, from which we obtain two additional coefficients $d_{\parallel}$ and $d_\perp$. This yields a match between the model and the measured spectra within the experimental accuracy, with a correct separation of the $m_F=\pm2$ states, while the $m_F = \pm3$ and $m_F = \pm4$ doublets remain degenerate. The RMS error of the fit decreases further to \SI{0.06}{\mega\hertz}.

% Octupole fit of 173Yb
We have also extracted the $\alpha_{31}$ octupole coefficient for \ybiononeseventhree, which is found to be twice as small as that of \eriononesixseven. Smaller octupole or hexadecapole coefficients could not be confidently determined due to the limited resolution of the \ybiononeseventhree dataset. All parameters obtained from the fit to \Cref{eq:full_ham} (including strain parameters discussed in \Cref{sec:strain_zfac}) are summarized in \Cref{table:HamParametersErYb}.

\section{Strain-induced avoided crossings}
\label{sec:strain_zfac}

% Line shapes of J transitions
While absorption profiles of most hyperfine transitions exhibit narrow linewidths of $\sim\SI{0.1}{\mega\hertz}$, transitions involving the $m_F=\pm1$ states, such as J$_1$ and J$_2$, show asymmetric broadening of the linewidth by more than an order of magnitude (see \Cref{fig:splitting_Er_Yb}(a)). The corresponding line shapes exhibit an avoided-crossing-like behavior near zero magnetic field. This suggests that strain-induced random local lattice distortions provide a dominant contribution to both the frequency shifts and the distributed line shapes.

% Distribution parameters
A key indication that strain is indeed responsible for the observed behavior is provided by the \ybiononesevenone spin ensemble. Here, both the electronic and nuclear spins are $S=\sfrac{1}{2}$ and $I=\sfrac{1}{2}$, respectively, such that higher-rank multipolar interactions can be excluded. The observed ZFAC of the $m_F=\pm1$ states (\Cref{fig:splitting_Er_Yb}(e)), therefore, provides direct evidence for a strain-induced contribution. In particular, strain induces local lattice distortions, reducing the crystalline symmetry and thereby generating asymmetry in the $A_\perp$ components and off-diagonal hyperfine tensor elements that couple the $m_F=\pm1$ states. We model this contribution by assuming a distribution of strain amplitudes as described in \Cref{ssec:local_strain}. This distribution reproduces the asymmetric line shapes and avoided-crossing behavior observed experimentally. The overlay of simulations including both higher-rank interactions and strain for \eriononesixseven and \ybiononesevenone is shown in the Supplementary Information~\cite{Supplement}, while the corresponding results for the F\textsubscript{1,2} and J\textsubscript{1,2} groups of erbium are shown in \Cref{fig:splitting_Er_Yb}(a). We obtained width parameters $\xi(\textsuperscript{167}\text{Er})=2.3\cdot10^{-3}$ and $\xi(\textsuperscript{171}\text{Yb})=6.1\cdot10^{-4}$ for Sam2 and Sam4, respectively. For the discrete Hamiltonian parameters listed in \Cref{table:HamParametersErYb}, $\delta A_2$ represents the value that reproduces the transition frequency related to the maxima of the spectral envelope. Since the width of the strain distribution is associated with the defect concentration, this parameter is expected to be sample-dependent.

% Dominant strain terms
These observations identify local strain in the $ab$-plane as the dominant strain component for the $m_F=\pm1$ transitions. Other strain terms considered for the scheelite structure in Ref.~\cite{Malkin.2017} either couple only quadratically to the relevant states or predominantly produce a global shift of the energy levels, and therefore cannot account for the observed ZFACs and asymmetric broadening. Strain components involving the $c$-axis can affect states with $m_F\neq\pm1$ and may consequently dominate the broadening and symmetry of the line shapes of other transitions. However, at comparable strain amplitudes, their effect is expected to be approximately three to four orders of magnitude smaller (<\,\SI{1}{\kilo\hertz}). Therefore, all transitions except the ones involving the $m_F=\pm1$ states (J\textsubscript{1,2} and H\textsubscript{1,2}) remain narrow.

% Summary
The most rigorous treatment of interactions with crystalline strain would employ the full crystal-field Hamiltonian. Here, we instead use an effective spin model in which local lattice deformations are incorporated as a reduced-symmetry extension of the hyperfine tensor. Strain-induced modifications of the electronic $g$-tensor and the quadrupole tensor (as well as higher-rank terms) are neglected, as their contributions are expected to be small in the low-field regime investigated in this work. A detailed determination of the microscopic strain distribution is beyond the scope of this work. Nevertheless, the results demonstrate that the nominal $S_4$ point symmetry of REIs in \cwo is locally distorted. The resulting symmetry breaking produces the observed asymmetric line shapes and avoided crossings in an ensemble measurement. The magnitude of this distortion is approximately three orders of magnitude smaller than the axial hyperfine-tensor components, explaining why its effects become appreciable only at vanishingly small magnetic fields below $\SI{2}{\milli\tesla}$. At larger magnetic fields, the Zeeman interaction dominates over the strain-induced coupling, and the hyperfine tensor obtained under ideal $S_4$ symmetry provides a good approximation.

\section{Improved coherence at new ZEFOZ points}
\label{sec:new_zefoz_points}

\begin{figure*}[tbh!]
% \\badwwmi\user\groups\Quantum-Speicher\Austausch\Georg Mair\Figures\zefoz_transition\zefoz_transition_G_v4\zefoz_transition_G_labeled.png
    \centering
    \includegraphics[width=\linewidth]{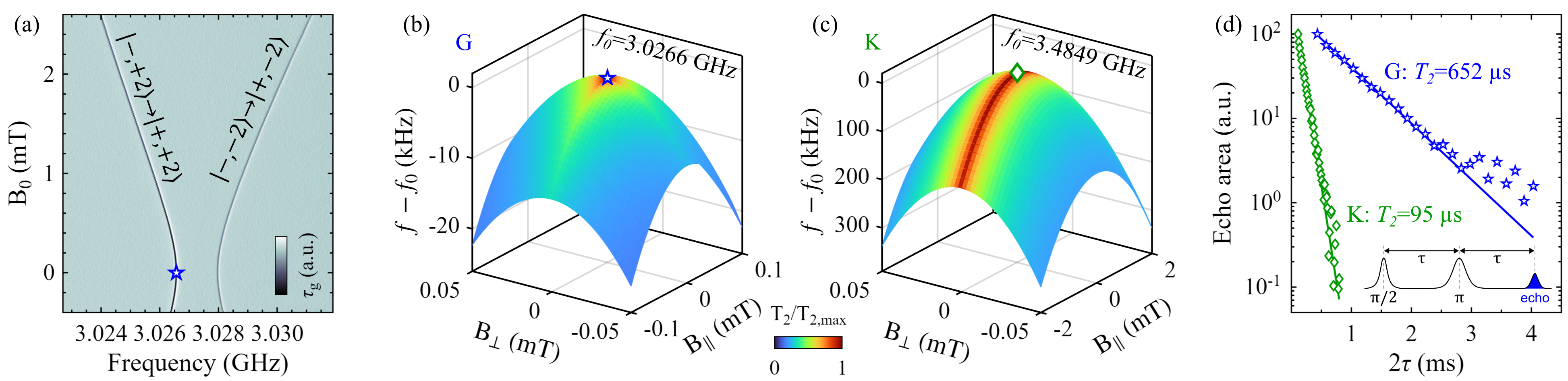}
    \caption{
    \textbf{Zero-field ZEFOZ points in \eriononesixseven:\cwo:}
    (a)~Field-swept group delay spectrum of transition group G with $\bm{B}_0\parallel c$. Due to the avoided crossings between the $m_F=\pm2$ states arising from the octupole interaction, the G transition frequency becomes first-order insensitive to magnetic fields along the $c$-axis at $B_0=\SI{0}{\milli\tesla}$. The emerging ZEFOZ points are expected to exhibit superior coherence properties compared to the ZEFOZ point in the K transition.
    (b),~(c)~Calculated transition frequency $f$ for fields applied parallel and perpendicular to the $c$-axis. The color encodes the estimated coherence time. The curvature at the ZEFOZ point (marked by a star (G) and a diamond (K)) is 17 times smaller for the G transition compared to K.
    (d)~Hahn echo decay of the G and K transitions measured in the \SI{34}{ppm} sample (Sam2) confirms improved coherence of the G transition.
    }
    \label{fig:new_zefoz_G}
\end{figure*}

% ZEFOZ transitions introduction
Precise knowledge of the hyperfine parameters enables the identification of ZEFOZ points. They are defined by a vanishing gradient of the transition frequency $f$ with respect to the magnetic field, $\boldsymbol{S}_1=\nabla_{\boldsymbol{B}}f=0$. As a consequence, the sensitivity to magnetic noise $\delta\boldsymbol{B}$, such as field fluctuations arising from nuclear spins in the host lattice, is significantly reduced. These operating points are particularly desired to enhance the coherence of spin ensembles~\cite{Zhong.2015, Ortu.2018, Tiranov.2026}. The decoherence rate of the spin system at the ZEFOZ point can be estimated as $(\pi T_2)^{-1} = \boldsymbol{S}_1\cdot\delta\boldsymbol{B} + \delta\boldsymbol{B}\cdot\boldsymbol{S}_2\cdot\delta\boldsymbol{B}$. The coherence time $T_2$ is then mainly limited by $\boldsymbol{S}_2^{i,j}=\frac{1}{2}\partial^2 f/\partial B_i\partial B_j$ with $i,j\in\{x,y,z\}$. This curvature tensor is commonly reduced to the largest absolute eigenvalue $\left|S_2\right|$ as the main dephasing contribution. Within the conventional spin Hamiltonian (\Cref{eq:conventional_ham}), the $\ket{\pm, 0}$ transition (K) represents the only ZEFOZ point at zero magnetic field.

% New ZEFOZ points in G transition group and 173Yb
The avoided crossings generated by the octupole interaction and coupling to random local strain give rise to additional ZEFOZ points that are absent in the conventional $S_4$-symmetric spin Hamiltonian. For \eriononesixseven, two such ZEFOZ points emerge at zero magnetic field in transition group G, as shown in \Cref{fig:new_zefoz_G}. Within the conventional Hamiltonian (\Cref{eq:conventional_ham}), the first-order Zeeman sensitivity $\boldsymbol{S}_1$ vanishes for magnetic fields perpendicular to the crystalline $c$-axis, but remains finite for $\bm{B}_0\parallel c$. The octupole-induced avoided crossings between the states $\ket{+,\mp2}$ and $\ket{-,\pm2}$ remove this remaining first-order sensitivity. Consequently, the transitions $\ket{\pm,-2}$ and $\ket{\pm,+2}$ become first-order insensitive to magnetic fields along the $c$-axis, such that the full gradient $\boldsymbol{S}_1$ vanishes and true ZEFOZ conditions are established. The corresponding second-order Zeeman coefficients are $|S_2^{\ket{\pm,-2}}|=\SI{3.96}{\mega\hertz\per\milli\tesla\squared}$ and $|S_2^{\ket{\pm,+2}}|=\SI{3.82}{\mega\hertz\per\milli\tesla\squared}$. Similarly, ZEFOZ conditions arising from the octupole interaction are expected to emerge for the corresponding transitions in \ybiononeseventhree.

% Coherence time estimate
For \eriononesixseven, the newly identified ZEFOZ points in group G are particularly promising. As already discussed, the $\ket{\pm,0}$ transition in group K constitutes the only zero-field ZEFOZ point predicted by the conventional Hamiltonian, but its larger second-order Zeeman sensitivity implies inferior magnetic-noise-limited coherence. Comparing the corresponding curvatures predicts a coherence time approximately 17 times longer for the group G transitions. Using the magnetic-field fluctuations of \SI{7}{\micro\tesla} estimated for host and dopants of a \SI{100}{ppm} \erion:\cwo crystal by Monte Carlo simulations~\cite{Marsh.2026}, the $\ket{\pm,+2}$ transition yields an estimated coherence time of $T_2=\SI{425}{\micro\second}$.

% Coherence times measurement
We perform pulsed EPR with Hahn echo sequences using our broadband setup (see \Cref{fig:intro}(d) and Ref.~\cite{Supplement}) to measure the decay of the echo area as a function of pulse separation (\Cref{fig:new_zefoz_G}(d)) at zero magnetic field. For the \SI{34}{ppm} sample (Sam2), we obtained coherence times of $T_2=\SI{95}{\micro\second}$ for K and $T_2=\SI{652}{\micro\second}$ for G. In the \SI{50}{ppm} sample (Sam3), we measured $T_2=\SI{36}{\micro\second}$ for the K and $T_2=\SI{414}{\micro\second}$ for the G transition. This verifies the improved coherence properties of the newly identified ZEFOZ point. The observed deviation of the coherence-time ratio from the predicted value may be attributed to an uncompensated static magnetic bias field in the experimental setup. Our analysis indicates that a residual field of \qtyrange{9}{17}{\micro\tesla} along the crystalline $c$-axis is sufficient to account for the discrepancy. The maximum coherence times could therefore be achieved by using, for example, a vector magnet to compensate for field components along all three spatial axes, or by operating in a magnetically isolated environment.

% Strain-induced ZEFOZ and 171Yb
A similar mechanism generates additional zero-field ZEFOZ points via strain-induced avoided crossings in the $m_F=\pm1$ doublet. In \eriononesixseven, this mechanism affects the J\textsubscript{1,2} and H\textsubscript{1,2} transition groups. The avoided crossings introduce first-order insensitivity to magnetic fields parallel to $c$, while the corresponding transitions already exhibit vanishing first-order sensitivity for fields perpendicular to $c$. The combination, therefore, produces zero-field ZEFOZ conditions. The same mechanism occurs in \ybiononesevenone, where the strain-induced avoided crossing of the $m_F=\pm1$ doublet generates additional zero-field ZEFOZ points. The coherence time measurement in \ybiononesevenone under conditions similar to those for \eriononesixseven yields $T_2=\SI{880}{\micro\second}$ on transition J\textsubscript{1}\textsuperscript{(\ybiononesevenone)} ($\ket{-,0}\leftrightarrow\ket{-,-1}$) and $T_2=\SI{8.0}{\milli\second}$ on transition K\textsuperscript{(\ybiononesevenone)}. In this case, the ZEFOZ point of the  $m_F=\pm1$ doublet does not provide lower second-order Zeeman sensitivity and thus does not offer a corresponding advantage in magnetic-noise-limited coherence, as compared to the $\ket{\pm,0}$ ZEFOZ transition. However, it provides a substantial enhancement in coherence at zero magnetic field over the unstrained conditions.

% Further possibilities
The ZEFOZ conditions may also occur at finite magnetic fields. In particular, transitions that have a vanishing first-order Zeeman sensitivity for fields perpendicular to $c$ can become full ZEFOZ points when an avoided crossing additionally suppresses the sensitivity to $\bm{B}_0\parallel c$. This mechanism substantially increases the number of candidate ZEFOZ transitions compared with the conventional spin Hamiltonian. It may also provide access to ZEFOZ transitions involving higher crystal-field or electronic manifolds, although we do not investigate these possibilities in the present work.

\section{Conclusion and outlook}

% General summary
We have investigated the low-field hyperfine structure of \eriononesixseven, \ybiononesevenone, and \ybiononeseventhree ions in \cwo using broadband EPR spectroscopy at millikelvin temperatures. The achieved spectral resolution enabled a quantitative determination of the spin-Hamiltonian parameters and yielded the first direct measurement of the quadrupole interaction of \ybiononeseventhree in this host.

% Observation of high-rank interactions and strain
Beyond the conventional Zeeman, hyperfine, and quadrupole descriptions, we observed reproducible avoided crossings in zero magnetic field that remain unexplained within the conventional effective-spin Hamiltonian. By comparing isotopes with different nuclear spins and exploiting the symmetry of the observed level splittings, we have identified two distinct microscopic origins. The avoided crossings involving $m_F=\pm2$ states can be consistently described by pseudospin-nuclear octupole and hexadecapole interactions, while those involving $m_F=\pm1$ states arise from local symmetry lowering caused by random strain. The observation of octupole-induced splittings in both \eriononesixseven and \ybiononeseventhree provides, to the best of our knowledge, the first direct spectroscopic evidence of high-rank interactions of REIs in a solid-state system.

% New ZEFOZ points
An important consequence of these interactions is the formation of additional ZEFOZ transitions. In contrast to the conventional Hamiltonian, where only a single ZEFOZ point exists at zero magnetic field, the octupole- and strain-induced avoided crossings suppress the remaining first-order magnetic-field sensitivity of further hyperfine transitions and generate additional protected transitions. Their coherence time is estimated to provide up to a 17-fold improvement. Spin echo measurements confirm that these newly identified operating points can substantially outperform the conventional zero-field ZEFOZ transition, establishing higher-rank interactions as a practical resource for coherence enhancement rather than merely a small correction to the spin Hamiltonian.

% Outlook
In a broader context, our results demonstrate that weak interactions hidden within conventional EPR measurements become directly accessible in the strong electron-nuclear mixing regime near zero magnetic field. This opens a pathway to experimentally investigate higher-rank multipolar interactions in rare-earth systems and to assess their influence on coherence, control, and quantum device operation. The methodology developed here is readily transferable to other rare-earth hosts and paramagnetic solids. It provides a framework for systematically extending effective-spin Hamiltonians beyond the conventional rank-two description. Future studies combining lower linewidths, improved isotopic control, and coherence spectroscopy may enable the exploration of even weaker higher-rank interactions and their role in protected quantum states.

%__________________________________________________________________

%_______________________________________________________________________________
\begin{acknowledgments}

We acknowledge support from the German Research Foundation via Germany’s Excellence Strategy (EXC-2111-390814868) and the German Federal Ministry of Education and Research via the project QuaMToMe (Grant No.~16KISQ036).

GM thanks P.~Oehrl for help with the pulsed EPR setup.

NI was supported by JST NEXUS, Grant Number JPMJNX26C1, Japan.

TC thanks R. Bouland, J. Debray, and S. Zhang (Inst. Néel) for their help in preparing the samples and acknowledges support from the Plan France 2030 project QMEMO No. ANR-22-PETQ-0010 and the project CHORIZO No. ANR-24-CE47-1190.

\end{acknowledgments}

%\begin{widetext}
%   \hrulefill
%\end{widetext}

% \input{Sections/X-Appendix}

\newpage
%\bibliography{References_citavi,References_new} %,References_old}
%\bibliography{Bibliography}

\clearpage
\onecolumngrid             % optional
\begin{center}\textbf{\large Supplementary Information: Observation of high-rank interactions of rare earths in the solid state}\end{center}
\setcounter{equation}{0}\setcounter{figure}{0}\setcounter{table}{0}\setcounter{section}{0}
\renewcommand{\theequation}{S\arabic{equation}}
\renewcommand{\thefigure}{S\arabic{figure}}
\renewcommand{\thetable}{S\arabic{table}}

\section{Theory}

\subsection{Diagonalization of the conventional Hamiltonian}
\label{apx:hamiltoniandiagonalization}

\subsubsection{Energy levels}

For a magnetic field $B_0$ applied along the $z$-axis parallel to the axis of anisotropy ($c$-axis in CaWO$_4$), the hyperfine level structure can be derived by diagonalization of the effective spin Hamiltonian. In the high-field limit, the eigenstates are well described by the Zeeman basis $\ket{m_S,m_I}$, where $m_S$ and $m_I$ are the electronic and nuclear spin projections, respectively. In the low-field regime, where the Zeeman interaction is no longer dominant, these quantum numbers are not conserved due to state mixing. We therefore introduce the coupled basis $\ket{\pm, m_F}$, denoted by the sign of $m_S$ in the high-field regime, and the total angular momentum $m_F=m_S+m_I$ along the $z$-axis. The following derivation extends the analysis of electro-nuclear spin states of bismuth dopants in silicon~\cite{Mohammady.2010, Mohammady.2012} by the addition of the quadrupolar interaction.

% Hamiltonian decomposition
Within the conventional model, the Hamiltonian $\hatH_\mathrm{eff,0}$ is block diagonal and can be decomposed into 1- and 2-dimensional sub-Hamiltonians $\hatH_{m_F}^\mathrm{1d,2d}$. The Hamiltonian $\hatH_{m_F}^\mathrm{1d}$ corresponds to the uncoupled states with $m_F=\pm(S+I)$ whereas $\hatH_{m_F}^\mathrm{2d}$ is spanned by $\ket{m_S=\pm\tfrac{1}{2},m_I=m_F\mp\tfrac{1}{2}}$ with $m_F\neq\pm(S+I)$. The latter is a real-valued symmetric $2\times2$ matrix and can be expanded in the Pauli basis as
\begin{equation}
    \hatH_{m_F}^\mathrm{2d} = \Delta_{m_F}\hat{\sigma}_z +\Omega_{m_F}\hat{\sigma}_x - \epsilon_{m_F}\mathbb{1},
\end{equation}
where $\mathbb{1}$ denotes the identity matrix, $\hat{\sigma}_i$ are the Pauli matrices and
\begin{equation}
    \begin{split}
        \Delta_{m_F} =& \frac{B_0}{2}(\mu_\mathrm{B}g_\parallel + \mu_\mathrm{N}g_\mathrm{N}) + \frac{m_F}{2}(A_\parallel - 2P_\parallel) \\
        \Omega_{m_F} =& \frac{A_\perp}{2}\sqrt{I(I+1) + \frac{1}{4} - m_F^2} \\
        \epsilon_{m_F} =& \mu_\mathrm{N}g_\mathrm{N}B_0m_F + \frac{A_\parallel}{4} + P_\parallel\left(\frac{1}{3}I(I+1) - m_F^2 - \frac{1}{4}\right).
    \end{split}
\end{equation}

% Eigenenergies
The corresponding eigenenergies are found by calculating the roots of the second-degree characteristic polynomial of $\hatH_{m_F}^\mathrm{2d}$.
\begin{equation}
\begin{split}
    E_{\ket{\pm, m_F\neq\pm(S+I)}} =
    & -\mu_\mathrm{N}g_\mathrm{N}B_0 m_F - \frac{A_\parallel}{4} - P_\parallel\left(\frac{1}{3}I(I + 1) - m_F^2 - \frac{1}{4}\right) \\
    & \pm\frac{1}{2}\bigg[\left((\mu_\mathrm{B}g_\parallel + \mu_\mathrm{N}g_\mathrm{N})B_0 + (A_\parallel - 2P_\parallel)m_F\right)^2 + A^2_\perp\left(I(I+1) + \frac{1}{4} - m^2_F\right)\bigg]^{1/2}
\end{split}
\label{eq:levelenergy_coupled}
\end{equation}
For uncoupled states, the one-dimensional Hamiltonians yield energies
\begin{equation}
    E_{\ket{\pm, m_F=\pm(S+I)}} = \pm\Delta_{m_F} - \epsilon_{m_F},
    \label{eq:levelenergy_uncoupled}
\end{equation}
consistent with the expressions reported in Ref.~\cite{AbragamBleaney.1970}.

\begin{figure}
\begin{minipage}{0.33\textwidth}
% \\badwwmi\user\groups\Quantum-Speicher\Austausch\Georg Mair\Figures\level_plot\level_plot_171Yb_v1\level_plot_171Yb_labeled.png
    (a)
    \par
    \includegraphics[height=11cm]{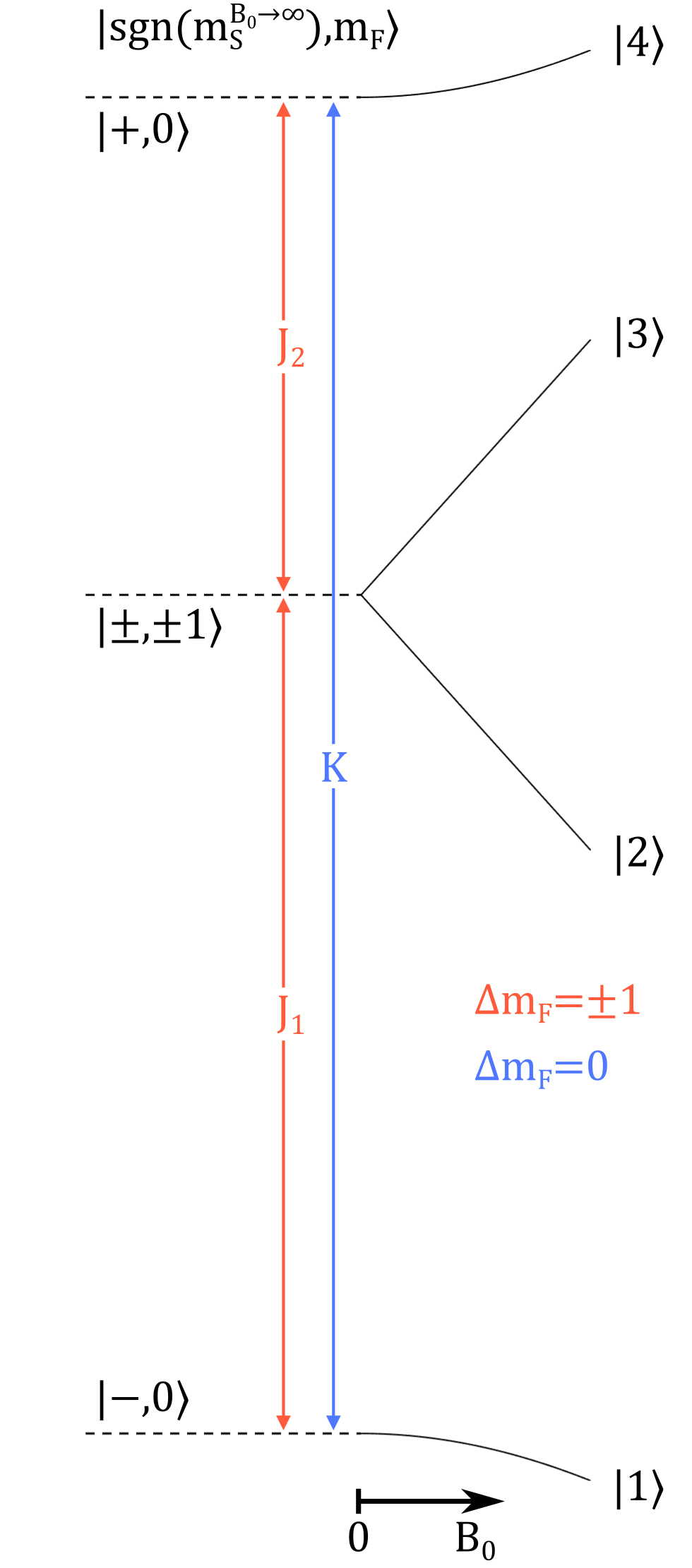}
\end{minipage}
\quad
\begin{minipage}{0.52\textwidth}
% \\badwwmi\user\groups\Quantum-Speicher\Austausch\Georg Mair\Figures\level_plot\level_plot_173Yb_v1\level_plot_173Yb_labeled.png
    (b)
    \par
    \includegraphics[height=11cm]{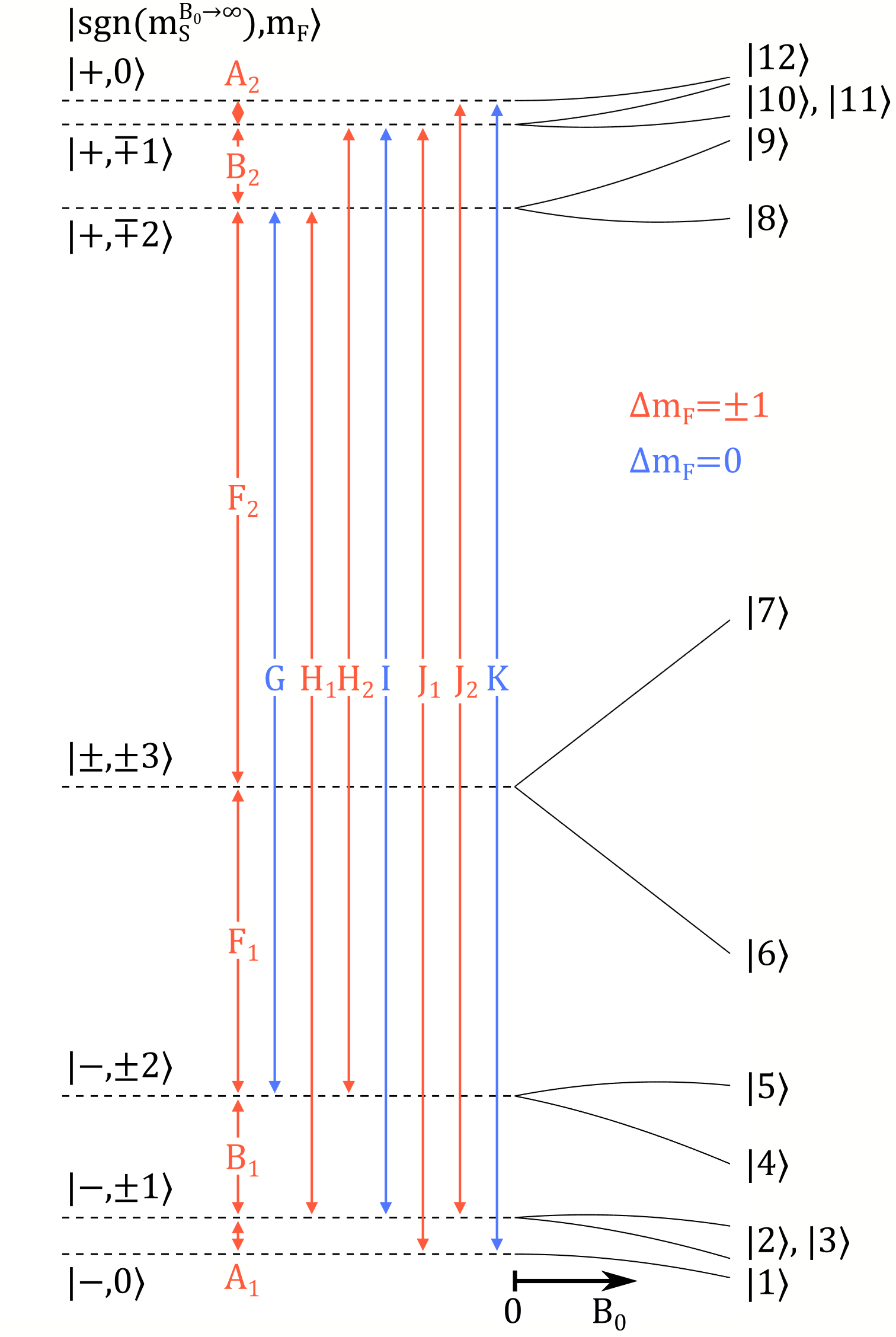}
\end{minipage}
    \caption{\textbf{Level diagrams of \ybion:}
    Energy levels and corresponding state labels for \ybiononesevenone (a) and \ybiononeseventhree (b). The transition groups are labeled following the nomenclature introduced for \eriononesixseven. The colored arrows represent allowed transitions following the selection rules.
    }
\label{fig:levelplot_171Yb_173Yb}
\end{figure}

% Mixing coefficients   
The eigenstates of the coupled subspaces can be expressed as superpositions of the uncoupled basis states. Introducing the mixing angle $\tfrac{1}{2}\theta_{m_F}$, the corresponding coefficients for the states $\ket{\pm\tfrac{1}{2},m_F\mp\tfrac{1}{2}}$ are
\begin{equation}
\label{eq:mixingcoeff}
    a_{m_F} = \cos\biggl(\frac{\theta_{m_F}}{2}\biggr), \quad b_{m_F} = \sin\biggl(\frac{\theta_{m_F}}{2}\biggr).
\end{equation}
The diagonalization of $\hatH_{m_F}^\mathrm{2d}$ under rotation by the mixing angle yields
\begin{equation}
    \cos(\theta_{m_F}) = \frac{\Delta_{m_F}}{\sqrt{\Delta_{m_F}^2 + \Omega_{m_F}^2}}.
\end{equation}
The mixing angle is evidently equal to zero for the uncoupled $m_F = \pm(S+I)$ states.
By restricting $\theta_{m_F}$ to $[0, \pi]$, we ensure $a_{m_F},b_{m_F}>0$. The relative sign $\zeta$ of the coefficients is then determined by their ratio
\begin{equation}
\label{eq:mixingcoeffsign}
%\begin{split}
    \zeta = \sgn\biggl(\frac{b_{m_F}^\pm}{a_{m_F}^\pm}\biggr)
    = \sgn\Biggl(\frac{-\Delta_{m_F}\pm\sqrt{\Delta_{m_F}^2 + \Omega_{m_F}^2}}{\Omega_{m_F}}\Biggr) \\
    = \pm\sgn(A_\perp)
%\end{split}
\end{equation}
obtained from the eigenvalue problem
\begin{equation}
    \hatH_{m_F}^\mathrm{2d}
    \begin{pmatrix} a_{m_F}^\pm \\ b_{m_F}^\pm \end{pmatrix}
    = E_{\ket{\pm, m_F}}
    \begin{pmatrix} a_{m_F}^\pm \\ b_{m_F}^\pm \end{pmatrix}.
\end{equation}
Note that the sign of $A_\perp$ unambiguously defines $\zeta$. The states are finally expressed by the Zeeman basis states as
\begin{equation}
    \ket{\pm, m_F} = a_{m_F}\ket{\pm\frac{1}{2}, m_F\mp\frac{1}{2}} + \zeta b_{m_F}\ket{\mp\frac{1}{2}, m_F\pm\frac{1}{2}},
\end{equation}
The coupled basis states of \eriononesixseven, \ybiononesevenone and \ybiononeseventhree are summarized in~\Cref{table:CoupledStates}. Level diagrams of both ytterbium isotopes are shown in \Cref{fig:levelplot_171Yb_173Yb}.

\subsubsection{Hamiltonian parameters}

The hyperfine and quadrupole coefficients can be extracted from the transition frequencies observed in the broadband EPR spectrum at zero magnetic field. In this regime, the Zeeman interaction vanishes, and the level structure is governed solely by the hyperfine and quadrupole interactions. Consequently, the transition energies are determined by the parameters of the interaction tensors $\bm{A}$ and $\bm{P}$ entering the conventional Hamiltonian $\hatH_\mathrm{eff,0}$ as defined in the main text.

The perpendicular component of the hyperfine tensor $A_\perp$ can be obtained directly from the splitting $\Delta E_{\ket{\pm,0}}$ between the two singlet states with $m_F=0$. Using Eq.~(\ref{eq:levelenergy_coupled}) this relation can be written as
\begin{equation}
    \lvert A_\perp\rvert = \frac{\Delta E_{\ket{\pm, 0}}}{\sqrt{I(I+1) + \frac{1}{4}}}.
\end{equation}

\begin{figure}[th]
% \\badwwmi\user\groups\Quantum-Speicher\Austausch\Georg Mair\Figures\quadrupole_splitting_FHJ\quadrupole_splitting_FHJ_v1\quadrupole_splitting_FHJ_labeled.png
    \centering
    \includegraphics[width=0.99\linewidth]{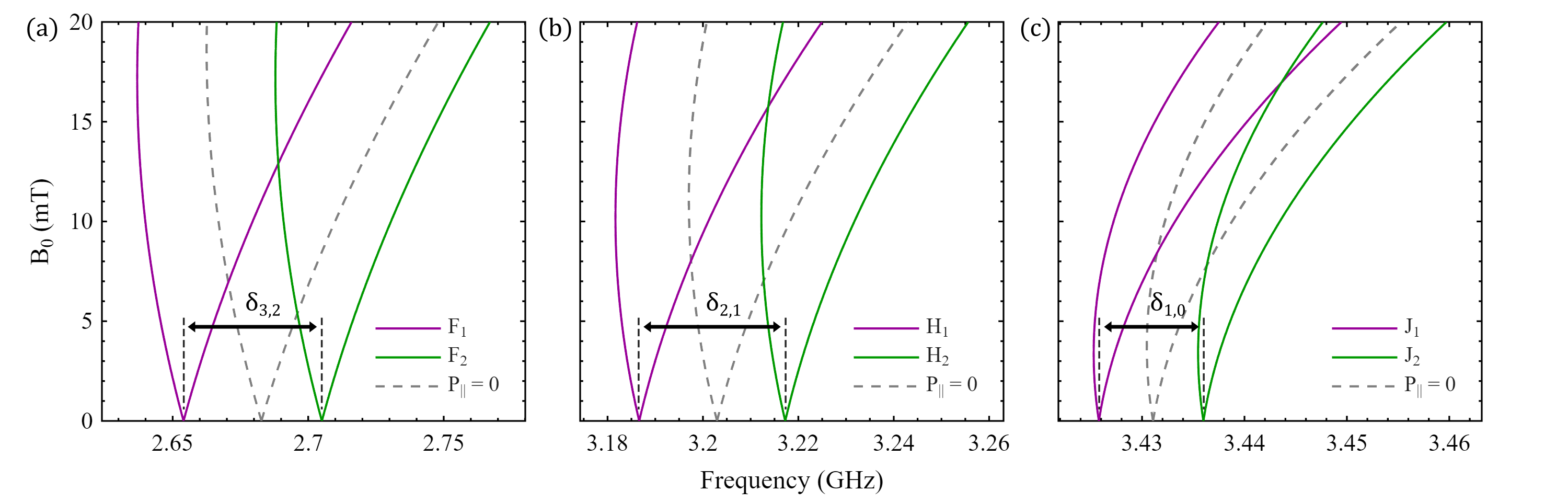}
    \caption{\textbf{Quadrupole splitting:}
    The quadrupole interaction splits pairs of $\Delta m_F=\pm1$ transitions, which would otherwise coincide (dashed lines). The resulting transition groups are labeled by a common letter and an index, such as the F\textsubscript{1,2} (a), H\textsubscript{1,2} (b), and J\textsubscript{1,2} (c) transition groups of \eriononesixseven. The separation $\delta_{m_F^{(1)}, m_F^{(2)}}$ is proportional to the interaction coefficient and involved $m_F$ numbers (\Cref{eq:Pparallel_formula}). $P_\parallel$ can thus be determined from measurement of zero-field transition frequencies.
    }
    \label{fig:quadrupole_splitting}
\end{figure}

The magnitude of the quadrupole interaction can be inferred from pairs of $\Delta m_F=\pm1$ transitions. Let $m_F^{(1)}=m_F^{(2)}+1$ and $\big|m_F^{(1,2)}\big|\neq S+I$. For every transition of the form $|-,m_F^{(1)}\rangle\leftrightarrow|+,m_F^{(2)}\rangle$, a corresponding transition $|-,m_F^{(2)}\rangle\leftrightarrow|+,m_F^{(1)}\rangle$ exists. In the broadband EPR spectrum, such a pair is separated by a frequency difference
\begin{equation}
    \delta_{m_F^{(1)}, m_F^{(2)}} = 
    \left(E_{|+,m_F^{(2)}\rangle} - E_{|-,m_F^{(1)}\rangle}\right) - 
    \left(E_{|+,m_F^{(1)}\rangle} - E_{|-,m_F^{(2)}\rangle}\right),
\end{equation}
as illustrated in \Cref{fig:quadrupole_splitting}. This splitting depends only on the quadrupole interaction and defines the corresponding coefficient as
\begin{equation}
\label{eq:Pparallel_formula}
    P_\parallel = -\frac{\delta_{m_F^{(1)}, m_F^{(2)}}}{2\big(m_F^{(1)} + m_F^{(2)}\big)}.
\end{equation}
Quadrupole-split transition pairs are labeled by a common capital letter with indices 1 and 2.

Once $\left|A_\perp\right|$ and $P_\parallel$ are known, the remaining parameter $A_\parallel$ can be calculated from the energy difference of any transition involving at least one level with $m_F\neq0$. The resulting expression is particularly compact for transitions with $\Delta m_F=0$.
\begin{equation}
\label{eq:Aparallel_formula}
    A_\parallel = 2P_\parallel\pm\sqrt{\frac{\Delta E_{\ket{\pm, m_F\neq0}}^2 - A_\perp^2\left(I(I+1) + \frac{1}{4} - m_F^2\right)}{m_F^2}}
\end{equation}
This leaves us with two solutions for $A_\parallel$ that have opposite signs (assuming $|A_\parallel|\gg|P_\parallel|$). Since the transition frequencies alone do not define $\sgn(A_\parallel)$ unambiguously, we consider differences in transition intensities of low-temperature spectra in \Cref{ssec:hyperfine_sign}.

For $I=\sfrac{1}{2}$ nuclei (like \ybiononesevenone), \Cref{eq:Pparallel_formula} yields no result. Instead, $A_\parallel$ can be calculated directly from the energy difference (transition frequency) of a $\Delta m_F=\pm1$ transition using \Cref{eq:levelenergy_coupled} and \Cref{eq:levelenergy_uncoupled}. For the $\ket{-, 0}\rightarrow\ket{-, -1}$ transition, we obtain
\begin{equation}
    A_\parallel^{(I=\sfrac{1}{2})} = 2\Delta E_{\ket{-,0}\rightarrow\ket{-,-1}}-\left|A_\perp\right|.
\end{equation}
In this way, the complete set of hyperfine and quadrupole parameters of the effective Hamiltonian can be determined solely from the zero-field EPR transition spectrum.

\begin{table*}[tbh]
% \\badwwmi\user\groups\Quantum-Speicher\Austausch\Georg Mair\Matlab\Transition_levels\mixing_coefficients.mlx
\centering
\caption{Energy levels of \eriononesixseven, \ybiononesevenone, and \ybiononeseventhree in \cwo at $B_0=\SI{0}{\milli\tesla}$ in (un)coupled basis using the conventional Hamiltonian $\hatH_\mathrm{eff,0}$. Exact mixing coefficients are given by \Cref{eq:mixingcoeff} and \Cref{eq:mixingcoeffsign} up to a global phase. The energy $E_\mathrm{fit}$ is calculated from the parameters of the extended Hamiltonian $\hatH_\mathrm{full}$ presented in this work. $E_\mathrm{exp}$ is calculated from the experimentally observed transition frequencies. All energies are given in frequency units (MHz).}
\begin{tabular}{ l|lllrr }
\hline \hline
Isotope & Label & Coupled basis & Zeeman representation & $E_\mathrm{fit}$ ($\hatH_\mathrm{full}$) & $E_\mathrm{exp}$ \\
    &   & $\ket{\pm, m_F}$  &  $\ket{m_s, m_I}$ &  & \\
    
\hline
$^{167}$Er& $\ket{1}$ & $\ket{-, 0}$ & $\tfrac{1}{\sqrt{2}}\left(\ket{-\sfrac{1}{2}, +\sfrac{1}{2}} + \ket{+\sfrac{1}{2}, -\sfrac{1}{2}}\right)$ & -1685.13 & -1685.13 \\
& $\ket{2}$ & $\ket{-, -1}$ & $0.719\ket{-\sfrac{1}{2}, -\sfrac{1}{2}} + 0.695\ket{+\sfrac{1}{2}, -\sfrac{3}{2}}$ & -1636.45 & -1636.49 \\
& $\ket{3}$ & $\ket{-, +1}$ & $0.695\ket{-\sfrac{1}{2}, +\sfrac{3}{2}} + 0.719\ket{+\sfrac{1}{2}, +\sfrac{1}{2}}$ & -1635.49 & -1635.52 \\
& $\ket{4}$ & $\ket{-, -2}$ & $0.734\ket{-\sfrac{1}{2}, -\sfrac{3}{2}} + 0.679\ket{+\sfrac{1}{2}, -\sfrac{5}{2}}$ & -1477.24 & -1477.23 \\
& $\ket{5}$ & $\ket{-, +2}$ & $0.679\ket{-\sfrac{1}{2}, +\sfrac{5}{2}} + 0.734\ket{+\sfrac{1}{2}, +\sfrac{3}{2}}$ & -1476.38 & -1476.37 \\
& $\ket{6}$ & $\ket{-, -3}$ & $0.758\ket{-\sfrac{1}{2}, -\sfrac{5}{2}} + 0.652\ket{+\sfrac{1}{2}, -\sfrac{7}{2}}$ & -1154.66 & -1154.66 \\
& $\ket{7}$ & $\ket{-, +3}$ & $0.652\ket{-\sfrac{1}{2}, +\sfrac{7}{2}} + 0.758\ket{+\sfrac{1}{2}, +\sfrac{5}{2}}$ & -1154.66 & -1154.66 \\
& $\ket{8}$ & $\ket{-, -4}$ & $\ket{-\sfrac{1}{2}, -\sfrac{7}{2}}$ & -257.97 & -257.97 \\
& $\ket{9}$ & $\ket{+, +4}$ & $\ket{+\sfrac{1}{2}, +\sfrac{7}{2}}$ & -257.97 & -257.97 \\
& $\ket{10}$ & $\ket{+, +3}$ & $0.652\ket{+\sfrac{1}{2}, +\sfrac{5}{2}} - 0.758\ket{-\sfrac{1}{2}, +\sfrac{7}{2}}$ & 1177.31 & 1177.32 \\
& $\ket{11}$ & $\ket{+, -3}$ & $0.758\ket{+\sfrac{1}{2}, -\sfrac{7}{2}} - 0.652\ket{-\sfrac{1}{2}, -\sfrac{5}{2}}$ & 1177.31 & 1177.32 \\
& $\ket{12}$ & $\ket{+, +2}$ & $0.679\ket{+\sfrac{1}{2}, +\sfrac{3}{2}} - 0.734\ket{-\sfrac{1}{2}, +\sfrac{5}{2}}$ & 1550.19 & 1550.19 \\
& $\ket{13}$ & $\ket{+, -2}$ & $0.734\ket{+\sfrac{1}{2}, -\sfrac{5}{2}} - 0.679\ket{-\sfrac{1}{2}, -\sfrac{3}{2}}$ & 1550.74 & 1550.74 \\
& $\ket{14}$ & $\ket{+, +1}$ & $0.695\ket{+\sfrac{1}{2}, +\sfrac{1}{2}} - 0.719\ket{-\sfrac{1}{2}, +\sfrac{3}{2}}$ & 1739.89 & 1739.92 \\
& $\ket{15}$ & $\ket{+, -1}$ & $0.719\ket{+\sfrac{1}{2}, -\sfrac{3}{2}} - 0.695\ket{-\sfrac{1}{2}, -\sfrac{1}{2}}$ & 1740.78 & 1740.80 \\
& $\ket{16}$ & $\ket{+, 0}$ & $\tfrac{1}{\sqrt{2}}\left(\ket{+\sfrac{1}{2}, -\sfrac{1}{2}} - \ket{-\sfrac{1}{2}, +\sfrac{1}{2}}\right)$ & 1799.74 & 1799.74 \\

\hline
$^{171}$Yb& $\ket{1}$ & $\ket{-, 0}$ & $\tfrac{1}{\sqrt{2}}\left(\ket{+\sfrac{1}{2}, -\sfrac{1}{2}} - \ket{-\sfrac{1}{2}, +\sfrac{1}{2}}\right)$ & -1738.63 & -1738.63 \\
& $\ket{2}$ & $\ket{-, -1}$ & $\ket{-\sfrac{1}{2}, -\sfrac{1}{2}}$ & 196.38 & 196.38 \\
& $\ket{3}$ & $\ket{+, +1}$ & $\ket{+\sfrac{1}{2}, +\sfrac{1}{2}}$ & 197.05 & 197.05 \\
& $\ket{4}$ & $\ket{+, 0}$ & $\tfrac{1}{\sqrt{2}}\left(\ket{+\sfrac{1}{2}, -\sfrac{1}{2}} + \ket{-\sfrac{1}{2}, +\sfrac{1}{2}}\right)$ & 1345.21 & 1345.21 \\

\hline
$^{173}$Yb& $\ket{1}$ & $\ket{-, 0}$ & $\tfrac{1}{\sqrt{2}}\left(\ket{-\sfrac{1}{2}, +\sfrac{1}{2}} + \ket{+\sfrac{1}{2}, -\sfrac{1}{2}}\right)$ & -1257.5 & -1257.6 \\
& $\ket{2}$ & $\ket{-, -1}$ & $0.742\ket{-\sfrac{1}{2}, -\sfrac{1}{2}} + 0.670\ket{+\sfrac{1}{2}, -\sfrac{3}{2}}$ & -1177.1 & -1177.0 \\
& $\ket{3}$ & $\ket{-, +1}$ & $0.670\ket{-\sfrac{1}{2}, +\sfrac{3}{2}} + 0.742\ket{+\sfrac{1}{2}, +\sfrac{1}{2}}$ & -1176.7 & -1176.8 \\
& $\ket{4}$ & $\ket{-, -2}$ & $0.790\ket{-\sfrac{1}{2}, -\sfrac{3}{2}} + 0.612\ket{+\sfrac{1}{2}, -\sfrac{5}{2}}$ & -908.2 & -908.1 \\
& $\ket{5}$ & $\ket{-, +2}$ & $0.612\ket{-\sfrac{1}{2}, +\sfrac{5}{2}} + 0.790\ket{+\sfrac{1}{2}, +\sfrac{3}{2}}$ & -908.0 & -908.0 \\
& $\ket{6}$ & $\ket{-, -3}$ & $\ket{-\sfrac{1}{2}, -\sfrac{5}{2}}$ & -224.6 & -224.6 \\
& $\ket{7}$ & $\ket{+, +3}$ & $\ket{+\sfrac{1}{2}, +\sfrac{5}{2}}$ & -224.6 & -224.8 \\
& $\ket{8}$ & $\ket{+, +2}$ & $0.612\ket{+\sfrac{1}{2}, +\sfrac{3}{2}} - 0.790\ket{-\sfrac{1}{2}, +\sfrac{5}{2}}$ & 1053.9 & 1053.8 \\
& $\ket{9}$ & $\ket{+, -2}$ & $0.790\ket{+\sfrac{1}{2}, -\sfrac{5}{2}} - 0.612\ket{-\sfrac{1}{2}, -\sfrac{3}{2}}$ & 1054.0 & 1054.1 \\
& $\ket{10}$ & $\ket{+, +1}$ & $0.670\ket{+\sfrac{1}{2}, +\sfrac{1}{2}} - 0.742\ket{-\sfrac{1}{2}, +\sfrac{3}{2}}$ & 1238.6 & 1238.6 \\
& $\ket{11}$ & $\ket{+, -1}$ & $0.742\ket{+\sfrac{1}{2}, -\sfrac{3}{2}} - 0.670\ket{-\sfrac{1}{2}, -\sfrac{1}{2}}$ & 1238.9 & 1238.9 \\
& $\ket{12}$ & $\ket{+, 0}$ & $\tfrac{1}{\sqrt{2}}\left(\ket{+\sfrac{1}{2}, -\sfrac{1}{2}} - \ket{-\sfrac{1}{2}, +\sfrac{1}{2}}\right)$ & 1291.4 & 1291.5 \\
\hline \hline
\end{tabular}
\label{table:CoupledStates}
\end{table*}

\subsection{Sign of hyperfine coefficients}
\label{ssec:hyperfine_sign}

\begin{figure}[th]
% \\badwwmi\user\groups\Quantum-Speicher\Austausch\Georg Mair\Figures\hyperfine_parameter_sign\hyperfine_parameter_sign_171Yb_CWO_v1\hyperfine_parameter_sign_171Yb_CWO_labeled.png
    \centering
    \includegraphics[width=0.99\linewidth]{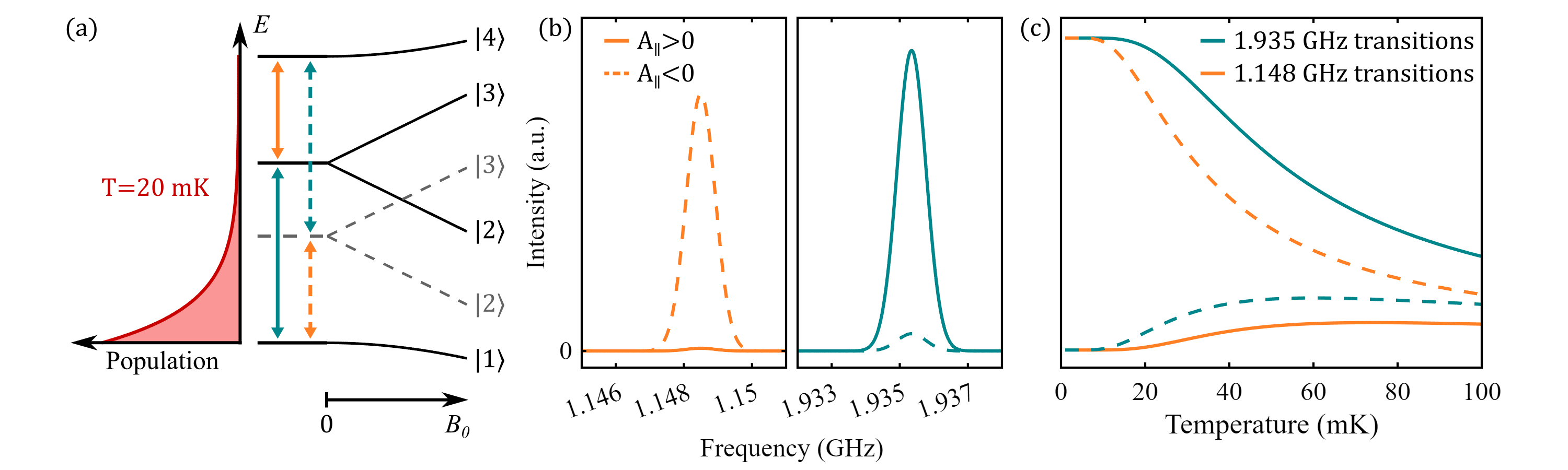}
    \caption{\textbf{Sign of $A_\parallel$ for \ybiononesevenone:}
    (a)~The sign of $A_\parallel$ determines the relative position of energy levels. For $A_\parallel<0$ (dashed lines), levels $\ket{2}$ and $\ket{3}$ have lower energy than for $A_\parallel>0$, which affects the relative thermal population at millikelvin temperatures. (b)~The sign-dependent population difference of the $\Delta m_F=\pm1$ transitions translates to the spectroscopically observable transition intensity. (c)~Determining $\sgn(A_\parallel)$ from the relative intensities is valid independent of the temperature of the spin system. The populations equilibrate across all levels in the high-temperature limit.
    }
    \label{fig:hyperfinesign}
\end{figure}

The literature has often omitted the definition of the signs of hyperfine coefficients. While they do not affect transition frequencies, they are crucial for determining correct transition intensities, which becomes most apparent at low temperatures. As described in Ref.~\cite{Nehrkorn.2015}, the continuous wave EPR line intensity for a transition $\ket{a}\rightarrow\ket{b}$ is defined by the population difference $\chi$ of the two levels, the driving field $\bm{B}_1$, and the transition matrix element of the total magnetic dipole moment operator $\bm{\mu}=\mel{a}{\hat{\bm{\mu}}}{b}$, with $\hat{\bm{\mu}}=\mu_\mathrm{N}g_\mathrm{N}\hat{\bm{I}}-\mu_\mathrm{B}\boldsymbol{g}\hat{\bm{S}}$. For a CPW geometry (driving field in the plane normal to the propagation vector $k$ of the microwave), the intensity can be approximated as
\begin{equation}
    I_{ab} = \chi\frac{\abs{\bm{B}_1}^2}{8}(\lvert\boldsymbol{\mu}\rvert^2 - \lvert\boldsymbol{n}^T_k\boldsymbol{\mu}\rvert^2),
    \label{eq:transitionintensity}
\end{equation}
where $\boldsymbol{n}_k$ is the unit propagation vector. Broadband EPR allows comparison of relative transition intensities across the spectrum. A qualitative analysis is sufficient to unambiguously determine the sign of the coefficients. The effect can most easily be described with the example of \ybiononesevenone, due to its simple four-level structure, as depicted in \Cref{fig:hyperfinesign}. We consider the transition groups connecting the highest ($\ket{4}$) and lowest ($\ket{1}$) to the middle levels ($\ket{2}$ and $\ket{3}$).
The sign of $A_\parallel$ shifts the relative energy of levels $\ket{2}$ and $\ket{3}$ as indicated in \Cref{fig:hyperfinesign}(a). Due to the strong population gradient across the levels at millikelvin temperatures, the $\chi$ differs significantly for the considered transitions. Consequently, measurement of relative transition intensities reveals the corresponding sign of the coefficient (\Cref{fig:hyperfinesign}(b,c)). Analogously, the sign of $A_\parallel$ can be inferred for \eriononesixseven. Note that the sign of $A_\perp$ remains ambiguous. In this work, we find $A_\parallel<0$ for \eriononesixseven and \ybiononeseventhree, while $A_\parallel>0$ for \ybiononesevenone.

\subsection{Analytical estimate of quadrupole coefficients}

The complete analytical calculations of quadrupole coefficients have been summarized in multiple works~\cite{Ram.1985, Erickson.1986}, which include the calculation of the pseudo-quadrupole coefficients introduced by Teplov~\cite{Teplov.1968} and true quadrupole coefficients stated by Bleaney et al.~\cite{Bleaney.1982}. 
Following their work, a more general second-rank interaction spin Hamiltonian in orthorhombic symmetry is represented by 
\begin{equation}
    \hatH_{q} = D \left[ I_z^2 -\frac{1}{3}I(I+1) \right] + E \left[ I_x^2 - I_y^2 \right],
\end{equation}
where $D$ is the quadrupole coefficient of the standard trace-zero part of the quadrupole interaction, which is applicable to higher tetragonal symmetry. If the symmetry is reduced within the a-b plane, an additional term with a coefficient $E$ is present. Both these coefficients are composed of the following contributions:
\begin{align}
    D =& P_{4f} + P_\mathrm{lat} + D_\mathrm{pseudo},\\
    E =& \frac{1}{3}\eta \left( P_{4f} + P_\mathrm{lat} \right) + E_\mathrm{pseudo}.
\end{align}
Here, the moment of the $4f$ electrons acting on the nucleus induced by the crystal field (with shielding by the closed shells) is given by the coefficient $P_{4f}$:
\begin{equation}
    P_{4f} = - \frac{3 e^2Q}{4I(2I-1)} \langle r^{-3} \rangle_{4f} (1-R_0) \langle J||\alpha||J\rangle \langle|3J_z^2-J(J+1)|\rangle,
\end{equation}
and the electric-field gradient of the lattice, anti-shielded by the closed-shell electrons, acts on the quadrupole moment of the nucleus via a coefficient
\begin{equation}
    P_\mathrm{lat} = - \frac{3 Q}{2I(2I-1)} \langle r^{-2} \rangle_{4f}
    \frac{(1-\gamma_\infty)}{(1-\sigma_2)} B^0_2.
\end{equation}
Here, $Q$ is the free ion quadrupole moment and $B^0_2$ is the crystal field coefficient. 
$\langle r^n \rangle_{4f}$ are the radial integrals over the $4f$ electrons,
$\gamma_\infty\simeq-60$, $\sigma_2\simeq 0.60$ and $R_0\simeq0.127$ are the Sternheimer shielding-antishielding factors~\cite{Sternheimer.1986, Gupta.1973}. The latter take into account the polarization of the core electrons due to field-gradient contributions from lattice charges and the valence electrons. $\langle J||\alpha||J\rangle$ and $\langle|3J_z^2-J(J+1)|\rangle$ are the matrix elements tabulated in Ref.~\cite{AbragamBleaney.1970}. The asymmetric contribution $\frac{1}{3}\eta \left( P_{4f} + P_\mathrm{lat} \right) $ turns into zero in case of the perfect $S_4$ symmetry, since $\eta = (V_{xx}-V_{yy})/V_{zz}$, where $V_{ii}$ are the principal axes of the electric field gradient. In the presence of random strain deformations, which reduce the point symmetry to $C_2$, $\eta$ becomes non-zero. However, since the deformation is small, the resulting asymmetry is also small and can be neglected.

The second-order pseudo-quadrupole coefficient strongly depends on the interaction with the crystal field 
\begin{equation}
    \Lambda_{ii} = \sum_{n\neq0} \frac{A_J | \langle 0|J_i|n\rangle |^2}{E_n-E_0},
\end{equation}
which we approximate to
\begin{equation}
    \Lambda_{ii} = A_{ii} \sum_{n\neq0} \frac{1}{E_n-E_0},
\end{equation}
where $A_{ii}$ are the ground-state hyperfine tensor coefficients and $E_n$ the energies of the crystal field levels, which then form the pseudo-coefficients as follows
\begin{align}
    D_\mathrm{pseudo} &= A_J \left[ \frac{1}{2} (\Lambda_{xx}+\Lambda_{yy})  - \Lambda_{zz} \right] ,\\
    E_\mathrm{pseudo} &= A_J \left[ \frac{1}{2} (\Lambda_{xx}-\Lambda_{yy})  \right].
\end{align}
Analogously to the true quadrupole coefficients, $E_\mathrm{pseudo}$ is zero in perfect $S_4$ symmetry. Also, in perfect symmetry $\Lambda_{xx}=\Lambda_{yy}$ therefore $D_\mathrm{pseudo} = A_J \left( \Lambda_{xx} - \Lambda_{zz} \right) = A_J \left( \Lambda_{\perp} - \Lambda_{\parallel} \right) $.

With these equations, we can estimate the expected magnitudes of both true and pseudo-quadrupole coefficients. The lowest crystal-field excitation within the ground multiplet is \(\Delta_{\mathrm{CF}}=20.13~\mathrm{cm}^{-1}=603.48~\mathrm{GHz}\) for erbium~\cite{Becker.2025}, whereas a representative separation of \(\Delta_{\mathrm{CF}}=220~\mathrm{cm}^{-1}=6.6~\mathrm{THz}\) is used for ytterbium~\cite{Zheng.2004}. The considerably smaller denominator for erbium enhances its expected pseudo-multipolar contribution relative to that of ytterbium.

Using the \(A_{\perp}\) values listed in the main paper gives
\begin{equation}
    D_\mathrm{pseudo,\eriononesixseven}
    \sim
    -2.71~\mathrm{MHz}
\end{equation}
for \eriononesixseven, and
\begin{equation}
    D_\mathrm{pseudo,\ybiononeseventhree}
    \sim
    -0.53~\mathrm{MHz}
\end{equation}
for \ybiononeseventhree, subject to the single-level and diagonal-matrix-element approximations described above. For \ybiononesevenone, \(I=\sfrac{1}{2}\), and an irreducible rank-two nuclear-spin operator vanishes identically. A nuclear pseudo-quadrupole term is therefore not an independent operator in the \ybiononesevenone spin Hamiltonian.

These estimates suggest that the rank-two interaction in \eriononesixseven may contain a comparatively large pseudo-quadrupole contribution. For \ybiononeseventhree, the larger crystal-field separation suppresses the pseudo term, so that the effective quadrupole interaction is expected to be more strongly governed by the intrinsic electric-quadrupole contribution.

\subsection{Derivation of higher-rank terms in $S_4$ symmetry}

%\subsection{Ranks in spin Hamiltonian}

Interaction between the electronic and nuclear degrees of freedom can be expanded in irreducible tensor operators:
\begin{equation}
    \hatH_\mathrm{e\text{-}n}
    =
    \sum_k
    \hat{\bm{T}}^\mathrm{e}_{k}
    \cdot
    \hat{\bm{T}}^\mathrm{n}_{k}
    =
    \sum_k \sum_{q,q'=-k}^{k}
    (-1)^{q+q'}
    \hat{T}_{kq}^\mathrm{n}
    \hat{T}_{kq'}^\mathrm{e},
    \label{eq:general_en_tensor}
\end{equation}
where \(\hat{\bm{T}}_{k}^\mathrm{e}\) and \(\hat{\bm{T}}_{k}^\mathrm{n}\) are irreducible tensor operators of rank \(k\) acting on the electronic and nuclear coordinates, respectively.
For an isotropic system with electronic angular momentum \(S\), nuclear spin \(I\), and total angular momentum \(F\), the first-order energy shift can be written as
\begin{equation}
\begin{split}
    W_F^{(1)}
    &=
    \bra{(IS)FM_F}
    \hatH_{e\text{-}n}
    \ket{(IS)FM_F}
    \\
    &=
    \sum_k
    (-1)^{I+S+F}
    \begin{Bmatrix}
        S & I & F \\
        I & S & k
    \end{Bmatrix}
    \bra{S}\!\left\|
        \hat{\bm{T}}_{k}^\mathrm{e}
    \right\|\!\ket{S}
    \bra{I}\!\left\|
        \hat{\bm{T}}_{k}^\mathrm{n}
    \right\|\!\ket{I},
    \label{eq:first_order_tensor_energy}
\end{split}
\end{equation}
where the braces denote a Wigner \(6j\) symbol, equivalently expressible in terms of Racah coefficients, and the double bars denote reduced matrix elements. This tensor expansion underlies the conventional hierarchy of magnetic-dipole, electric-quadrupole, magnetic-octupole, and electric-hexadecapole hyperfine interactions, corresponding to ranks \(k=1,2,3,\) and \(4\), respectively \cite{Schwartz.1955, Beloy.2008a}. 

For the lowest Kramers doublet, the electronic degrees of freedom are represented by an effective spin \(S=\sfrac{1}{2}\). Any operator projected onto this two-dimensional electronic subspace can therefore be expanded in the identity and the three effective-spin operators \(\hat{S}_x,\hat{S}_y,\hat{S}_z\). Consequently, higher-rank effective electron-nuclear interactions do not correspond to rank-\(k>1\) tensor operators constructed solely from the effective electronic spin.
Instead, they appear as products of a rank-one electronic pseudospin operator and higher-rank nuclear spin operators, subject to the symmetry of the crystallographic site. This distinction is important because the rank of the microscopic electronic multipole is not, in general, retained as the rank of an operator acting within the effective \(S=\sfrac{1}{2}\) subspace \cite{Rudowicz.2001,Rudowicz.2015}.

Conservation of even parity and time-reversal symmetry requires that the even ranks are allowed only for electric multipole terms written as        
\begin{align}
 \hatH_\text{zfs}^I &= \sum_{k=2, 4, 6} \sum_{q=-k}^k B_{kq} \hat{T}^I_{kq},
 \label{eq:electric_ranks}
\end{align}
while the odd ranks are hyperfine magnetic higher-rank terms including electron spin operators
\begin{align}
 \hatH_{\text{hf}} &= \sum_{kq} \sum_{q'=-1}^1 I_{kq, 1q'} \hat{T}^I_{kq} \hat{T}^S_{1q'}.
 \label{eq:magnetic_ranks}
\end{align}

Higher-rank terms in the effective spin Hamiltonian may have two distinct origins. The first is an intrinsic, first-order coupling to a genuine higher nuclear multipole moment. The second is a pseudo-multipolar contribution arising from virtual transitions to excited crystal-field states through products of lower-rank interactions. The latter is described by second-order perturbation theory:
\begin{equation}
\begin{split}
    W_g^{(2)}
    &=
    -\sum_{n\neq g}
    \sum_{k_1,k_2}
    \frac{
        \bra{g}
        \hatH_\mathrm{e\text{-}n}^{(k_1)}
        \ket{n}
        \bra{n}
        \hatH_\mathrm{e\text{-}n}^{(k_2)}
        \ket{g}
    }{
        E_n-E_g
    }.
    \label{eq:pseudo_general}
\end{split}
\end{equation}
Here, \(\ket{g}\) denotes a state in the ground Kramers doublet and \(\ket{n}\) denotes an excited crystal-field state. The product of two tensor interactions of ranks \(k_1\) and \(k_2\) can generate effective operators of ranks
\begin{equation}
    \left|k_1-k_2\right|
    \leq k
    \leq k_1+k_2.
    \label{eq:tensor_product_ranks}
\end{equation}
For example, the product \(M1\times M1\) can generate contributions up to rank \(k=2\), whereas the cross term \(M1\times E2\) can generate a rank \(k=3\) contribution with the same symmetry as a magnetic-octupole interaction. Similarly, \(E2\times E2\) can generate contributions up to rank \(k=4\) \cite{Schwartz.1955, Beloy.2008a}.
Such effective terms are conventionally described as \emph{pseudo-multipolar} interactions. Unlike intrinsic first-order multipolar terms, pseudo terms may depend on the magnetic field because
the field changes the electronic eigenstates, energy denominators, and matrix elements entering Eq.~\eqref{eq:pseudo_general}
\cite{Baker.1969}. At zero magnetic field, however, intrinsic and
pseudo-multipolar terms transform according to the same irreducible representations of the site-symmetry group. They, therefore, cannot be separated using zero-field spectroscopic data alone. 
Experimentally determined coefficients of rank \(k\geq2\) should consequently be interpreted as effective coefficients containing both intrinsic and higher-order contributions.

\subsubsection{Nuclear zero-field splitting Hamiltonian}

% Even ranks of the zero-field Hamiltonian take the following form:
%
% \begin{align}
%  \hat{H}_\text{ZFS}^I &= \sum_{k=2, 4, 6} \sum_{q=-k}^k B_{kq} \hat{T}^I_{kq}.
% \end{align}
%
True even high-rank Hamiltonian terms, as given by \Cref{eq:electric_ranks}, produce nuclear zero-field shifts in the energy levels of paramagnetic electrons. 
The Hamiltonian should be self-conjugate:
\begin{align}
 \left(\hatH_\text{zfs}^I\right)^\dagger
  &= \sum_{k=2, 4, 6} \sum_{q=-k}^k B_{kq}^* \left(\hat{T}^I_{kq}\right)^\dagger
  \nonumber\\
  &= \sum_{k=2, 4, 6} \sum_{q=-k}^k B_{kq}^* (-1)^q \hat{T}^I_{k,-q}
  \nonumber\\
  &= \hatH_\text{zfs}^I.
\end{align}
The Hamiltonian also should be invariant under the $\hat{S}_{4z}^+$ operation and time inversion.
\begin{align}
 \hat{S}_{4z}^+ \hatH_\text{zfs}^I \left( \hat{S}_{4z}^+ \right)^\dagger 
 &= \sum_{k=2, 4, 6} \sum_{q=-k}^k B_{kq} 
 \hat{S}_{4z}^+ \hat{T}^I_{kq} \left( \hat{S}_{4z}^+ \right)^\dagger 
 \nonumber\\
 &= \sum_{k=2, 4, 6} \sum_{q=-k}^k B_{kq} e^{i \frac{\pi}{2} q} \hat{T}^I_{kq}
 \nonumber\\
 &= \hatH_\text{zfs}^I,
\end{align}
and
\begin{align}
 \Theta \hatH_\text{zfs}^I \Theta^{-1} &= \sum_{k=2, 4, 6} \sum_{q=-k}^k B_{kq}^* (-1)^{k-q} \hat{T}^I_{k,-q}.
\end{align}
Therefore,
\begin{align}
 B_{kq} &= B_{k,-q}^* (-1)^{q}.
 \label{Eq_B_SC}
 \\
 B_{kq} &= B_{k,-q}^* (-1)^{k+q},
 \label{Eq_B_TR}
 \\
 B_{kq} &= B_{kq} e^{i \frac{\pi}{2} q}.
 \label{Eq_B_S4}
\end{align}

The combinations of $(k,q)$ simultaneously satisfying Eqs.~(\ref{Eq_B_SC}), (\ref{Eq_B_TR}), and (\ref{Eq_B_S4}) are 
\begin{align}
 (k, q) &= (2, 0), (4, 0), (4, \mp 4), (6, 0), (6, \mp 4).
 \label{Eq_kq_S4}
\end{align}
Therefore,
\begin{align}
 \hatH_\text{zfs}^I &= %\sum_{k=2, 4, 6} \sum_{q=-k}^k B_{kq} \hat{T}^I_{kq}.
 B_{20} \hat{T}^I_{20} 
 + B_{40} \hat{T}^I_{40} + B_{4,-4} \hat{T}^I_{4,-4} + B_{4,+4} \hat{T}^I_{4,+4}
 \nonumber\\
 &+ B_{60} \hat{T}^I_{60} + B_{6,-4} \hat{T}^I_{6,-4} + B_{6,+4} \hat{T}^I_{6,+4},
 \label{Eq_Hzfs_S4}
\end{align}
with
\begin{align}
 B_{4,-4} = B_{4,+4}^*, \quad 
 B_{6,-4} = B_{6,+4}^*.
\end{align}
Within the $S_4$ symmetry, $B_{4,\mp 4}$ and $B_{6,\mp 4}$ can be complex, but the imaginary parts are small for the system close to $D_{2d}$ symmetry ($S_4$ is a subgroup of $D_{2d}$). 
In terms of the polynomials of $\hat{I}$~\cite{Lindgard.1974},
\begin{align}
 \hat{T}^I_{20} &\propto \frac{1}{2} \left( 3 \hat{I}_z^2 - \hat{\bm{I}}^2 \right),
 \\
 \hat{T}^I_{40} &\propto \frac{1}{8} \left( 35 \hat{I}_z^4 - (30 \hat{\bm{I}}^2 - 25) \hat{I}_z^2 + 3 (\hat{\bm{I}}^2)^2 - 6 \hat{\bm{I}}^2 \right),
 \\
 \hat{T}^I_{4,\mp 4} &\propto \sqrt{\frac{35}{128}} \hat{I}_\mp^4,
\end{align}
where $\hat{\bm{I}}^2 = \hat{I}_x^2 + \hat{I}_y^2 + \hat{I}_z^2 = I(I+1)$, and $\hat{I}_{\mp} = \hat{I}_x \mp i \hat{I}_y$.

\subsubsection{Hyperfine interaction}

% The hyperfine Hamiltonian take the following form:
%
% \begin{align}
%  \hat{H}_{\text{hf}} &= \sum_{kq} \sum_{q'=-1}^1 I_{kq, 1q'} \hat{T}^I_{kq} \hat{T}^S_{1q'}.
% \end{align}
%
True odd high-rank Hamiltonian terms, as given by \Cref{eq:magnetic_ranks}, produce hyperfine electron-nuclear coupled terms. 
This Hamiltonian should be as well self-conjugate:
\begin{align}
 \hatH_{\text{hf}}^\dagger &= \sum_{kq} \sum_{q'=-1}^1 I_{kq, 1q'}^* \left(\hat{T}^I_{kq}\right)^\dagger \left(\hat{T}^S_{1q'}\right)^\dagger
 \nonumber\\
 &= \sum_{kq} \sum_{q'=-1}^1 I_{k,-q, 1,-q'}^* (-1)^{q+q'} 
 \hat{T}^I_{kq} \hat{T}^S_{1q'}
 \nonumber\\
 &= \hatH_\text{hf}.
\end{align}
The Hamiltonian should be invariant under the $S_4$ operation and time inversion:
\begin{equation}
 \hat{S}_{4z}^+ \hatH_{\text{hf}} \left( \hat{S}_{4z}^+ \right)^\dagger 
 = \sum_{kq} \sum_{q'=-1}^1 I_{kq, 1q'} e^{i\frac{\pi}{2}(q+q')}\hat{T}^I_{kq} \hat{T}^S_{1q'}
 \nonumber\\
 = \hatH_{\text{hf}},
\end{equation}
and
\begin{equation}
 \Theta \hatH_{\text{hf}} \Theta^{-1} = \sum_{kq} \sum_{q'=-1}^1 I_{kq, 1q'}^* (-1)^{k-q} (-1)^{1-q'} \hat{T}^I_{k,-q} \hat{T}^S_{1,-q'}
 \nonumber\\
 = \hatH_{\text{hf}}.
\end{equation}
The hyperfine interaction parameters have to fulfill
\begin{align}
 I_{kq,1q'} &= I_{k,-q, 1,-q'}^* (-1)^{q+q'},
 \label{Eq_I_SC}
 \\
 I_{kq,1q'} &= I_{kq, 1q'} e^{i\frac{\pi}{2}(q+q')},
 \label{Eq_I_S4}
 \\
 I_{kq, 1q'} &= I_{k,-q, 1,-q'}^* (-1)^{k+q} (-1)^{1+q'}.
 \label{Eq_I_TR}
\end{align}
The combinations of $(k,q,q')$ simultaneously satisfying Eqs.~(\ref{Eq_I_SC}), (\ref{Eq_I_S4}), and (\ref{Eq_I_TR}) are
\begin{align}
 (k, q, q') =& (1, 0, 0), (1, \mp 1, \pm 1), 
 \nonumber\\
             & (3, 0, 0), (3, \mp 1, \pm 1), (3, \mp 3, \mp 1),  
 \nonumber\\
             & (5, 0, 0), (5, \mp 4, 0), (5, \mp 1, \pm 1), (5, \mp 3, \mp 1), 
 \nonumber\\
             & (5, \mp 5, \pm 1), 
 \nonumber\\
             & (7, 0, 0), (7, \mp 4, 0), (7, \mp 1, \pm 1), (7, \mp 3, \mp 1), 
 \nonumber\\
             & (7, \mp 5, \pm 1), (7, \mp 7, \mp 1).
 \label{Eq_kq_S4}
\end{align}
We use Racah-normalized operator equivalents $\tilde{C}^{(J)}_{k, q}$~\cite{Lindgard.1974} (listed up to rank $k=4$ in \Cref{tab:racah_operator_equivalents}) to write the explicit form of the pseudospin-nuclear octupole interaction term
\begin{equation}
\begin{split}
 \hatH_\text{hf}^{(3)} &= 
 \alpha_{00} 
 \hat{S}_z 
 %\left[ \frac{1}{2} \left(5\hat{I}_z^3 - (3X-1) \hat{I}_z\right)\right]
 \tilde{C}^{(I)}_{3,0} \\
 &+\hat{S}_x 
   \left[
      \alpha_{31} \frac{1}{\sqrt{2}} \left(\tilde{C}^{(I)}_{3,-3} - \tilde{C}^{(I)}_{3,+3} \right)
     + \beta_{31}  \frac{1}{i\sqrt{2}}\left(\tilde{C}^{(I)}_{3,-3} + \tilde{C}^{(I)}_{3,+3} \right) \right. \\
    & \left. \qquad
     - \alpha_{11} \frac{1}{\sqrt{2}} \left(\tilde{C}^{(I)}_{3,-1} - \tilde{C}^{(I)}_{3,+1} \right)
     - \beta_{11}  \frac{1}{i\sqrt{2}}\left(\tilde{C}^{(I)}_{3,-1} + \tilde{C}^{(I)}_{3,+1} \right)
   \right] \\
 &+\hat{S}_y 
   \left[
      \alpha_{31} \frac{1}{i\sqrt{2}}\left(\tilde{C}^{(I)}_{3,-3} + \tilde{C}^{(I)}_{3,+3} \right)
     - \beta_{31}  \frac{1}{\sqrt{2}} \left(\tilde{C}^{(I)}_{3,-3} - \tilde{C}^{(I)}_{3,+3} \right) \right. \\
    & \left. \qquad
     + \alpha_{11} \frac{1}{i\sqrt{2}}\left(\tilde{C}^{(I)}_{3,-1} + \tilde{C}^{(I)}_{3,+1} \right)
     - \beta_{11}  \frac{1}{\sqrt{2}} \left(\tilde{C}^{(I)}_{3,-1} - \tilde{C}^{(I)}_{3,+1} \right)
   \right] \\
   &=
 \alpha_{00} 
 \hat{S}_z 
 \left[ \frac{1}{2} \left(5\hat{I}_z^3 - (3\hat{\bm{I}}^2-1) \hat{I}_z\right)\right] \\
 &+
  \alpha_{31} \sqrt{\frac{5}{8}}
  \left[
 \hat{S}_x \left(\hat{I}_x^3-3 \overline{\hat{I}}_{xyy}\right)
 +
 \hat{S}_y \left(\hat{I}_y^3-3 \overline{\hat{I}}_{yxx}\right)
 \right] \\
 &-
  \alpha_{11} \sqrt{\frac{3}{32}} 
  \left[
    \hat{S}_x  \left[\hat{I}_x \left(5\hat{I}_z^2 - \hat{\bm{I}}^2 - \frac{1}{2}\right) + \left(5\hat{I}_z^2 - \hat{\bm{I}}^2 - \frac{1}{2}\right) \hat{I}_x \right]
    +
    \hat{S}_y  \left[\hat{I}_y \left(5\hat{I}_z^2 - \hat{\bm{I}}^2 - \frac{1}{2}\right) + \left(5\hat{I}_z^2 - \hat{\bm{I}}^2 - \frac{1}{2}\right) \hat{I}_y \right]
  \right] \\
  &+ \beta_{31} \sqrt{\frac{5}{8}}
  \left[
   \hat{S}_x \left(\hat{I}_y^3-3 \overline{\hat{I}}_{yxx}\right)
   -
   \hat{S}_y \left(\hat{I}_x^3-3 \overline{\hat{I}}_{xyy}\right)
  \right] \\
  &+ \beta_{11} \sqrt{\frac{3}{32}}
    \left[
     \hat{S}_x \left[\hat{I}_y \left(5\hat{I}_z^2 - \hat{\bm{I}}^2 - \frac{1}{2}\right) + \left(5\hat{I}_z^2 - \hat{\bm{I}}^2 - \frac{1}{2}\right) \hat{I}_y \right]
     - 
     \hat{S}_y \left[\hat{I}_x \left(5\hat{I}_z^2 - \hat{\bm{I}}^2 - \frac{1}{2}\right) + \left(5\hat{I}_z^2 - \hat{\bm{I}}^2 - \frac{1}{2}\right) \hat{I}_x \right]
     \right],
\label{Eq_Hhf3}
\end{split}
\end{equation}
where
\begin{align*}
 I_{30,10} &= \alpha_{00},
 \\
 I_{3,\mp 1,1,\pm 1} &= \alpha_{11} \mp i \beta_{11},
 \\
 I_{3,\mp 3,1,\mp 1} &= \alpha_{31} \mp i \beta_{31},
\end{align*}
$\alpha_{00}$, $\alpha_{11}$, $\alpha_{31}$, $\beta_{11}$, and $\beta_{31}$ are real, and $\hat{\bm{I}}^2 = I(I+1)$. We can assume that the interaction parameters are real ($\beta = 0$), which is a good approximation when the system is close to the $D_{2d}$ symmetry ($S_4$ is a subgroup of $D_{2d}$).

\begin{table}
  \centering
  \caption{Racah-normalized operator equivalents $\tilde C^{(J)}_{k,q}$ up to rank $k = 4$,
    with $X = J(J+1)$, $J_\pm = J_x \pm iJ_y$ and $\{A,B\} = AB + BA$.}
    \smallskip
  \label{tab:racah_operator_equivalents}
  \renewcommand{\arraystretch}{1.4}
  \setlength{\tabcolsep}{8pt}
  \begin{tabular}{ccl}
    \hline\hline
    $k$ & $q$ & $\tilde C^{(J)}_{k,q}$ \\
    \hline
    0 & $0$     & $1$ \\
    \hline
    1 & $0$     & $J_z$ \\
      & $\pm 1$ & $\mp \frac{1}{\sqrt{2}}\, J_\pm$ \\
    \hline
    2 & $0$     & $\frac{1}{2}\left(3J_z^2 - X\right)$ \\
      & $\pm 1$ & $\mp \frac{\sqrt{6}}{4} \left\{J_z, J_\pm\right\}$ \\
      & $\pm 2$ & $\frac{\sqrt{6}}{4}\, J_\pm^2$ \\
    \hline
    3 & $0$     & $\frac{1}{2}\left[5J_z^3 - (3X - 1)J_z\right]$ \\
      & $\pm 1$ & $\mp \frac{\sqrt{3}}{8} \left\{5J_z^2 - X - \tfrac{1}{2},\, J_\pm\right\}$ \\
      & $\pm 2$ & $\frac{\sqrt{30}}{8} \left\{J_z, J_\pm^2\right\}$ \\
      & $\pm 3$ & $\mp \frac{\sqrt{5}}{4}\, J_\pm^3$ \\
    \hline
    4 & $0$     & $\frac{1}{8}\left[35J_z^4 - (30X - 25)J_z^2 + 3X^2 - 6X\right]$ \\
      & $\pm 1$ & $\mp \frac{\sqrt{5}}{8} \left\{7J_z^3 - (3X + 1)J_z,\, J_\pm\right\}$ \\
      & $\pm 2$ & $\frac{\sqrt{10}}{16} \left\{7J_z^2 - X - 5,\, J_\pm^2\right\}$ \\
      & $\pm 3$ & $\mp \frac{\sqrt{35}}{8} \left\{J_z, J_\pm^3\right\}$ \\
      & $\pm 4$ & $\frac{\sqrt{70}}{16}\, J_\pm^4$ \\
    \hline\hline
  \end{tabular}
\end{table}

\subsection{Origins of degeneracy lifting}

As discussed in the main text, the strain component that induces a rhombic lattice distortion can be included in the hyperfine tensor and leads to a splitting of the $m_F=\pm1$ doublets in zero magnetic field. This can be understood by treating the strain effect as a perturbation to the conventional Hamiltonian. $\hatH_\mathrm{strain}$ can be rewritten in terms of the ladder operators $\hat{S}_\pm$ for electronic and $\hat{I}_\pm$ for nuclear spin as
\begin{equation}
    \hatH_\mathrm{strain} = \varrho_1\hat{S}_+\hat{I}_+ + \varrho_2\hat{S}_-\hat{I}_- = \hat{V},
\end{equation}
where $\varrho_1=\frac{\delta A_1-2i\delta A_2}{4}$ and $\varrho_2=\frac{\delta A_1+2i\delta A_2}{4}=\varrho_1^*$. The terms simultaneously raise or lower electron and nuclear spins and therefore couple states differing by $\Delta m_F=\pm2$, including the $m_F=\pm1$ doublet.

Within non-degenerate perturbation theory, the first-order energy shift of a state $\ket{i}$ vanishes due to the absence of diagonal terms, that is
\begin{equation}
    E^{(1)}_i = \hat{V}_{ii} = 0 ,
\end{equation}
where $\hat{V}_{ij}=\bra{i}\hat{V}\ket{j}$ corresponds to the transition matrix element between states $\ket{i}$ and $\ket{j}$. In the second order, the state experiences a quadratic shift in $\varrho_{1,2}$:
\begin{equation}
    E^{(2)}_i = \sum_{j\neq i}\frac{\big|\hat{V}_{ij}\big|^2}{E^{(0)}_i - E^{(0)}_j} \propto \varrho_{1,2}^2 .
\end{equation}
Here, $E^{(0)}_i$ is the unperturbed eigenenergy of state $\ket{i}$. While all states within the \eriononesixseven manifold are affected, their contribution does not induce any measurable lifting of degeneracy, considering the observed linewidths. However, asymmetric line shapes can result in an ensemble spectrum averaged over a distribution of strain (\Cref{fig:lineshape_octupole_strain_erbium}).

In contrast, for degenerate doublets $m_F=\pm1$, the correction can be found by diagonalizing the perturbed Hamiltonian of the degenerate subspace spanned by states $\ket{i}$ and $\ket{j}$ with energies $E^{(0)}_i=E^{(0)}_j=E_0$.
\begin{equation}
    \hatH^{(i,j)} = \left(
    \begin{matrix}
        E_0 & \hat{V}_{ij} \\
        \hat{V}^*_{ij} & E_0
    \end{matrix}
    \right)
\end{equation}
The resulting eigenvalues are $E_\pm=E_0\pm\big|\hat{V}_{ij}\big|$ and therefore, the energy correction is linearly proportional to $\varrho_{1,2}$.

A similar argument applies to the octupole interaction term led by the coefficient $\alpha_{31}$. It contains $\hat{S}_\pm\hat{I}_\pm^3$ terms that couple states differing by $\Delta m_F=\pm4$. Analogously to the derivation above, the energy levels in the $m_F=\pm2$ doublets shift linearly with $\alpha_{31}$ and generate a measurable splitting, while others are affected only in second order.

\section{Experimental}

\subsection{Setup and samples}

\begin{figure}
% \\badwwmi\user\groups\Quantum-Speicher\Austausch\Georg Mair\Figures\measurement_setup\measurement_setup_v5-Pulsing\Measurement_setup_pulsing.png
    \centering
    \includegraphics[width=0.8\linewidth]{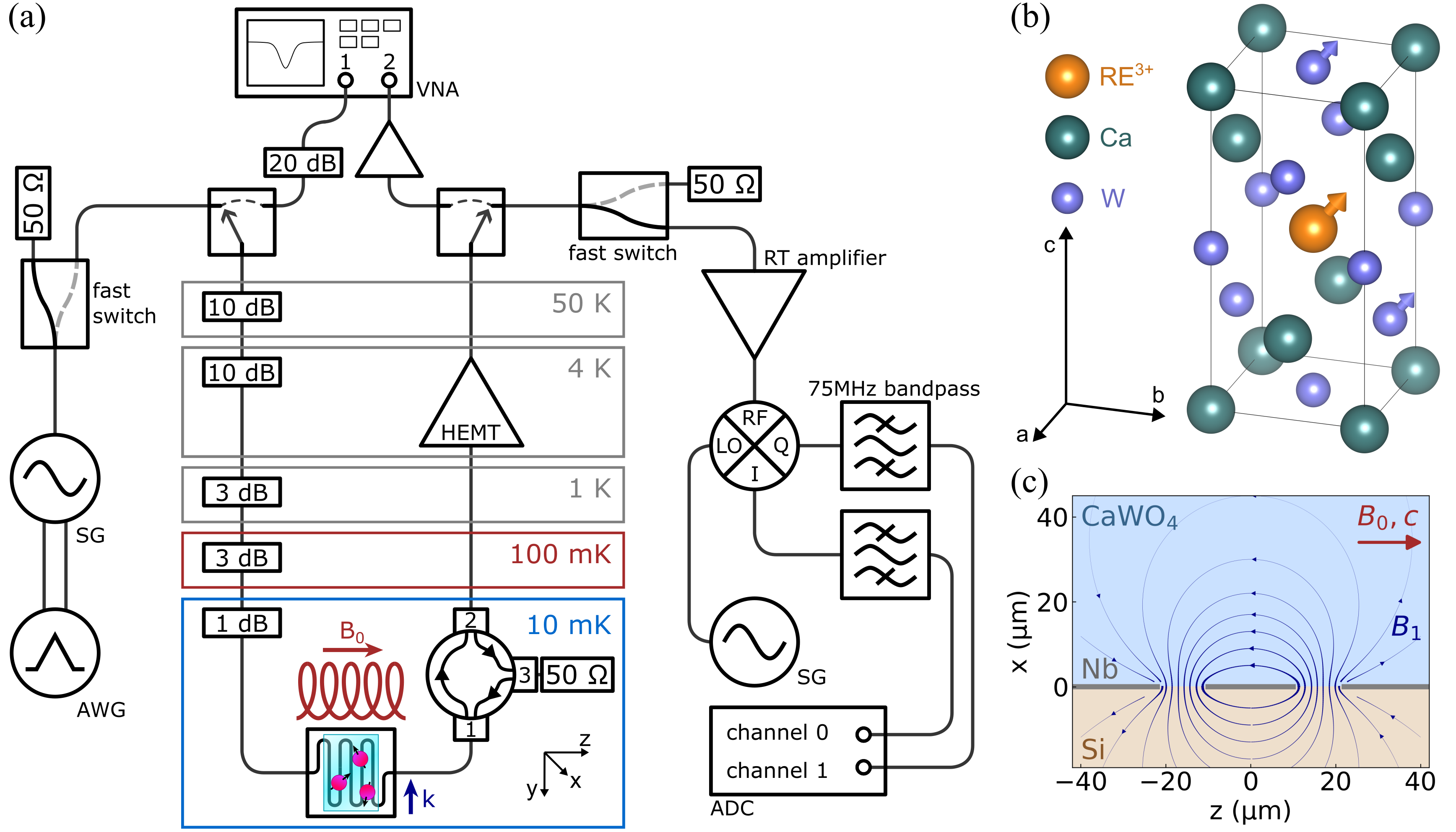}
    \caption{\textbf{Experimental setup:}
    (a)~Schematic of the measurement setup, indicating the thermal stages of the dilution refrigerator and the microwave circuitry. The crystal under investigation is mounted on a meander-shaped CPW located within the bore of a superconducting coil that generates the static magnetic field $B_{0}$. A VNA is used to measure the complex transmission parameter $S_{21}$. The setup can be switched to pulsed mode via two switches at the cryostat input and output. An arbitrary waveform generator (AWG) modulates the output of a signal generator (SG). After down-conversion, signals from the spin ensemble can be digitized.
    (b)~Coordination of a trivalent rare-earth ion RE\textsuperscript{3+} within the unit cell of \cwo~\cite{Momma.2011}.
    (c)~Simulated spatial distribution of the microwave magnetic field $B_{1}$ around the central conductor of the CPW. The upper region of the mode extends into the crystal, enabling coupling to the embedded rare-earth ions.}
    \label{fig:expsetup_crystalstruct}
\end{figure}

\begin{table}
    \centering
    \caption{Specifications of the investigated samples}
\begin{tabular}{l|cccc}
    Name & Dopants & Concentration (ppm) & Manufacturer & Size ($a\times b\times c\,\mathrm{mm}^3$) \\
    \hline
    Sam1 & \textsuperscript{n.a.}\erion & 50 & SurfaceNet & $0.2\times5\times5$ \\
    Sam2 & \textsuperscript{n.a.}\erion & 34 & SurfaceNet & $0.6\times8\times9$ \\
    Sam3 & \textsuperscript{n.a.}\erion, \textsuperscript{n.a.}\ybion*, \textsuperscript{n.a.}Mn\textsuperscript{2+}\xspace* & 50 & Scientific Materials & $0.6\times5\times5$ \\
    Sam4 & \ybiononesevenone & 34 & SurfaceNet & $0.6\times8\times9$ \\
\end{tabular}
    \medskip
    \\ *Unintentional dopants of unknown concentration.
    \label{tab:sample_specs}
\end{table}

% Superconducting CPW
The spin transitions are probed via microwave absorption measurements. The rare-earth ions couple to the oscillating magnetic field propagating in a superconducting coplanar waveguide (CPW) transmission line fabricated from niobium on a silicon substrate. In contrast to approaches in which the conductor is patterned directly onto the crystal surface~\cite{Billaud.2025}, the crystal is attached to the top of the CPW chip using copper tape~\cite{Strinic.2025}. This configuration is chosen to minimize additional strain associated with direct patterning or rigid adhesive bonding, thereby optimally preserving the intrinsic properties of the host matrix of the ions. The microwave propagation vector is oriented along the $y$-axis in the laboratory frame and is thus perpendicular to the external magnetic field $\bm{B}_0$ (along $z$-axis), corresponding to a Voigt-like geometry (see \Cref{fig:expsetup_crystalstruct}). Owing to the radial distribution of the microwave magnetic field $\bm{B}_1$ within the propagating mode, both field components parallel and perpendicular to the crystalline $c$-axis are present. Although only half of the microwave mode overlaps with the crystal, high absorption amplitude is achieved by employing a meandering CPW geometry with a total transmission line length of approximately \SI{26}{\centi\meter}. The actual absorption length depends on the dimensions of the sample under test and its overlap with the waveguide. This design enhances absorption depth by maximizing the interaction volume and enables the detection of bulk spin ensemble properties.

% Samples
In this work, we investigated ensembles of \eriononesixseven, \ybiononesevenone, and \ybiononeseventhree ions in \cwo single crystals with varying dopant concentrations. Details of the investigated samples are summarized in \Cref{tab:sample_specs}. Different dopant concentrations allow a trade-off between the absorption signal intensity and the spectral linewidth. For the precise determination of hyperfine parameters, samples with lower dopant concentrations are preferred, as they exhibit narrower linewidths and enable more accurate extraction of transition frequencies. All crystals are polished on at least one surface to ensure optimal contact with the waveguide and to maximize the interaction volume. The lateral dimensions of the samples range from \SI{5}{\milli\meter} to \SI{9}{\milli\meter}.

% Cryostat and magnet
The crystal–waveguide assembly is operated at a base temperature of \SI{9}{\milli\kelvin} in a \textsuperscript{3}He/\textsuperscript{4}He dilution refrigerator. The sample is thermally anchored to the mixing chamber stage of the cryostat, while positioned at the center of a superconducting solenoid connected to the \SI{100}{\milli\kelvin} stage. A schematic of the experimental configuration is shown in Fig.~\ref{fig:expsetup_crystalstruct}. The magnet provides a static magnetic field $B_0$ of up to \SI{\pm50}{\milli\tesla}. The crystal is oriented so that its $c$-axis is aligned with the applied magnetic field. To suppress external magnetic perturbations, the setup is enclosed within three layers of magnetic shielding, consisting of two Cryophy layers and one aluminum layer, thereby minimizing the influence of the Earth’s magnetic field. A more detailed description of the experimental setup is provided by Vogl~et~al.~\cite{Vogl.2026}.

% Measurement equipment cw spectroscopy
For continuous wave (CW) spectroscopy, a vector network analyzer (VNA) is used to measure the complex transmission parameter $S_{21}$, enabling detection of microwave absorption by the rare-earth ions. To suppress thermal noise from higher-temperature stages, the input line of the cryostat is attenuated by \SI{47}{dB}. The transmitted signal is subsequently amplified by a low-noise high-electron-mobility transistor (HEMT) amplifier operating at \SI{4}{\kelvin}. The broadband characteristics of the microwave components enable measurements over the frequency range of \qtyrange{0.1}{5}{\giga\hertz}, covering all relevant spin transitions. By performing repeated frequency sweeps as a function of the applied magnetic field, a three-dimensional absorption spectrum is obtained, providing direct access to the hyperfine level structure of the dopants.

% Pulsed setup
Time-domain measurements are performed using a pulsed microwave setup and a dedicated detection chain, which can be readily interchanged with the CW setup via switches at the input and output of the cryostat. An arbitrary waveform generator (AWG) is used to modulate a signal generator (SG) operating at the spin-transition frequency, thereby generating the microwave pulse sequence. A fast switch (switching time of approximately \SI{35}{\nano\second}) disconnects the cryostat input line from the SG, preventing leakage signals from reaching the sample. It is only engaged during pulse application. Simultaneously, a second switch terminates the detection path with a \SI{50}{\ohm} load, protecting the detection circuitry. A spin-echo signal from the output line of the cryostat is down-converted to an intermediate frequency of \SI{75}{\mega\hertz}. Its in-phase ($I$) and quadrature ($Q$) components are subsequently digitized.

\subsection{Group delay and data normalization}

The absorption features associated with the spin transitions are weak compared with the background features in the transmission spectrum. A normalization scheme can be used to enhance their visibility. In field-swept spectra, the normalized group delay $\Delta\tau_g$ was found to provide the best contrast. $\tau_{g} = -\tfrac{1}{2\pi}\mathrm{d}\phi/\mathrm{d}f$ is obtained as the negative derivative of the phase $\phi$ of $S_{21}$ with respect to frequency $f$. Although normalization can slightly distort the intrinsic line shapes, it reliably reveals the transition positions. The precise transition frequencies were subsequently determined from higher-resolution measurements containing only a single transition group. If the background phase response is sufficiently linear in such narrow-band spectra, normalization is not required due to the derivative in the group delay calculation.

For a field-swept spectrum, each frequency sweep is first normalized by subtracting its average at a given magnetic field, followed by subtracting the average over the magnetic field axis. This procedure can be written as
\begin{equation}
    \Delta\tau_{g}(f^\prime, B_0^\prime) = \tau_{g}(f^\prime, B_0^\prime) - \langle\tau_{g}\rangle_f(B_0^\prime) - \langle\tau_{g}\rangle_{B_0}(f^\prime) + \langle\langle\tau_{g}\rangle_f\rangle_{B_0} ,
\label{eq:delaynormalization}
\end{equation}
thereby removing contributions that are independent of frequency or magnetic field. Since the spin transitions shift with the applied magnetic field and are significantly narrower than the overall measurement range, they remain largely unaffected by this averaging procedure. The method is particularly effective for broadband measurements, where background variations in transmission originating from microwave components exceed the comparatively weak absorption signals of the paramagnetic species.

\subsection{Spin temperature}

Due to the limited thermal conductivity of the experimental setup, the actual spin temperature may deviate from the temperature $T_\mathrm{MXC}$ measured by the thermometer mounted at the mixing chamber (MXC) stage of the cryostat. To account for this discrepancy, we employ the empirical model proposed in Refs.~\cite{Kukharchyk.2020, Kukharchyk.2019}, which describes the effective spin temperature as
\begin{equation}
    T_\mathrm{eff} = T_\mathrm{min}\left(1 + \left(\frac{T_\mathrm{MXC}}{T_\mathrm{min}}\right)^2\right)^\frac{1}{2}
\end{equation}
where $T_\mathrm{min}$ denotes the minimum achievable spin temperature and $T_\mathrm{eff}$ is the effective temperature of the spin system.

\begin{figure}[th]
% \\badwwmi\user\groups\Quantum-Speicher\Austausch\Georg Mair\Figures\absorption_vs_temperature_fit_Tmin\absorption_vs_temperature_fit_Tmin_Er_CWO_XLD1_r47_v1\absorption_vs_temperature_fit_Tmin_Er_CWO_XLD1_r47.png
    \centering
    \includegraphics[width=0.99\linewidth]{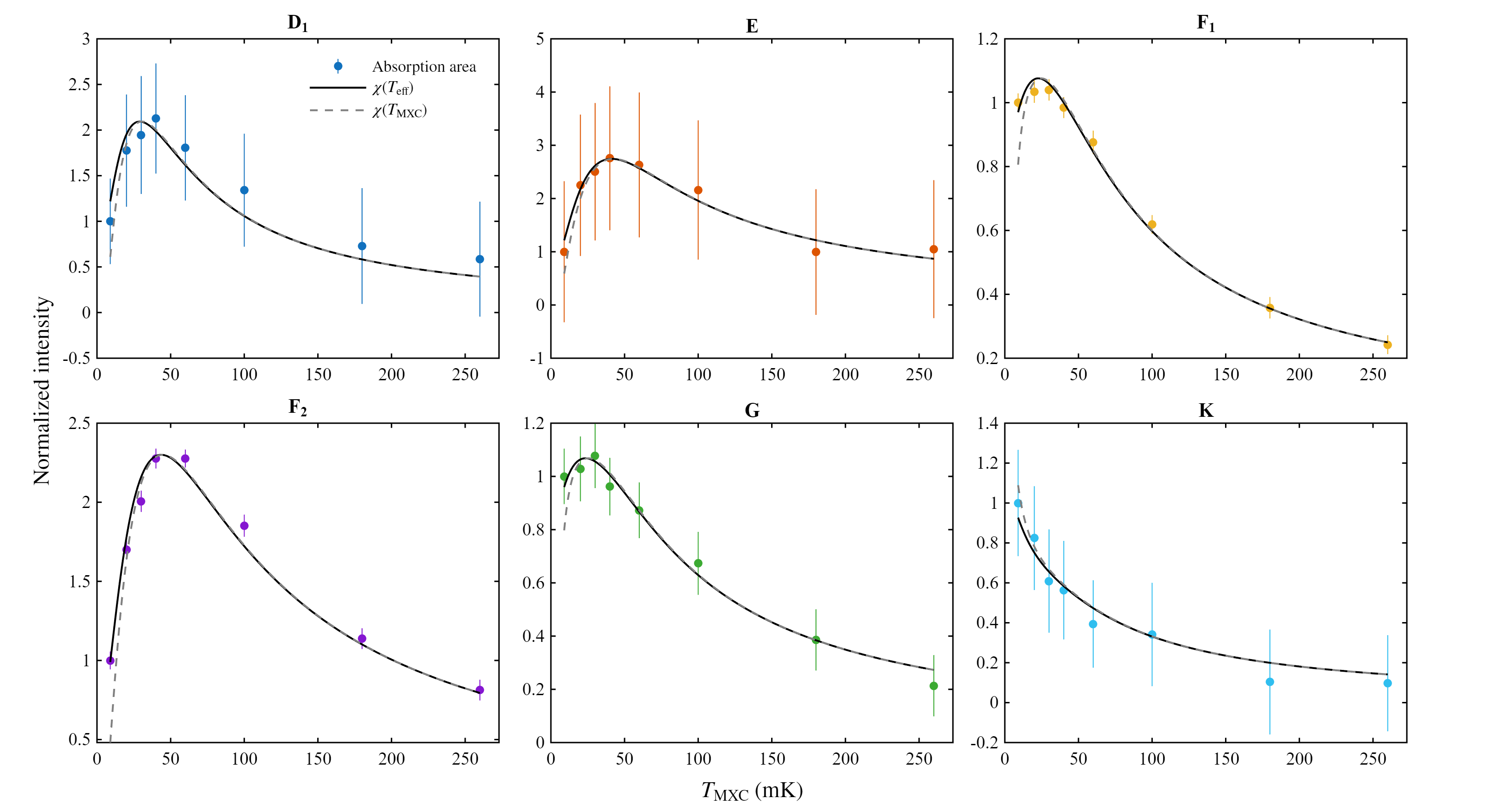}
    \caption{\textbf{Effective spin temperature:}
    The temperature of the spin system is inferred from the temperature dependence of the absorption area of different transition groups. Since the true spin temperature may deviate from the sensor reading at the MXC stage, we use an effective temperature model. The intensities $\chi(T_\mathrm{MXC})$ for perfect thermalization are shown for comparison. A global fit yields $T_\mathrm{eff}=\SI{13.1(5)}{\milli\kelvin}$. Data points of each transition group are normalized to the value at $T_\mathrm{MXC}=\SI{9}{\milli\kelvin}$.
    }
    \label{fig:spin_temperature}
\end{figure}

To determine the effective spin temperature, we measured the relative absorption of the D\textsubscript{1}, E, F\textsubscript{1}, F\textsubscript{2}, G, and K transitions over a mixing chamber temperature ($T_\mathrm{MXC}$) range of \qtyrange{9}{260}{\milli\kelvin}. The relative absorption was calculated from the transmission coefficients measured on and off resonance with a spin transition, $S_{21}^\mathrm{res}$ and $S_{21}^\mathrm{off}$, respectively, according to $1-\sfrac{\big|S_{21}^\mathrm{res}\big|^2}{\big|S_{21}^\mathrm{off}\big|^2}$. The integrated spectral line shape of each transition is proportional to the absorbed power, $P_\mathrm{abs}\propto\chi$, where $\chi$ denotes the population difference between the two levels involved in the transition. These population differences were calculated assuming a Boltzmann distribution at the effective spin temperature. We then performed a global weighted fit of $\chi(T_\mathrm{eff})$ across all transition groups (\Cref{fig:spin_temperature}) to determine the minimum achievable spin temperature, yielding $T_\mathrm{min}=\SI{9.5(6)}{\milli\kelvin}$. This corresponds to an effective spin temperature of $T_\mathrm{eff}=\SI{13.1(5)}{\milli\kelvin}$ at the base temperature of the cryostat. The population differences calculated directly from the measured mixing-chamber temperature are shown as dashed lines in \Cref{fig:spin_temperature}. The improved agreement obtained with the effective-temperature model demonstrates that the spin system does not fully thermalize with the mixing-chamber stage at the lowest temperatures.

\subsection{Probing power}

\begin{figure}[th]
% \\badwwmi\user\groups\Quantum-Speicher\Austausch\Georg Mair\Figures\probing_power_saturation_contrast\probing_power_saturation_contrast_v1\probing_power_saturation_contrast_labeled.png
    \centering
    \includegraphics[width=0.99\linewidth]{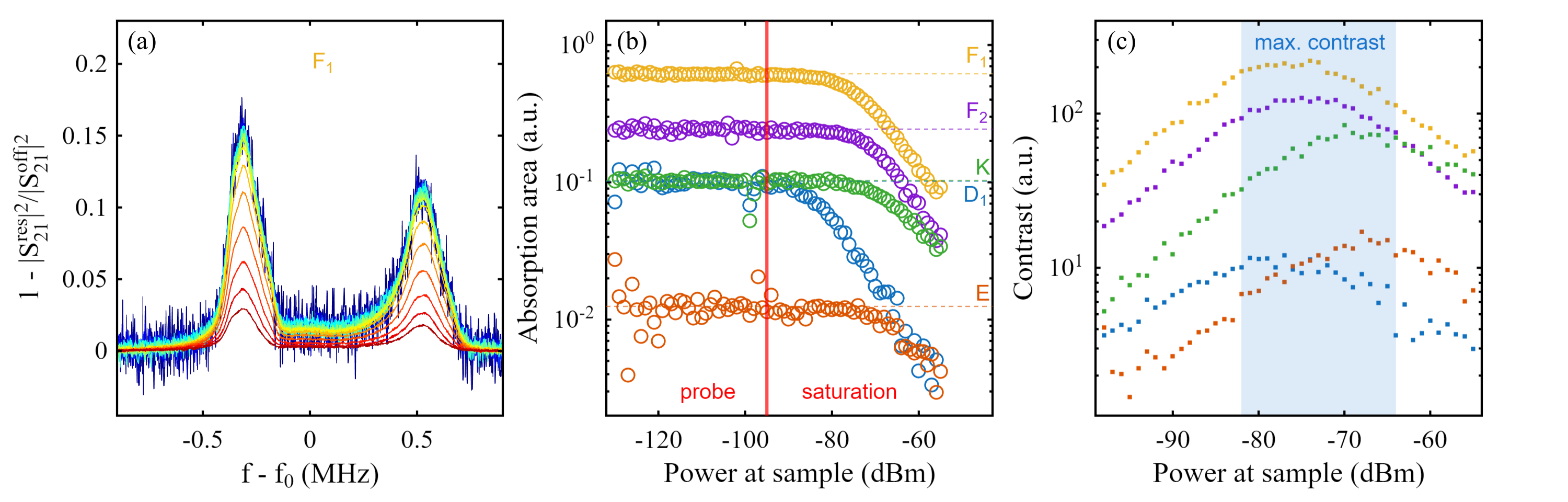}
    \caption{\textbf{Probing power:}
    (a) Power-dependent spectra of the F\textsubscript{1} transition group of \eriononesixseven at $B_0=\SI{0}{\milli\tesla}$. As microwave power increases (blue to red), absorption decreases due to saturation of the spin system. Upon equilibration of populations, absorption would vanish completely.
    (b) Integrated relative absorption areas are constant for powers below the saturation limit, and are therefore assumed to maintain thermal equilibrium during probing.
    (c) Due to the improvement in noise figure at higher measurement powers, the contrast (maximum absorption amplitude divided by RMS noise) peaks in the partially saturating regime.
    }
    \label{fig:saturation}
\end{figure}

The microwave probing power influences the population dynamics of the spin system and can drive the transitions into saturation. As the population difference between the involved levels is reduced, the effective number of spins contributing to the absorption decreases, resulting in a reduced absorption signal in the transmission spectrum, as shown in \Cref{fig:saturation}(a,b). To probe the spin system close to thermal equilibrium, the spin-lattice relaxation must be sufficiently fast compared to the excitation rate. Experimentally, this condition is reached in the low-power regime where the relative absorption becomes independent of the applied microwave power. We therefore used this regime for the high-resolution zero-field measurements, from which the transition frequencies were extracted to determine the Hamiltonian parameters.

For detecting weak absorption signals from low-concentration impurities or transitions with thermally suppressed population differences, higher probing powers can nevertheless be advantageous. First, increasing the microwave power improves the signal-to-noise performance of the measurement devices, such that the maximum spectral contrast, defined as the absorption amplitude divided by the root-mean-square noise, is obtained at powers above the probing regime (see \Cref{fig:saturation}(c)). Second, partial saturation can redistribute population among higher-energy levels that would otherwise be only weakly populated at thermal equilibrium, thereby enhancing the visibility of transitions involving these states. Within the investigated power range, neither power broadening nor power-induced frequency shifts were observed. Higher probing powers were therefore employed for broadband measurements covering the complete hyperfine spectrum.

\subsection{Additional broadband spectra}

\begin{figure}[th]
% \\badwwmi\user\groups\Quantum-Speicher\Austausch\Georg Mair\Figures\broadband_ESR_sim_overlay\broadband_ESR_sim_overlay_B_perp_c_v1\broadband_ESR_B_perp_c_XLD1_r35_r43_167Er-corr_171Yb-corr_labeled.png
    \centering
    \includegraphics[trim={0 0 {.08\textwidth} 0},clip,width=1\textwidth]{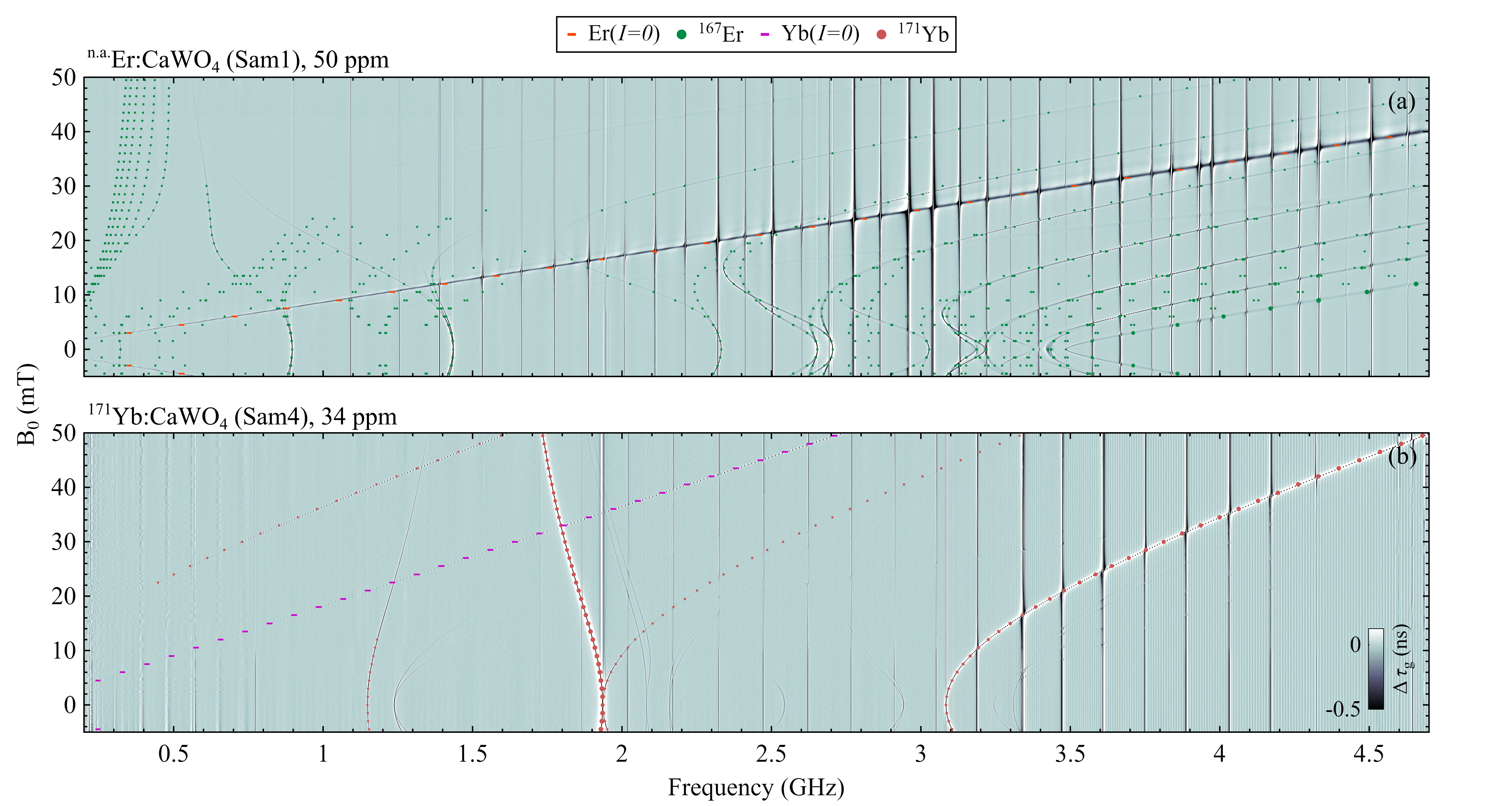}
    \caption{\textbf{Broadband spectra of Sam1 and Sam4:} The spectra are taken in Faraday configuration ($\bm{B}_0\parallel k$) and the external magnetic field $\bm{B}_0$ is oriented orthogonal to the crystalline $c$-axis. Parasitic background resonances originating from the experimental setup appear as vertical lines in the normalized spectra. The symbol size represents the relative transition intensity at a spin temperature of \SI{13}{\milli\kelvin}. An intensity threshold is applied in the simulation to exclude transitions below the experimental detection limit. Both spectra were acquired at signal powers in the saturation regime to detect weak transitions. Hence, simulated intensities match the measurement only qualitatively. (a)~\erion dopants of natural isotopic abundance at a concentration of \SI{50}{ppm}. (b)~\ybiononesevenone dopants at a concentration of \SI{34}{ppm}. Despite the isotopically purified doping, trace amounts of ytterbium isotopes without nuclear spin are detectable. In addition, several other absorptions of unknown origin are detected (likely hyperfine transitions of other paramagnetic impurities).}
    \label{fig:broadbandBperp}
\end{figure}

Broadband spectra of samples Sam1 and Sam4 were recorded in the Faraday configuration ($\bm{B}_0\parallel k$) with the external magnetic field oriented perpendicular to the crystalline $c$-axis ($\bm{B}_0 \perp c$). The measured spectra are in good agreement with simulations based on the hyperfine parameters determined in this work, providing additional validation of the extracted Hamiltonian parameters (\Cref{fig:broadbandBperp}). In this field orientation, the selection rules discussed in the main text do not apply, allowing additional transitions to become observable. For example, for \ybiononesevenone, the $\ket{2}\rightarrow\ket{3}$ transition becomes clearly visible for $B_0 > \SI{20}{\milli\tesla}$, appearing at a frequency of approximately \SI{0.4}{\giga\hertz} (\Cref{fig:broadbandBperp}(b)).

\subsection{High-resolution low-field spectra}

\subsubsection{\eriononesixseven}

\begin{figure}[tbh!]
% \\badwwmi\user\groups\Quantum-Speicher\Austausch\Georg Mair\Figures\zero_field_features_higher_order_terms_overlay\zero_field_features_higher_order_terms_overlay_XLD1_r47_v4\zero_field_features_higher_order_terms_overlay.png
    \centering
    \includegraphics[trim={{.14\textwidth} {.11\textwidth} {.14\textwidth} {.09\textwidth}},clip,width=0.8\textwidth]{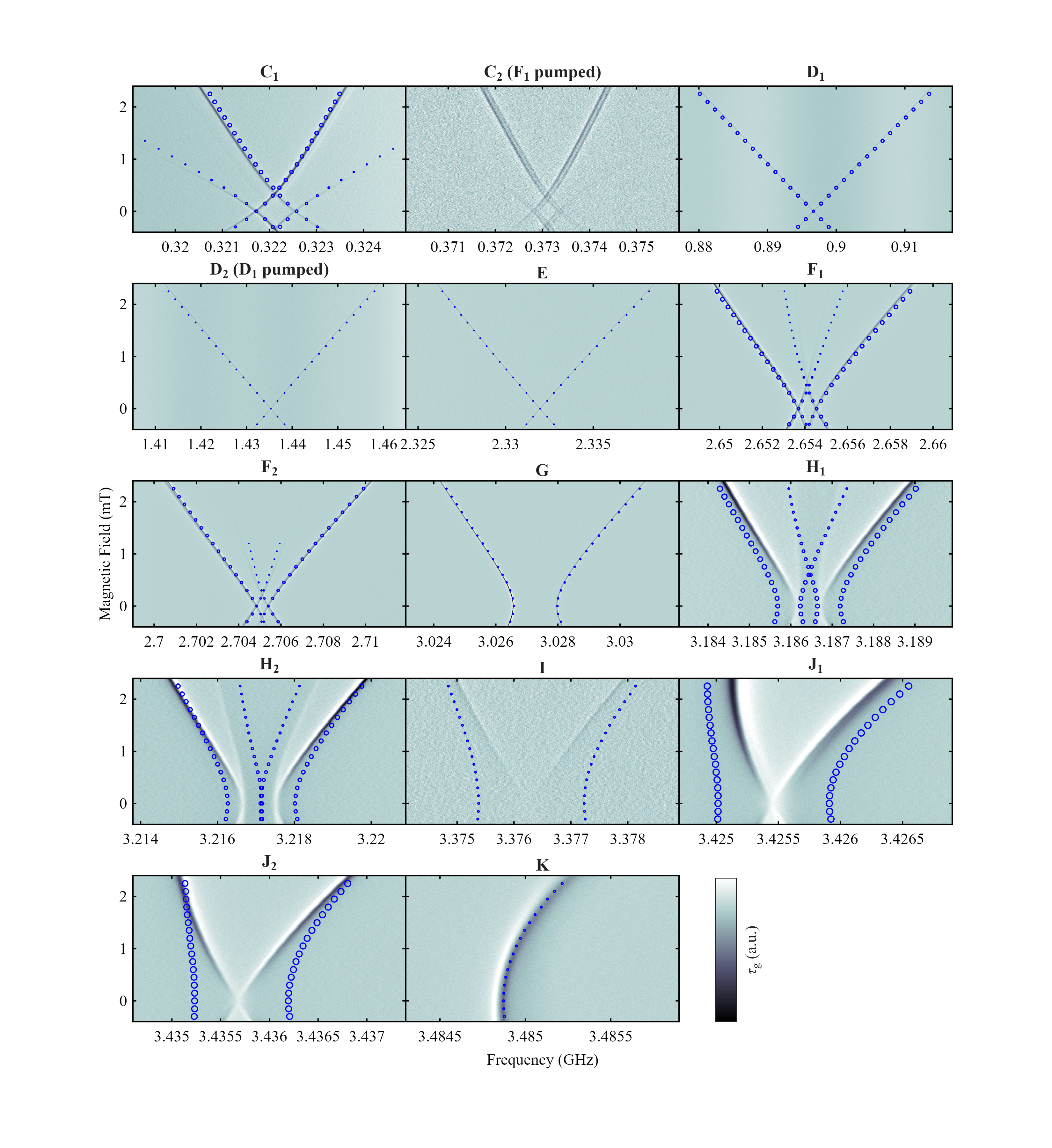}
    \caption{\textbf{Near-zero field spectra of \eriononesixseven:} High-resolution field-swept spectra of Sam2 around \SI{0}{\milli\tesla} for all observable transitions. Blue circles represent simulated transition frequencies of a spin system at \SI{13}{\milli\kelvin} including the higher-rank terms and strain. The size of the symbols indicates the transition probability. Due to a small thermal population gradient among the highest energy levels, the simulated intensity of transition group C\textsubscript{2} drops below the detection limit. To experimentally create a detectable population difference, group F\textsubscript{1} is pumped before recording the spectrum at each magnetic field. Similarly, group D\textsubscript{2} is detected by pumping group D\textsubscript{1}. $\abs{S_{21}}$ spectra at $\abs{\bm{B}_0}=\SI{0}{\milli\tesla}$ are shown in \Cref{fig:lineshape_octupole_strain_erbium}.}
    \label{fig:zero_field_overlay}
\end{figure}

\begin{figure}[th]
% \\badwwmi\user\groups\Quantum-Speicher\Austausch\Georg Mair\Figures\lineshape_sim_octupole_and_strain\lineshape_sim_octupole_and_strain_167Er_CWO_v7\lineshape_sim_octupole_and_strain_167Er_CWO_fast.png
    \centering
    \includegraphics[width=0.99\linewidth]{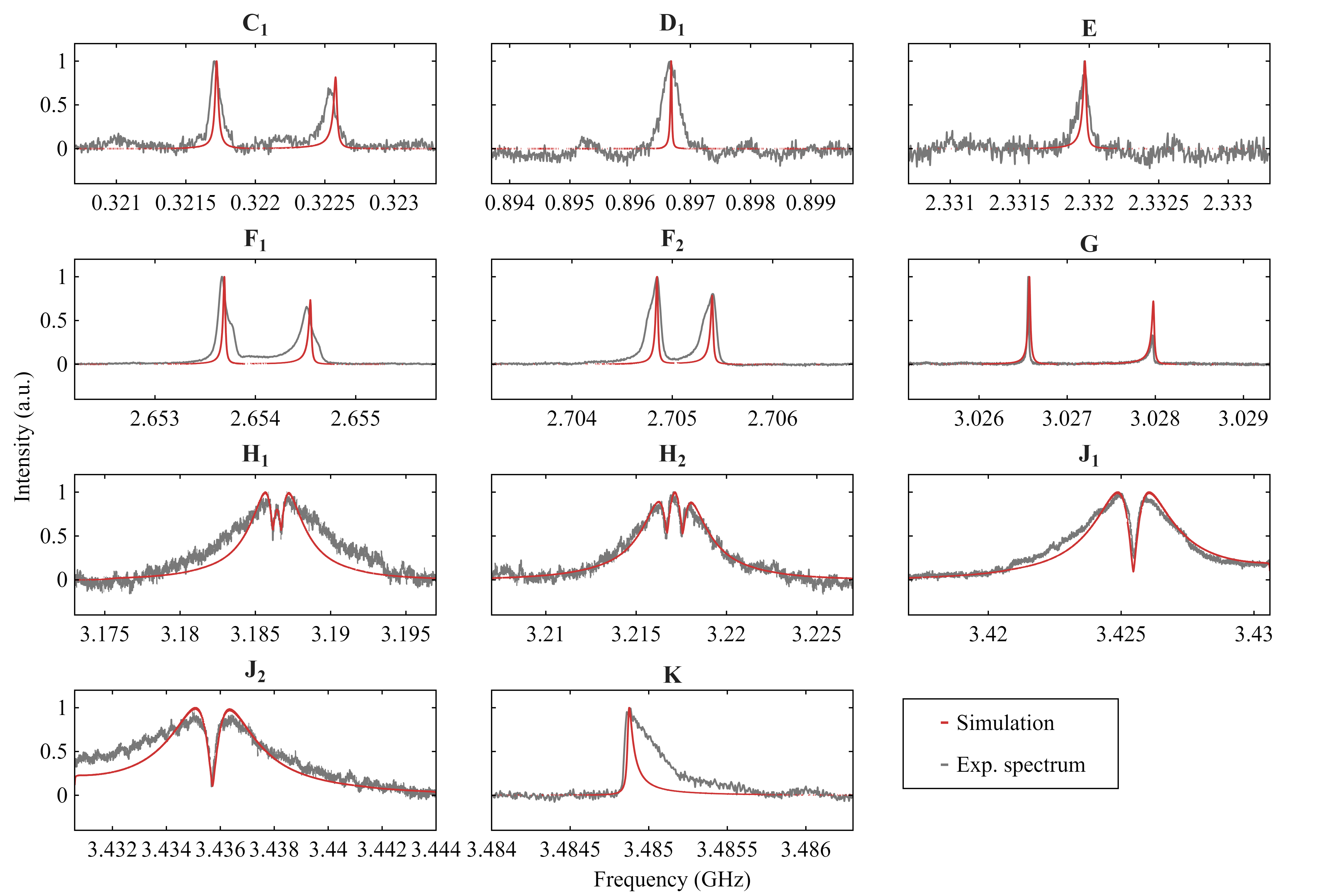}
    \caption{\textbf{Line shape simulation including higher-rank interactions and strain in \eriononesixseven:\cwo:}
    Asymmetric line shapes observed in transition groups involving the $m_F=\pm1$ states (H\textsubscript{1,2}, J\textsubscript{1,2}) are reproduced by the distribution of $\delta A_1$ and $\delta A_2$, while keeping other transitions largely unaffected. Simulation and experimental $|S_{21}|$ in decibels are separately normalized to their respective maximum for each transition group.
    }
    \label{fig:lineshape_octupole_strain_erbium}
\end{figure}

The agreement between the simulations and the high-resolution field-swept spectra of Sam2 (\Cref{fig:zero_field_overlay}) further validates the inclusion of higher-rank interaction terms in the spin Hamiltonian. In particular, the model reproduces characteristic spectral features, such as superimposed avoided crossings with forbidden transitions, in the C\textsubscript{1,2} and F\textsubscript{1,2} transition groups. These additional couplings are attributed to the octupole and hexadecapole interactions. The transition groups H\textsubscript{1,2} and J\textsubscript{1,2}, in contrast, are strongly affected by off-diagonal and asymmetric perpendicular components in the hyperfine tensor, introduced by local lattice deformations. These terms lift the degeneracy of the $m_F=\pm1$ doublets, resulting in avoided crossings accompanied by pronounced line broadening (\Cref{fig:lineshape_octupole_strain_erbium}). The energy levels associated with transition groups D\textsubscript{1} and D\textsubscript{2} are not split by the higher-rank interactions but instead experience small frequency shifts.

\subsubsection{\ybiononesevenone}

\begin{figure}[tbh!]
% \\badwwmi\user\groups\Quantum-Speicher\Austausch\Georg Mair\Figures\zero_field_features_higher_order_terms_overlay\zero_field_features_higher_order_terms_overlay_LD3_r18_171Yb_v1\zero_field_features_higher_order_terms_overlay_171Yb.png
    \centering
    \includegraphics[width=1\textwidth]{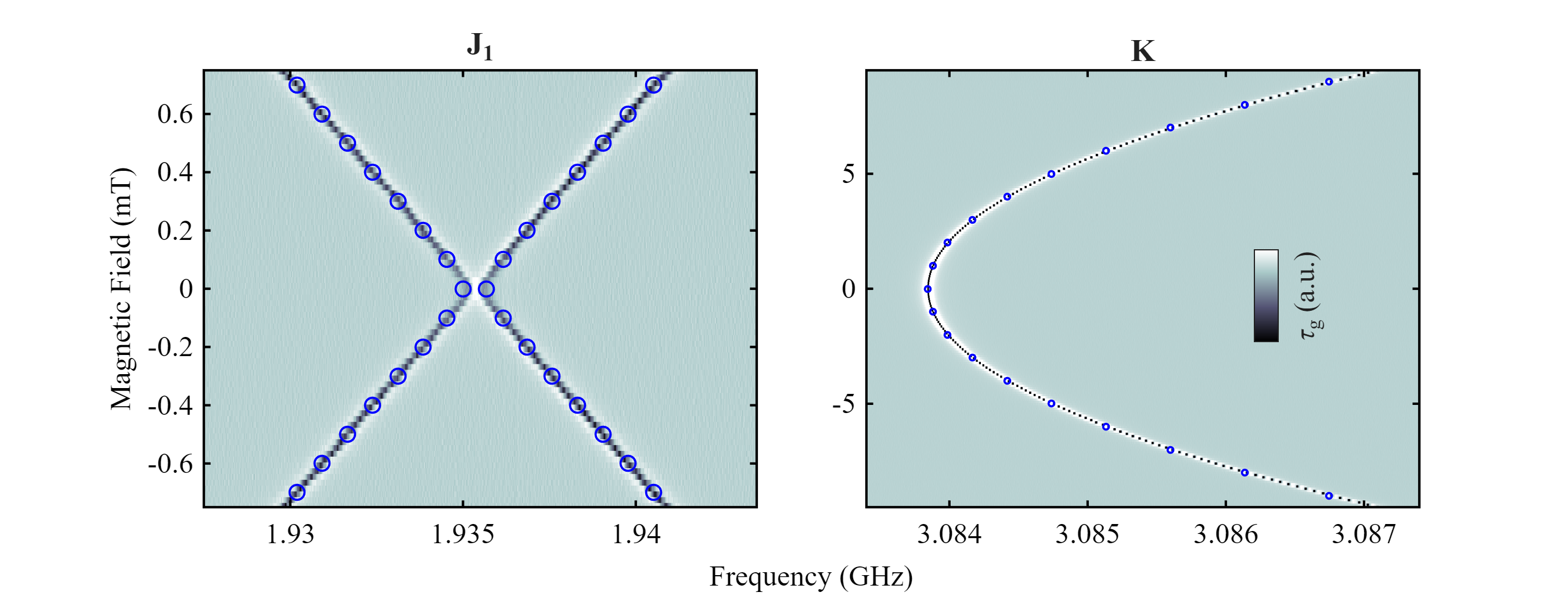}
    \caption{\textbf{Near-zero field spectra of \ybiononesevenone:} High-resolution field-swept spectra of Sam4 around \SI{0}{\milli\tesla} for the J\textsubscript{1} and K transitions overlaid with a Hamiltonian simulation of a spin system at \SI{13}{\milli\kelvin} including strain. The size of the symbols indicates the transition probability. The J\textsubscript{2} transition is not detectable at the base temperature without pumping population to the $\ket{\pm,\pm1}$ states.}
    \label{fig:zero_field_overlay_171Yb}
\end{figure}

\begin{figure}[th]
% \\badwwmi\user\groups\Quantum-Speicher\Austausch\Georg Mair\Figures\lineshape_sim_octupole_and_strain\lineshape_sim_octupole_and_strain_171Yb_CWO_v5\lineshape_sim_octupole_and_strain_171Yb_CWO_fast.png
    \centering
    \includegraphics[width=0.75\linewidth]{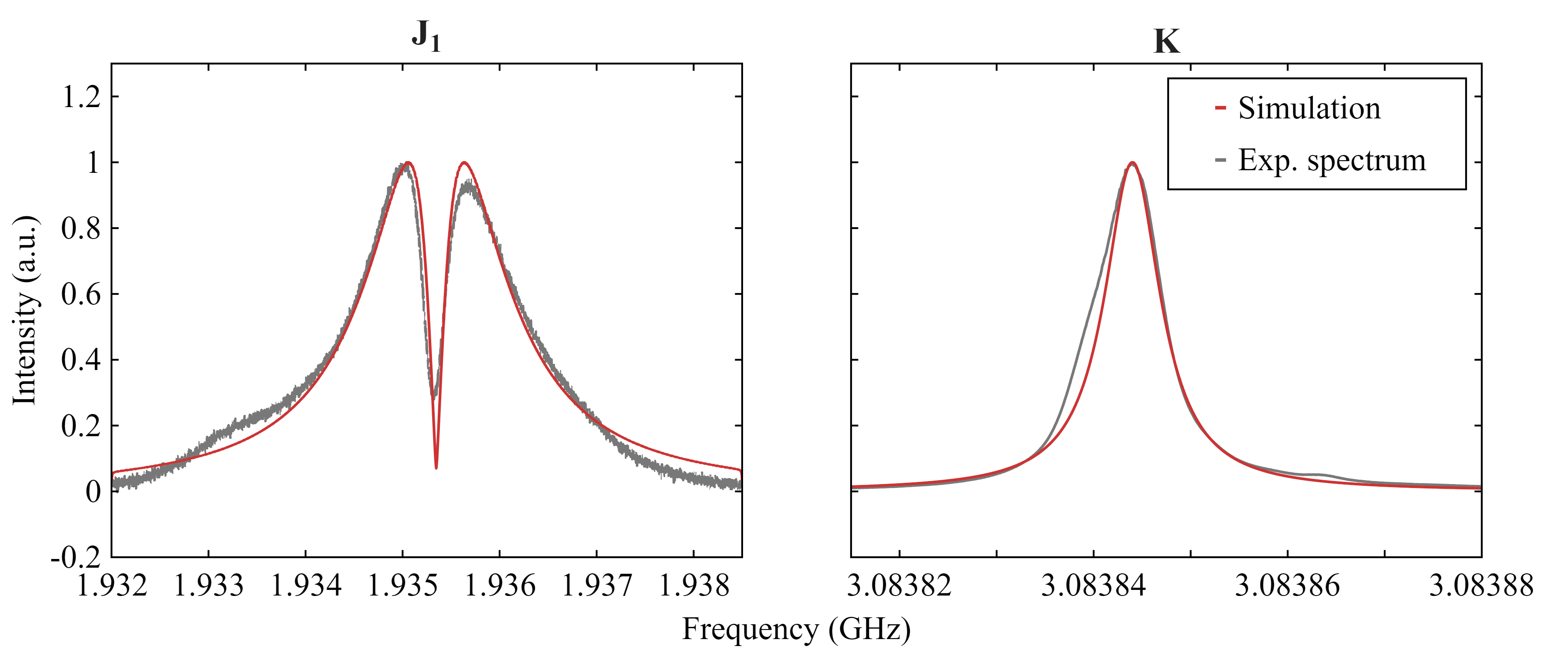}
    \caption{\textbf{Line shape simulation including strain in \ybiononesevenone:\cwo:}
    Asymmetric line shapes observed in transition groups involving the $m_F=\pm1$ states (J\textsubscript{1,2}) are reproduced by the distribution of $\delta A_1$ and $\delta A_2$, while keeping the K transition (in particular the line shape) unaffected. Simulation and experimental $|S_{21}|$ in decibels are separately normalized to their respective maximum for each transition group.
    }
    \label{fig:lineshape_octupole_strain_ytterbium}
\end{figure}

A simulation accounting for random lattice deformations reproduces the avoided crossing of the $m_F=\pm1$ doublet at zero magnetic field observed in Sam4 (\Cref{fig:zero_field_overlay_171Yb}). Considering the distribution of $\delta A_1$ and $\delta A_2$, the strongly asymmetric line shape of the J\textsubscript{1} transitions is recreated (\Cref{fig:lineshape_octupole_strain_ytterbium}).

\subsection{Reproducibility}

\begin{figure}[th]
% \\badwwmi\user\groups\Quantum-Speicher\Austausch\Georg Mair\Figures\reproducibility_G_transition\reproducibility_G_transition_v3\reproducibility_G_transition_labeled.png
    \centering
    \includegraphics[width=0.5\linewidth]{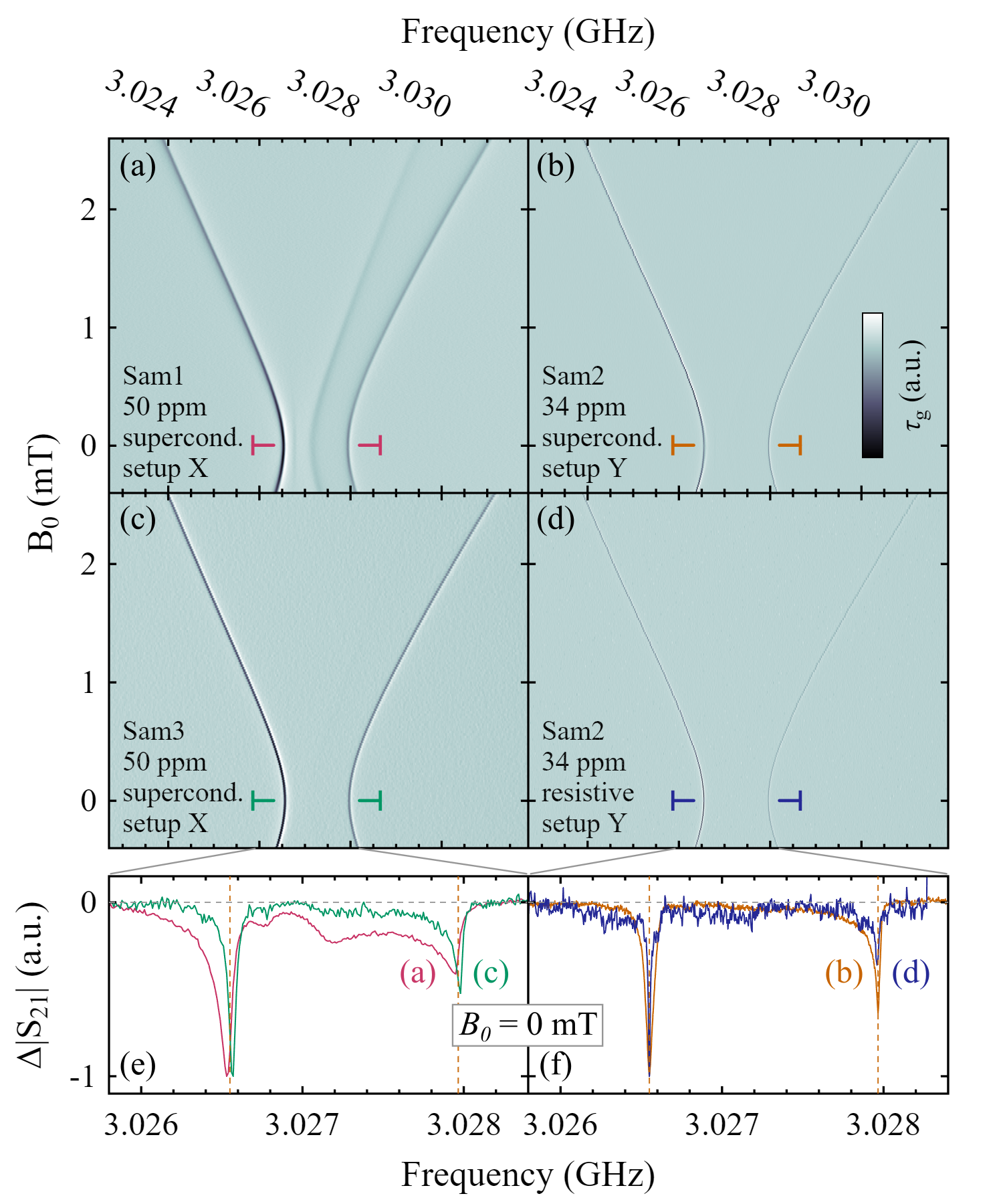}
    \caption{\textbf{Reproducibility of avoided crossings:}
    (a)-(d): High-resolution field-swept ($\bm{B}_0\parallel c$) group delay spectra of transition group G (allowed transitions: $\ket{-,+2}\leftrightarrow\ket{+,+2}$ and $\ket{-,-2}\leftrightarrow\ket{+,-2}$) demonstrate consistent behavior across samples from different suppliers, dopant concentrations, and multiple cool-down cycles in independent cryogenic setups with superconducting and resistive waveguides. The reduction in linewidth at lower doping concentrations is attributed to a suppression of spin-spin interactions. Both transitions exhibit previously unpredicted ZEFOZ conditions, which arise from the avoided crossings between $m_F=\pm2$ states (directly observable in transition groups F\textsubscript{1} and F\textsubscript{2}). Details are discussed in the main text. The color scale in each subplot is normalized to its respective minimum value. (e), (f): Magnitude spectra taken at $B_0=\SI{0}{\milli\tesla}$ to compare the line shapes. All spectra are independently normalized by background subtraction and scaled to $-1$.}
    \label{fig:splitting_reproducibility}
\end{figure}

All ZFACs, \Cref{fig:zero_field_overlay}, were reproducibly observed across all investigated samples, which originate from different crystal growths and suppliers with doping concentrations of \SI{50}{ppm} and \SI{34}{ppm}. The measurements were conducted in two independent cryogenic setups, yielding consistent results and showing no dependence on the orientation of the crystalline $c$-axis with respect to $k$ and Earth's magnetic field. Furthermore, replacing the superconducting waveguide with a gold-coated copper PCB did not result in measurable changes in the observed transition frequencies or splittings. The magnitude of the avoided crossings also remained consistent upon increasing the cryostat temperature up to \SI{800}{\milli\kelvin} at the mixing chamber sensor.

The reproducibility of the ZFACs under the conditions discussed above is exemplified in Fig.~\ref{fig:splitting_reproducibility} by transition group G of \eriononesixseven. One can see minuscule shifts of the absolute positions of absorption lines across the various samples, while the magnitude of the splittings remains unchanged. This applies to all other transition groups. In Sam1, we observed an additional peak associated with the $\ket{-,-2}\leftrightarrow\ket{+,-2}$ transition (high-frequency branch) shown in \Cref{fig:splitting_reproducibility}(a). A similar spectral feature was observed in the corresponding branches of transition groups C\textsubscript{1} and F\textsubscript{1}, both of which involve the energy level $\ket{-,-2}$. This feature was observed in other crystals at significantly lower intensity and is attributed to crystal quality. The reproducibility of the avoided crossing in \ybiononesevenone was likewise confirmed over several samples, further supporting the reproducibility of the observed phenomena.

% Experimental aspects
The experimental results are consistent across independent cryogenic systems and measurement configurations. The hypothesis that the observed splittings originate from an uncompensated magnetic offset field oriented perpendicular to the magnet axis can be rejected for several reasons: 
First, the magnitude of the Earth’s magnetic field is insufficient to account for the observed splittings. Furthermore, a static bias field would affect all transitions. In contrast, the splittings are observed only in selected transition groups, while others retain the expected energy-level degeneracy. This selective behavior is inconsistent with an external field offset. The strongly asymmetric line shapes observed at the avoided crossings involving states $\ket{-,\pm1}$ and $\ket{+,\mp1}$ (e.g., in transition groups J\textsubscript{1} and J\textsubscript{2}) indicate level repulsion due to an intrinsic interaction. A simple bias field would instead lead to broadened, overlapping spectral lines rather than the characteristic signatures of avoided crossings. Lastly, the effect persists for different orientations of the crystalline $c$-axis relative to the Earth’s field and remains unchanged in the presence of magnetic shielding.

% Superhyperfine interaction
The expected magnitude of the superhyperfine interaction, based on dipole-dipole interactions to neighboring spins in the \cwo lattice, is insufficient to account for the observed splittings. Moreover, the dominant source of nuclear spins in the host lattice is \textsuperscript{183}W with a natural abundance of 14\% and nuclear spin $I = \sfrac{1}{2}$. The superhyperfine interaction would hence affect only a subset of ions, which is inconsistent with the uniform appearance of the splittings across the ensemble. Lastly, superhyperfine coupling is expected to produce a $2I+1$-fold splitting superimposed on the existing transitions, which does not match the experimental observations.

\subsection{Zero-field coherence time}

\begin{figure}[th]
% \\badwwmi\user\groups\Quantum-Speicher\Austausch\Georg Mair\Figures\coherence_time\Coherence_time_fit_v3\coherence_time_fit_167Er.png
    \centering
    \includegraphics[width=0.7\linewidth]{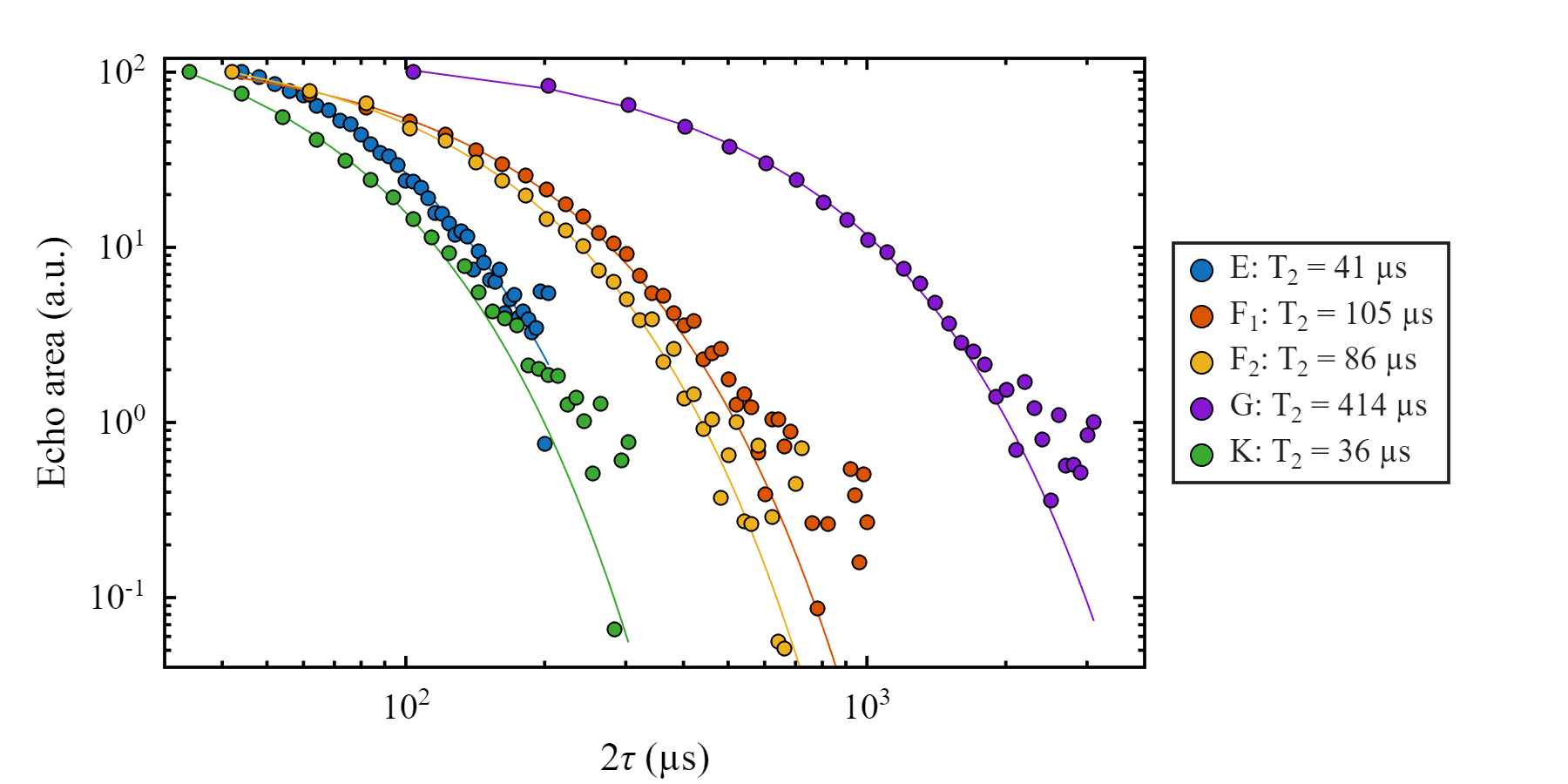}
    \caption{\textbf{Zero-field coherence times of \eriononesixseven:}
    Hahn echo coherence times for the E, F\textsubscript{1,2}, G, and K transitions. Only G and K satisfy true ZEFOZ conditions in zero magnetic field. The former is a result of the $m_F=\pm2$ doublet splitting arising from the octupole and hexadecapole interactions. The longest coherence time is obtained for the G transition group as expected from its second-order Zeeman sensitivity (see main text).}
    \label{fig:zerofield_coherence}
\end{figure}

The Hahn echo coherence times of several transition groups in zero magnetic field are presented in \Cref{fig:zerofield_coherence}. Among the true ZEFOZ points, the G transition group outperforms the K transition due to its smaller curvature of the transition frequency, thereby reducing sensitivity to magnetic field fluctuations. Notably, transitions with a non-vanishing gradient of the transition frequency (E, F\textsubscript{1,2}) achieve longer coherence times than the K transition. A possible explanation is the large second-order sensitivity of the K transition to fields perpendicular to the $c$-axis, combined with an uncompensated bias field, which creates an offset from the ideal ZEFOZ conditions.

\subsection{Manganese impurities}
\label{apx:manganese}

\begin{figure}[th]
% \\badwwmi\user\groups\Quantum-Speicher\Austausch\Georg Mair\Figures\broadband_ESR_sim_overlay\broadband_ESR_sim_overlay_55Mn_v1\broadband_ESR_XLD1_r46_55Mn.png
    \centering
    \includegraphics[width=0.99\linewidth]{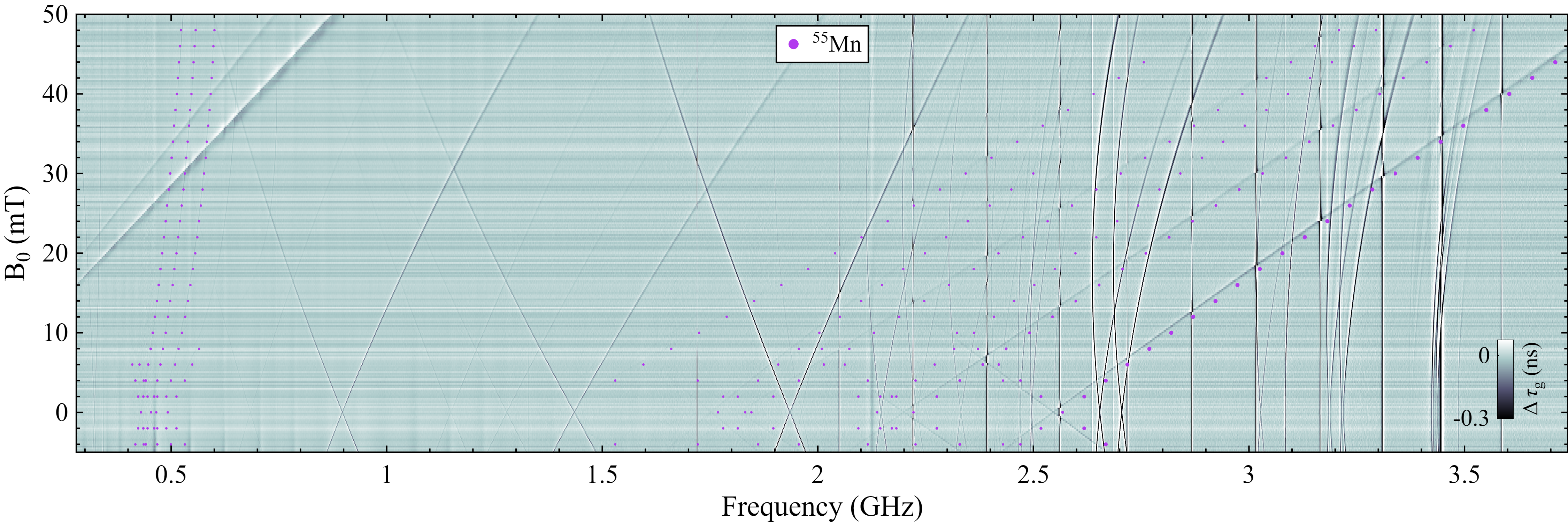}
    \caption{\textbf{Broadband spectrum and simulation of \textsuperscript{55}Mn\textsuperscript{2+}:\cwo:}
    The normalized broadband group delay spectrum reveals hyperfine transitions of unintentional \textsuperscript{55}Mn\textsuperscript{2+} dopants in Sam3. The external magnetic field $\bm{B}_0$ is oriented along the crystalline $c$-axis. The simulation of the hyperfine transitions at a spin temperature of \SI{13}{\milli\kelvin} uses Hamiltonian parameters from Ref.~\cite{Hempstead.1960}.}
    \label{fig:broadbandMn}
\end{figure}

As discussed in the main text, the sample from Scientific Materials has been contaminated with manganese, which is found in natural abundance of a single isotope \textsuperscript{55}Mn with $S=\sfrac{5}{2}$ and $I=\sfrac{5}{2}$.
The measured spectrum is reconstructed using the parameters in Ref.~\cite{Hempstead.1960} using a Hamiltonian based on extended Stevens operators $\hat{O}^q_k$.
\begin{equation}
%\begin{split}
    %\hatH_\text{eff} = & \hatH_\text{eZ} + \hatH_\text{hf} + \hatH_\text{zf} + \hatH_\text{nZ} \\
    %= & \mu_\text{B}\bm{B}_0\boldsymbol{g}\hat{\bm{S}}
    %+ \hat{\bm{S}}\boldsymbol{A}\hat{\bm{I}}
    %+ \hat{\bm{S}}\boldsymbol{D}\hat{\bm{S}}
    %- \mu_\text{N}\bm{B}_0 g_N\hat{\bm{I}},
    \hatH_\mathrm{Mn} =
    \mu_\mathrm{B}\bm{B}_0\bm{g}\hat{\bm{S}} + \hat{\bm{S}}\bm{A}\hat{\bm{I}} + 
    B^0_2 \hat{O}^0_2 + B^0_4 \hat{O}^0_4 + B^4_4 \hat{O}^4_4
%\end{split}
%\label{eq:hamiltonian}
\end{equation}
For \textsuperscript{55}Mn\textsuperscript{2+}, $g_\parallel = 1.99987$, $g_\perp = 1.99980$, $A_\parallel = \SI{-266.6}{\mega\hertz}$, $A_\perp = \SI{-268.4}{\mega\hertz}$, and the higher-order parameters of the zero-field splitting are given as $B^0_2=\SI{-137.5}{\mega\hertz}$, $B^0_4=\SI{-60}{\kilo\hertz}$, and $B^4_4=\SI{-575}{\kilo\hertz}$.

\clearpage
\bibliography{Bibliography}

@book{AbragamBleaney.1970,
  author    = {Abragam, A. and Bleaney, B.},
  title     = {Electron Paramagnetic Resonance of Transition Ions},
  publisher = {Clarendon Press},
  address   = {Oxford},
  year      = {1970}
}

@article{Afzelius.2013,
  author  = {Afzelius, M. and Sangouard, N. and Johansson, G. and Staudt, M. U. and Wilson, C. M.},
  title   = {Proposal for a coherent quantum memory for propagating microwave photons},
  journal = {New J. Phys.},
  volume  = {15},
  number  = {6},
  pages   = {065008},
  year    = {2013},
  doi     = {10.1088/1367-2630/15/6/065008}
}

@article{An.2025,
  author  = {An, Xiantong and Sun, Weiye and Xu, Zhehao and Xiao, Wanting and Ren, Miaomiao and Guo, Mucheng and Liu, Shuping and Wang, Fudong and Zhong, Manjin},
  title   = {Optical coherence and hyperfine structure of the $^{7}{F}_{0}\leftrightarrow{}^{5}{D}_{0}$ transition in {Eu}$^{3+}$:{CaWO}$_4$},
  journal = {J. Lumin.},
  volume  = {281},
  pages   = {121209},
  year    = {2025},
  doi     = {10.1016/j.jlumin.2025.121209}
}

@article{Antipin.1968,
  author  = {Antipin, A. A. and Katyshev, A. N. and Kurkin, I. N. and Shekun, L. Ya.},
  title   = {Paramagnetic resonance and spin-lattice relaxation of {Er}$^{3+}$ and {Tb}$^{3+}$ ions in the crystal lattice of {CaWO}$_4$},
  journal = {Sov. Phys. Solid State},
  volume  = {10},
  pages   = {468},
  year    = {1968}
}

@article{Baker.1969,
  author  = {Baker, J. M. and Blake, W. B. J. and Copland, G. M.},
  title   = {{ENDOR} of $^{171}${Yb}$^{3+}$ and $^{173}${Yb}$^{3+}$ on cubic sites in calcium fluoride},
  journal = {Proc. R. Soc. Lond. A},
  volume  = {309},
  number  = {1496},
  pages   = {119--139},
  year    = {1969},
  doi     = {10.1098/rspa.1969.0032}
}

@article{Baker.1970,
  author  = {Baker, J. M. and Blake, W. B. J.},
  title   = {{ENDOR} of $^{173}${Yb}$^{3+}$ on tetragonal sites in calcium fluoride: determination of the crystal field parameters},
  journal = {Proc. R. Soc. Lond. A},
  volume  = {316},
  number  = {1524},
  pages   = {63--80},
  year    = {1970},
  doi     = {10.1098/rspa.1970.0066}
}

@article{Becker.2025,
  author  = {Becker, Fabian and Curtin, Catherine L. and KC, Sudip and Schneider, Tim and Sauerzopf, Lorenz J. J. and Elzeiny, Ibrahim and M{\"u}ller, Kai},
  title   = {Spectroscopic investigations of multiple environments in {Er}:{CaWO}$_4$ through charge imbalance},
  journal = {Phys. Rev. Mater.},
  volume  = {9},
  number  = {7},
  pages   = {076203},
  year    = {2025},
  doi     = {10.1103/6srk-3k4n}
}

@article{Beloy.2008a,
  author  = {Beloy, K. and Derevianko, A. and Johnson, W. R.},
  title   = {Hyperfine structure of the metastable $^{3}{P}_{2}$ state of alkaline-earth-metal atoms as an accurate probe of nuclear magnetic octupole moments},
  journal = {Phys. Rev. A},
  volume  = {77},
  pages   = {012512},
  year    = {2008},
  doi     = {10.1103/PhysRevA.77.012512}
}

@article{Beloy.2008b,
  author  = {Beloy, K. and Derevianko, A. and Dzuba, V. A. and Howell, G. T. and Blinov, B. B. and Fortson, E. N.},
  title   = {Nuclear magnetic octupole moment and the hyperfine structure of the $5{D}_{3/2,5/2}$ states of the {Ba}$^{+}$ ion},
  journal = {Phys. Rev. A},
  volume  = {77},
  pages   = {052503},
  year    = {2008},
  doi     = {10.1103/PhysRevA.77.052503}
}

@article{Bertaina.2009,
  author  = {Bertaina, S. and Shim, J. H. and Gambarelli, S. and Malkin, B. Z. and Barbara, B.},
  title   = {Spin-orbit qubits of rare-earth-metal ions in axially symmetric crystal fields},
  journal = {Phys. Rev. Lett.},
  volume  = {103},
  pages   = {226402},
  year    = {2009},
  doi     = {10.1103/PhysRevLett.103.226402}
}

@article{Billaud.2025,
  author  = {Billaud, E. and Balembois, L. and Travesedo, J. and Le Dantec, M. and Ran{\v{c}}i{\'c}, M. and Albertinale, E. and Truong, R. and Bertaina, S. and Chaneli{\`e}re, T. and Goldner, P. and Est{\`e}ve, D. and Vion, D. and Flurin, E. and Bertet, P.},
  title   = {Electron paramagnetic resonance spectroscopy of a scheelite crystal using microwave-photon counting},
  journal = {Phys. Rev. Res.},
  volume  = {7},
  pages   = {013011},
  year    = {2025},
  doi     = {10.1103/PhysRevResearch.7.013011}
}

@article{Bleaney.1963,
  author  = {Bleaney, B.},
  title   = {Hyperfine interactions in rare-earth metals},
  journal = {J. Appl. Phys.},
  volume  = {34},
  number  = {4},
  pages   = {1024--1031},
  year    = {1963},
  doi     = {10.1063/1.1729355}
}

@article{Bleaney.1982,
  author  = {Bleaney, B. and Gregg, J. F. and de Oliveira, A. C. and Wells, M. R.},
  title   = {Nuclear magnetic resonance of $^{51}${V} ({$I=7/2$}) in lanthanide vanadates: {II}. {The} nuclear electric quadrupole interaction},
  journal = {J. Phys. C: Solid State Phys.},
  volume  = {15},
  number  = {25},
  pages   = {5293--5303},
  year    = {1982},
  doi     = {10.1088/0022-3719/15/25/019}
}

@article{Bohr.1953,
  author  = {Bohr, Aage and Mottelson, Ben R.},
  title   = {Collective and individual-particle aspects of nuclear structure},
  journal = {Mat. Fys. Medd. Dan. Vid. Selsk.},
  volume  = {27},
  number  = {16},
  pages   = {1--174},
  year    = {1953}
}

@article{Boldyrev.2022,
  author  = {Boldyrev, K. N. and Malkin, B. Z. and Popova, M. N.},
  title   = {Observation of the hyperfine structure and anticrossings of hyperfine levels in the luminescence spectra of {LiYF}$_4$:{Ho}$^{3+}$},
  journal = {Light Sci. Appl.},
  volume  = {11},
  pages   = {245},
  year    = {2022},
  doi     = {10.1038/s41377-022-00933-2}
}

@article{Clauss.2013,
  author  = {Clauss, Conrad and Bothner, Daniel and Koelle, Dieter and Kleiner, Reinhold and Bogani, Lapo and Scheffler, Marc and Dressel, Martin},
  title   = {Broadband electron spin resonance from 500~{MHz} to 40~{GHz} using superconducting coplanar waveguides},
  journal = {Appl. Phys. Lett.},
  volume  = {102},
  number  = {16},
  pages   = {162601},
  year    = {2013},
  doi     = {10.1063/1.4802956}
}

@article{Enrique.1971,
  author  = {Enrique, Bernal G.},
  title   = {Optical spectrum and magnetic properties of {Er}$^{3+}$ in {CaWO}$_4$},
  journal = {J. Chem. Phys.},
  volume  = {55},
  number  = {5},
  pages   = {2538--2549},
  year    = {1971},
  doi     = {10.1063/1.1676445}
}

@article{Erickson.1986,
  author  = {Erickson, L. E.},
  title   = {Lattice and electronic contributions to the quadrupole interaction of trivalent europium},
  journal = {Phys. Rev. B},
  volume  = {34},
  number  = {1},
  pages   = {36--40},
  year    = {1986},
  doi     = {10.1103/PhysRevB.34.36}
}

@article{Fesquet.2024,
  author  = {Fesquet, Florian and Kronowetter, Fabian and Renger, Michael and Yam, Wun Kwan and Gandorfer, Simon and Inomata, Kunihiro and Nakamura, Yasunobu and Marx, Achim and Gross, Rudolf and Fedorov, Kirill G.},
  title   = {Demonstration of microwave single-shot quantum key distribution},
  journal = {Nat. Commun.},
  volume  = {15},
  number  = {1},
  pages   = {7544},
  year    = {2024},
  doi     = {10.1038/s41467-024-51421-7}
}

@article{Garrett.2022,
  author  = {Garrett, Paul E. and Zieli{\'n}ska, Magda and Cl{\'e}ment, Emmanuel},
  title   = {An experimental view on shape coexistence in nuclei},
  journal = {Prog. Part. Nucl. Phys.},
  volume  = {124},
  pages   = {103931},
  year    = {2022},
  doi     = {10.1016/j.ppnp.2021.103931}
}

@article{Gerginov.2003,
  author  = {Gerginov, V. and Derevianko, A. and Tanner, C. E.},
  title   = {Observation of the nuclear magnetic octupole moment of $^{133}${Cs}},
  journal = {Phys. Rev. Lett.},
  volume  = {91},
  pages   = {072501},
  year    = {2003},
  doi     = {10.1103/PhysRevLett.91.072501}
}

@article{Gupta.1973,
  author  = {Gupta, R. P. and Sen, S. K.},
  title   = {{Sternheimer} shielding-antishielding; rare-earth ions},
  journal = {Phys. Rev. A},
  volume  = {7},
  number  = {3},
  pages   = {850--858},
  year    = {1973},
  doi     = {10.1103/PhysRevA.7.850}
}

@article{Hempstead.1960,
  author  = {Hempstead, C. F. and Bowers, K. D.},
  title   = {Paramagnetic resonance of impurities in {CaWO}$_4$. {I}. {Two} {S}-state ions},
  journal = {Phys. Rev.},
  volume  = {118},
  number  = {1},
  pages   = {131--134},
  year    = {1960},
  doi     = {10.1103/PhysRev.118.131}
}

@article{Jaccarino.1954,
  author  = {Jaccarino, V. and King, J. G. and Satten, R. A. and Stroke, H. H.},
  title   = {Hyperfine structure of {I}$^{127}$. {Nuclear} magnetic octupole moment},
  journal = {Phys. Rev.},
  volume  = {94},
  number  = {6},
  pages   = {1798--1799},
  year    = {1954},
  doi     = {10.1103/PhysRev.94.1798}
}

@article{Kirton.1964,
  author  = {Kirton, J. and Newman, R. C.},
  title   = {Paramagnetic resonance of trivalent ytterbium ions in calcium tungstate},
  journal = {Phys. Lett.},
  volume  = {10},
  number  = {3},
  pages   = {277--278},
  year    = {1964},
  doi     = {10.1016/0031-9163(64)90500-1}
}

@article{Kirton.1965,
  author  = {Kirton, J.},
  title   = {Paramagnetic resonance of trivalent holmium ions in calcium tungstate},
  journal = {Phys. Rev.},
  volume  = {139},
  number  = {6A},
  pages   = {A1930--A1933},
  year    = {1965},
  doi     = {10.1103/PhysRev.139.A1930}
}

@article{Klimin.2010,
  author  = {Klimin, S. A. and Pytalev, D. S. and Popova, M. N. and Malkin, B. Z. and Vanyunin, M. V. and Korableva, S. L.},
  title   = {High-resolution optical spectroscopy of {Tm}$^{3+}$ ions in {LiYF}$_4$: {Crystal}-field energies, hyperfine and deformation splittings, and the isotopic structure},
  journal = {Phys. Rev. B},
  volume  = {81},
  pages   = {045113},
  year    = {2010},
  doi     = {10.1103/PhysRevB.81.045113}
}

@misc{Kukharchyk.2019,
  author        = {Kukharchyk, N. and Sholokhov, D. and Kalachev, A. A. and Bushev, P. A.},
  title         = {Enhancement of optical coherence in $^{167}${Er}:{Y}$_2${SiO}$_5$ crystal at sub-{Kelvin} temperatures},
  year          = {2019},
  eprint        = {1910.03096},
  archivePrefix = {arXiv},
  primaryClass  = {quant-ph}
}

@article{Kukharchyk.2020,
  author  = {Kukharchyk, Nadezhda and Sholokhov, Dmitriy and Morozov, Oleg and Korableva, Stella L. and Kalachev, Alexey A. and Bushev, Pavel A.},
  title   = {Electromagnetically induced transparency in a mono-isotopic $^{167}${Er}:$^{7}${LiYF}$_4$ crystal below 1~{Kelvin}: Microwave photonics approach},
  journal = {Opt. Express},
  volume  = {28},
  number  = {20},
  pages   = {29166},
  year    = {2020},
  doi     = {10.1364/OE.400222}
}

@article{Kunkel.2018,
  author  = {Kunkel, Nathalie and Goldner, Philippe},
  title   = {Recent advances in rare earth doped inorganic crystalline materials for quantum information processing},
  journal = {Z. Anorg. Allg. Chem.},
  volume  = {644},
  number  = {2},
  pages   = {66--76},
  year    = {2018},
  doi     = {10.1002/zaac.201700425}
}

@article{LeDantec.2021,
  author  = {Le Dantec, Marianne and Ran{\v{c}}i{\'c}, Milo{\v{s}} and Lin, Sen and Billaud, Eric and Ranjan, Vishal and Flanigan, Daniel and Bertaina, Sylvain and Chaneli{\`e}re, Thierry and Goldner, Philippe and Erb, Andreas and Liu, Ren Bao and Est{\`e}ve, Daniel and Vion, Denis and Flurin, Emmanuel and Bertet, Patrice},
  title   = {Twenty-three-millisecond electron spin coherence of erbium ions in a natural-abundance crystal},
  journal = {Sci. Adv.},
  volume  = {7},
  number  = {51},
  pages   = {eabj9786},
  year    = {2021},
  doi     = {10.1126/sciadv.abj9786}
}

@article{Lewty.2012,
  author  = {Lewty, N. C. and Chuah, B. L. and Cazan, R. and Sahoo, B. K. and Barrett, M. D.},
  title   = {Spectroscopy on a single trapped $^{137}${Ba}$^{+}$ ion for nuclear magnetic octupole moment determination},
  journal = {Opt. Express},
  volume  = {20},
  number  = {19},
  pages   = {21379--21384},
  year    = {2012},
  doi     = {10.1364/OE.20.021379}
}

@article{Lindgard.1974,
  author  = {Lindg{\aa}rd, P.-A. and Danielsen, O.},
  title   = {{Bose}-operator expansions of tensor operators in the theory of magnetism},
  journal = {J. Phys. C: Solid State Phys.},
  volume  = {7},
  number  = {8},
  pages   = {1523--1535},
  year    = {1974},
  doi     = {10.1088/0022-3719/7/8/017}
}

@article{Longdell.2006,
  author  = {Longdell, J. J. and Alexander, A. L. and Sellars, M. J.},
  title   = {Characterization of the hyperfine interaction in europium-doped yttrium orthosilicate and europium chloride hexahydrate},
  journal = {Phys. Rev. B},
  volume  = {74},
  pages   = {195101},
  year    = {2006},
  doi     = {10.1103/PhysRevB.74.195101}
}

@article{Malkin.2012,
  author  = {Malkin, B. Z. and Pytalev, D. S. and Popova, M. N. and Baibekov, E. I. and Falin, M. L. and Gerasimov, K. I. and Khaidukov, N. M.},
  title   = {Random lattice deformations in rare-earth-doped cubic hexafluoroelpasolites: {High}-resolution optical spectroscopy and theoretical studies},
  journal = {Phys. Rev. B},
  volume  = {86},
  pages   = {134110},
  year    = {2012},
  doi     = {10.1103/PhysRevB.86.134110}
}

@article{Malkin.2017,
  author  = {Malkin, B. Z. and Abishev, N. M. and Baibekov, E. I. and Pytalev, D. S. and Boldyrev, K. N. and Popova, M. N. and Bettinelli, M.},
  title   = {Distribution function of random strains in an elastically anisotropic continuum and defect strengths of {Tm}$^{3+}$ impurity ions in crystals with zircon structure},
  journal = {Phys. Rev. B},
  volume  = {96},
  pages   = {014116},
  year    = {2017},
  doi     = {10.1103/PhysRevB.96.014116}
}

@article{Marsh.2026,
  author  = {Marsh, Lewin and Yikai, Yang and Mattiroli, Cesare and Sayat, Mikhael T. and Tr{\'{\i}}, {\DJ{}}{\`a}m Minh and R{\o}nnow, Henrik M. and Longdell, Jevon J. and Soh, Jian-Rui},
  title   = {Nuclear quadrupole interaction and zero first-order {Zeeman} transitions of $^{167}${Er}$^{3+}$ in {CaWO}$_4$},
  journal = {Phys. Rev. B},
  volume  = {113},
  number  = {5},
  pages   = {054417},
  year    = {2026},
  doi     = {10.1103/pg5m-hk2n}
}

@article{McAuslan.2012,
  author  = {McAuslan, D. L. and Bartholomew, J. G. and Sellars, M. J. and Longdell, J. J.},
  title   = {Reducing decoherence in optical and spin transitions in rare-earth-metal-ion--doped materials},
  journal = {Phys. Rev. A},
  volume  = {85},
  pages   = {032339},
  year    = {2012},
  doi     = {10.1103/PhysRevA.85.032339}
}

@article{Mims.1965,
  author  = {Mims, W. B.},
  title   = {Electric field shift in paramagnetic resonance for four ions in a calcium tungstate lattice},
  journal = {Phys. Rev.},
  volume  = {140},
  number  = {2A},
  pages   = {A531--A535},
  year    = {1965},
  doi     = {10.1103/PhysRev.140.A531}
}

@article{Mohammady.2010,
  author  = {Mohammady, M. H. and Morley, G. W. and Monteiro, T. S.},
  title   = {Bismuth qubits in silicon: The role of {EPR} cancellation resonances},
  journal = {Phys. Rev. Lett.},
  volume  = {105},
  pages   = {067602},
  year    = {2010},
  doi     = {10.1103/PhysRevLett.105.067602}
}

@article{Mohammady.2012,
  author  = {Mohammady, M. H. and Morley, G. W. and Nazir, A. and Monteiro, T. S.},
  title   = {Analysis of quantum coherence in bismuth-doped silicon: A system of strongly coupled spin qubits},
  journal = {Phys. Rev. B},
  volume  = {85},
  number  = {9},
  pages   = {094404},
  year    = {2012},
  doi     = {10.1103/PhysRevB.85.094404}
}

@article{Momma.2011,
  author  = {Momma, Koichi and Izumi, Fujio},
  title   = {{VESTA 3} for three-dimensional visualization of crystal, volumetric and morphology data},
  journal = {J. Appl. Crystallogr.},
  volume  = {44},
  number  = {6},
  pages   = {1272--1276},
  year    = {2011},
  doi     = {10.1107/S0021889811038970}
}

@article{Mostafanejad.2014,
  author  = {Mostafanejad, Mohammad},
  title   = {Basics of the spin {Hamiltonian} formalism},
  journal = {Int. J. Quantum Chem.},
  volume  = {114},
  number  = {22},
  pages   = {1495--1512},
  year    = {2014},
  doi     = {10.1002/qua.24721}
}

@article{Nassau.1963,
  author  = {Nassau, K. and Loiacono, G. M.},
  title   = {Calcium tungstate---{III}: {Trivalent} rare earth substitution},
  journal = {J. Phys. Chem. Solids},
  volume  = {24},
  number  = {12},
  pages   = {1503--1510},
  year    = {1963},
  doi     = {10.1016/0022-3697(63)90090-8}
}

@article{Nehrkorn.2015,
  author  = {Nehrkorn, Joscha and Schnegg, Alexander and Holldack, Karsten and Stoll, Stefan},
  title   = {General magnetic transition dipole moments for electron paramagnetic resonance},
  journal = {Phys. Rev. Lett.},
  volume  = {114},
  number  = {1},
  pages   = {010801},
  year    = {2015},
  doi     = {10.1103/PhysRevLett.114.010801}
}

@article{Ortu.2018,
  author  = {Ortu, Antonio and Tiranov, Alexey and Welinski, Sacha and Fr{\"o}wis, Florian and Gisin, Nicolas and Ferrier, Alban and Goldner, Philippe and Afzelius, Mikael},
  title   = {Simultaneous coherence enhancement of optical and microwave transitions in solid-state electronic spins},
  journal = {Nat. Mater.},
  volume  = {17},
  number  = {8},
  pages   = {671--675},
  year    = {2018},
  doi     = {10.1038/s41563-018-0138-x}
}

@article{Ourari.2023,
  author  = {Ourari, Salim and Dusanowski, {\L}ukasz and Horvath, Sebastian P. and Uysal, Mehmet T. and Phenicie, Christopher M. and Stevenson, Paul and Raha, Mouktik and Chen, Songtao and Cava, Robert J. and de Leon, Nathalie P. and Thompson, Jeff D.},
  title   = {Indistinguishable telecom band photons from a single {Er} ion in the solid state},
  journal = {Nature},
  volume  = {620},
  pages   = {977--981},
  year    = {2023},
  doi     = {10.1038/s41586-023-06281-4}
}

@article{Probst.2015,
  author  = {Probst, S. and Rotzinger, H. and Ustinov, A. V. and Bushev, P. A.},
  title   = {Microwave multimode memory with an erbium spin ensemble},
  journal = {Phys. Rev. B},
  volume  = {92},
  pages   = {014421},
  year    = {2015},
  doi     = {10.1103/PhysRevB.92.014421}
}

@article{Rakonjac.2020,
  author  = {Rakonjac, Jelena V. and Chen, Yu-Hui and Horvath, Sebastian P. and Longdell, Jevon J.},
  title   = {Long spin coherence times in the ground state and in an optically excited state of $^{167}${Er}$^{3+}$:{Y}$_2${SiO}$_5$ at zero magnetic field},
  journal = {Phys. Rev. B},
  volume  = {101},
  pages   = {184430},
  year    = {2020},
  doi     = {10.1103/PhysRevB.101.184430}
}

@article{Ram.1985,
  author  = {Ram, K. and Sharma, K. K.},
  title   = {Enhanced nuclear magnetic moments in {HoF}$_3$},
  journal = {J. Phys. C: Solid State Phys.},
  volume  = {18},
  number  = {3},
  pages   = {619--624},
  year    = {1985},
  doi     = {10.1088/0022-3719/18/3/012}
}

@article{Ranon.1964,
  author  = {Ranon, U. and Volterra, V.},
  title   = {Paramagnetic resonance of {Yb}$^{3+}$ in {CaWO}$_4$},
  journal = {Phys. Rev.},
  volume  = {134},
  number  = {6A},
  pages   = {A1483--A1485},
  year    = {1964},
  doi     = {10.1103/PhysRev.134.A1483}
}

@article{Rochman.2023,
  author  = {Rochman, Jake and Xie, Tian and Bartholomew, John G. and Schwab, K. C. and Faraon, Andrei},
  title   = {Microwave-to-optical transduction with erbium ions coupled to planar photonic and superconducting resonators},
  journal = {Nat. Commun.},
  volume  = {14},
  pages   = {1153},
  year    = {2023},
  doi     = {10.1038/s41467-023-36799-0}
}

@article{Rudowicz.2001,
  author  = {Rudowicz, Czes{\l}aw and Misra, Sushil K.},
  title   = {Spin-{Hamiltonian} formalisms in electron magnetic resonance ({EMR}) and related spectroscopies},
  journal = {Appl. Spectrosc. Rev.},
  volume  = {36},
  number  = {1},
  pages   = {11--63},
  year    = {2001},
  doi     = {10.1081/ASR-100103089}
}

@article{Rudowicz.2015,
  author  = {Rudowicz, Czes{\l}aw and Karbowiak, Miros{\l}aw},
  title   = {Disentangling intricate web of interrelated notions at the interface between the physical (crystal field) {Hamiltonians} and the effective (spin) {Hamiltonians}},
  journal = {Coord. Chem. Rev.},
  volume  = {287},
  pages   = {28--63},
  year    = {2015},
  doi     = {10.1016/j.ccr.2014.12.006}
}

@article{Saglamyurek.2015,
  author  = {Saglamyurek, Erhan and Jin, Jeongwan and Verma, Varun B. and Shaw, Matthew D. and Marsili, Francesco and Nam, Sae Woo and Oblak, Daniel and Tittel, Wolfgang},
  title   = {Quantum storage of entangled telecom-wavelength photons in an erbium-doped optical fibre},
  journal = {Nat. Photonics},
  volume  = {9},
  number  = {2},
  pages   = {83--87},
  year    = {2015},
  doi     = {10.1038/nphoton.2014.311}
}

@article{Sattler.1970a,
  author  = {Sattler, J. P. and Nemarich, J.},
  title   = {Electron paramagnetic resonance of {Yb}$^{3+}$ in scheelite single crystals},
  journal = {Phys. Rev. B},
  volume  = {1},
  number  = {11},
  pages   = {4249--4256},
  year    = {1970},
  doi     = {10.1103/PhysRevB.1.4249}
}

@article{Sattler.1970b,
  author  = {Sattler, J. P. and Nemarich, J.},
  title   = {Unusual electron paramagnetic resonance hyperfine spectra of {Yb}$^{3+}$ in scheelites},
  journal = {Phys. Rev. B},
  volume  = {1},
  number  = {11},
  pages   = {4256--4261},
  year    = {1970},
  doi     = {10.1103/PhysRevB.1.4256}
}

@article{Schwartz.1955,
  author  = {Schwartz, Charles},
  title   = {Theory of hyperfine structure},
  journal = {Phys. Rev.},
  volume  = {97},
  number  = {2},
  pages   = {380--395},
  year    = {1955},
  doi     = {10.1103/PhysRev.97.380}
}

@book{Schweiger.2001,
  author    = {Schweiger, Arthur and Jeschke, Gunnar},
  title     = {Principles of Pulse Electron Paramagnetic Resonance},
  publisher = {Oxford University Press},
  address   = {Oxford},
  year      = {2001},
  doi       = {10.1093/oso/9780198506348.001.0001}
}

@article{Shakurov.2014,
  author  = {Shakurov, G. S. and Chukalina, E. P. and Popova, M. N. and Malkin, B. Z. and Tkachuk, A. M.},
  title   = {Random strain effects in optical and {EPR} spectra of electron-nuclear excitations in {CaWO}$_4$:{Ho}$^{3+}$ single crystals},
  journal = {Phys. Chem. Chem. Phys.},
  volume  = {16},
  number  = {45},
  pages   = {24727--24738},
  year    = {2014},
  doi     = {10.1039/C4CP03437F}
}

@article{Sternheimer.1986,
  author  = {Sternheimer, R. M.},
  title   = {Shielding and antishielding of nuclear quadrupole moments},
  journal = {Z. Naturforsch. A},
  volume  = {41},
  number  = {1--2},
  pages   = {24--26},
  year    = {1986},
  doi     = {10.1515/zna-1986-1-207}
}

@techreport{Stone.2014,
  author      = {Stone, N. J.},
  title       = {Table of Nuclear Magnetic Dipole and Electric Quadrupole Moments},
  institution = {International Atomic Energy Agency, Nuclear Data Section},
  address     = {Vienna},
  type        = {Report},
  number      = {INDC(NDS)-0658},
  year        = {2014},
  doi         = {10.61092/iaea.zhrz-3g5j}
}

@article{Strinic.2025,
  author  = {Strini{\'c}, Ana and Oehrl, Patricia and Marx, Achim and Bushev, Pavel A. and Huebl, Hans and Gross, Rudolf and Kukharchyk, Nadezhda},
  title   = {Broadband electron paramagnetic resonance spectroscopy of $^{167}${Er}:$^{7}${LiYF}$_4$ at millikelvin temperatures},
  journal = {Phys. Rev. B},
  volume  = {111},
  pages   = {214430},
  year    = {2025},
  doi     = {10.1103/5j7f-wfyl}
}

@phdthesis{Strinic.2026,
  author = {Strini{\'c}, Ana},
  title  = {Broadband Electron Paramagnetic Resonance Spectroscopy of Rare-Earth Spin Ensembles},
  school = {Technical University of Munich},
  year   = {2026}
}

@misc{Supplement,
  note = {See Supplementary Information for details.}
}

@article{Teplov.1968,
  author  = {Teplov, M. A.},
  title   = {Magnetic resonance on $^{141}${Pr} nuclei in a {Pr}$_2$({SO}$_4$)$_3\cdot$8{H}$_2${O} single crystal},
  journal = {Sov. Phys. JETP},
  volume  = {26},
  number  = {5},
  pages   = {872},
  year    = {1968},
  note    = {[Zh. Eksp. Teor. Fiz. \textbf{53}, 1510 (1967)]}
}

@article{Tiranov.2026,
  author  = {Tiranov, Alexey and Green, Emanuel and Hermans, Sophie and Liu, Erin and Chiossi, Federico and Serrano, Diana and Loiseau, Pascal and Manoj Kumar, Achuthan and Bertaina, Sylvain and Faraon, Andrei and Goldner, Philippe},
  title   = {Sub-second spin and lifetime-limited optical coherences in $^{171}${Yb}$^{3+}$:{CaWO}$_4$},
  journal = {Nat. Commun.},
  volume  = {17},
  pages   = {4115},
  year    = {2026},
  doi     = {10.1038/s41467-026-70534-9}
}

@article{Travesedo.2025,
  author  = {Travesedo, Jaime and O'Sullivan, James and Pallegoix, Louis and Huang, Zhiyuan W. and Hogan, Patrick and Goldner, Philippe and Chaneli{\`e}re, Thierry and Bertaina, Sylvain and Est{\`e}ve, Daniel and Abgrall, Patrick and Vion, Denis and Flurin, Emmanuel and Bertet, Patrice},
  title   = {All-microwave spectroscopy and polarization of individual nuclear spins in a solid},
  journal = {Sci. Adv.},
  volume  = {11},
  number  = {10},
  pages   = {eadu0581},
  year    = {2025},
  doi     = {10.1126/sciadv.adu0581}
}

@article{Verney.2025,
  author  = {Verney, David},
  title   = {History of the concept of nuclear shape},
  journal = {Eur. Phys. J. A},
  volume  = {61},
  number  = {4},
  pages   = {82},
  year    = {2025},
  doi     = {10.1140/epja/s10050-025-01545-1}
}

@article{Vogl.2026,
  author  = {Vogl, Lukas and Huber, Gerhard B. P. and Strini{\'c}, Ana and Marx, Achim and Filipp, Stefan and Fedorov, Kirill G. and Gross, Rudolf and Kukharchyk, Nadezhda P.},
  title   = {Experimental setup for the combined study of spin ensembles and superconducting quantum circuits},
  journal = {Cryogenics},
  volume  = {159},
  pages   = {104378},
  year    = {2026},
  doi     = {10.1016/j.cryogenics.2026.104378}
}

@article{Yam.2026,
  author  = {Yam, W. K. and Gandorfer, S. and Fesquet, F. and Handschuh, M. and Honasoge, K. E. and Marx, A. and Gross, R. and Fedorov, K. G.},
  title   = {Quantum teleportation over thermal microwave network},
  journal = {Phys. Rev. Lett.},
  volume  = {136},
  number  = {18},
  pages   = {180801},
  year    = {2026},
  doi     = {10.1103/n8j9-63mz}
}

@article{Zheng.2004,
  author  = {Zheng, Wen-Chen and Dong, Hui-Ning and Wu, Xiao-Xuan and Tang, Sheng},
  title   = {Investigations of the optical spectra and {EPR} parameters in {CaWO}$_4$:{Yb}$^{3+}$},
  journal = {Spectrochim. Acta A},
  volume  = {60},
  number  = {13},
  pages   = {3169--3171},
  year    = {2004},
  doi     = {10.1016/j.saa.2004.02.032}
}

@article{Zhong.2015,
  author  = {Zhong, Manjin and Hedges, Morgan P. and Ahlefeldt, Rose L. and Bartholomew, John G. and Beavan, Sarah E. and Wittig, Sven M. and Longdell, Jevon J. and Sellars, Matthew J.},
  title   = {Optically addressable nuclear spins in a solid with a six-hour coherence time},
  journal = {Nature},
  volume  = {517},
  pages   = {177--180},
  year    = {2015},
  doi     = {10.1038/nature14025}
}

@article{Zhou.2023,
  author  = {Zhou, Zong-Quan and Liu, Chao and Li, Chuan-Feng and Guo, Guang-Can and Oblak, Daniel and Lei, Mi and Faraon, Andrei and Mazzera, Margherita and de Riedmatten, Hugues},
  title   = {Photonic integrated quantum memory in rare-earth doped solids},
  journal = {Laser Photonics Rev.},
  volume  = {17},
  number  = {10},
  pages   = {2300257},
  year    = {2023},
  doi     = {10.1002/lpor.202300257}
}

\end{document}